\documentclass[review]{elsarticle}
\usepackage{setspace}
\usepackage{graphicx,psfrag,textcomp, amsthm}
\usepackage{enumerate, dsfont}
\usepackage{url,xcolor,natbib} 
\usepackage{booktabs, subfig, bm, paralist,mathpazo, todonotes,longtable}
\usepackage{color,soul,amsfonts,setspace,mathrsfs,mathtools,longtable,multirow}
\usepackage{tikz}
\usetikzlibrary{shapes.geometric, arrows}
\usepackage{ragged2e}
\usepackage[pdftex,colorlinks=true]{hyperref}
\definecolor{darkblue}{rgb}{0,0,.6}
\hypersetup{citecolor=darkblue,linkcolor=darkblue,urlcolor=darkblue}

\newcommand{\X}{\mathcal{X}}
\newcommand{\Y}{\mathcal{Y}}

\graphicspath{{plots/}}

\journal{Journal of Hydrology}

\biboptions{authoryear}

\begin{document}

\begin{frontmatter}

\title{A functional autoregressive model based on exogenous hydrometeorological variables for river flow prediction}

\author[mymainaddress]{Ufuk Beyaztas}
\ead{ufuk.beyaztas@marmara.edu.tr}

\author[mysecondaddress]{Han Lin Shang}
\ead{hanlin.shang@mq.edu.au}

\author[mythirdaddress]{Zaher Mundher Yaseen\corref{mycorrespondingauthor}}
\cortext[mycorrespondingauthor]{Corresponding author}
\ead{yaseen@alayen.edu.iq}

\address[mymainaddress]{Department of Statistics, Marmara University, Istanbul, Turkey}
\address[mysecondaddress]{Department of Actuarial Studies and Business Analytics, Macquarie University, Sydney, Australia}
\address[mythirdaddress]{New era and development in civil engineering research group, Scientific Research Center, Al-Ayen University, Thi-Qar, 64001, Iraq}

\begin{abstract}
In this research, a functional time series model was introduced to predict future realizations of river flow time series. The proposed model was constructed based on a functional time series's correlated lags and the essential exogenous climate variables. Rainfall, temperature, and evaporation variables were hypothesized to have substantial functionality in river flow simulation. Because an actual time series model is unspecified and the input variables' significance for the learning process is unknown in practice, it was employed a variable selection procedure to determine only the significant variables for the model. A nonparametric bootstrap model was also proposed to investigate predictions' uncertainty and construct pointwise prediction intervals for the river flow curve time series. Historical datasets at three meteorological stations (Mosul, Baghdad, and Kut) located in the semi-arid region, Iraq, were used for model development. The prediction performance of the proposed model was validated against existing functional and traditional time series models. The numerical analyses revealed that the proposed model provides competitive or even better performance than the benchmark models. Also, the incorporated exogenous climate variables have substantially improved the modeling predictability performance. Overall, the proposed model indicated a reliable methodology for modeling river flow within the semi-arid region.
\end{abstract}


\end{frontmatter}


\section{Introduction}

\subsection{Research background}

A comprehensive range planning and management of water resources demand an imperative development of an optimum river flow prediction model since river flow prediction are one of the essential stochastic features in the hydrology cycle \citep{Yu2020, rathinasamy2014wavelet}. River flow prediction has benefits to several hydrological engineering aspects. These include water operation, management, maintenance, irrigation and agricultural management, flood \& drought warning systems \citep{Yaseen2018, maier2010models}. The evaluation of river flow models' performance requires a specific period, which may be prolonged (e.g., seasonal, monthly, or weekly) or a short period (e.g., daily or hourly). River flow prediction has attracted most researchers' attention in water resources over the last two decades \citep{Yaseen2016}. Achieving a precise and reliable river flow prediction model is difficult and challenging task due to the chaotic attributes of river flow \citep{Wagena2020}. River flow time series exhibits non-linearity, stochasticity, and non-stationarity, which controls its behavior and account for its complexity \citep{Chang2001}. Also, river flow is influenced by several other factors, such as climatic changes \citep{Stockinger2017}, local \& seasonal patterns, variation in local and regional temperature, and the annual rate of rainfall. The other factors include temporal and spatial watershed variability, features of the catchments, and human activities \citep{Maier2001}. A univariate time series forecasting for river flow is not always the ideal option for providing a reliable expert control system. In comparison with with the univariate models, multivariate modeling procedures that is incorporating several essential exogenous hydrological variables may produce better results in river flow prediction.

\subsection{River flow prediction complexity}

Despite these challenges facing the development of a precise model for river flow prediction, scientists are still striving to develop reliable and accurate river flow prediction models that could address the complexities in river flow prediction \citep{Yaseen2015b}. Hydrologists are mainly relying on river flow prediction to come up with the sustainable theoretical basis of water infrastructures in flood monitoring and measurement by observing river flow patterns \citep{Murphy2019}. Owing to the non-linear relationship between the targeted river flow ``output variable'' and the various hydrological variables ``input variables'', the traditional regression-based models have showed a noticable limitations for river flow data modeling \citep{Yaseen2015b}. Hence, it is necessary to evaluate and extract the non-linear relationships between the predictors and the predictand to improve the employed models' capability for river flow data prediction.

\subsection{Research motivation and enthusiasm}

Based on the reported literature, the conventional regression and Box-Jenkins-based statistical approaches are mainly employed for modeling and analysis hydrological time series \citep{Amisigo2008}. However, the past two decades witnessed an increased interest in the use of data-driven models. They depend mainly on artificial intelligence (AI) for data pattern extraction to predict future river flow data \citep{Afan2020}. These data-driven models have proven reliable prediction tools to generate the estimated river flow data representing the actual river flow data \citep{Diop2018, Fu2020, Zhu2020}. Several AI-based techniques have been applied to various prediction studies for several reasons. For instance, AI-based models have shown good performance with low complexity and are applicable for solving highly stochastic problems \citep{yaseen2021insight, tao2021intelligent}. 

Furthermore, AI-based models are data inexpensive and can be easily used to design prediction models and other related applications \citep{Yaseen2020}. The merits and weaknesses of the AI-models have recently been reviewed and explored in the water resources field \citep{Nourani2014, Yaseen2015b, zounemat2020neurocomputing}. From the reviews, it has been concluded that no absolute AI model exists, which is applicable to model the hydrological processes in the forms (i.e., classification, optimization, estimation, and prediction). Further, there is no single data-driven technique that has been certified appropriately for all definite problems. Nevertheless, the AI models (with data pre-processing techniques) can be improved by combining hybrid AI models with data pre-processing approaches \citep{cui2020newly, yaseen2020hourly}. Wavelet transformation is the commonly used data pre-processing technique in water resources applications \citep{Nourani2014}. The wavelet transformation can enhance the certainty associated with predicted river flow by integrating two or more data pattern modeling approaches. Hence, it is considered a hybrid approach \citep{Fahimi2017}. Despite improving the accuracy of prediction models using some hybrid models \citep{Yaseen2017}, studies are still ongoing and aimed at coming up with a suitable approach that can effectively establish the optimal prediction solutions using the AI models with pre-processing techniques \citep{Roudier2014, fu2020deep}.

The accurate prediction of river flow occurrence with enough lead time has become necessary owing to the high watershed stochasticity. Although river flow prediction can be performed using physical and conceptual models, the conceptual models are the most employed in hydrologic prediction. They are accurate in various hydrological applications, require less information, and can be built with ease \citep{Shortridge2015}. Data collection processes are becoming ambidextrous due to the recent advancements in technology; theses advancements have increased the complexity and dimensionality of data structure \citep{Lange2020}. Having sampled such data type over time and space, they can be considered functional data and can be analyzed, modeled, visualized, and predicted via Functional Data Analysis (FDA) techniques \citep{Rio2018}. As a new area in Statistics, the FDA considers only data presented as curves. This feature removes the problem associated with a high number of variables as the approach focuses mostly on the temporal dependence within the curves during the analysis of the functional data \citep{Stadtmuller2015}. The FDA's usage has several advantages over the classical multivariate statistical analysis models \citep{ferraty2006}. For instance, a functional observation is considered as a whole in the FDA rather than a single dependent variable. The other advantages of the FDA are as follows: \begin{inparaenum} \item[i)] provision of further data-related information (such as data smoothness and derivatives); \item[ii)] addresses the issue of missing data by engaging smoothing and interpolation techniques, which significantly reduces the rate of noisy data; and, \item[iii)] applicability to time-series data that is irregularly sampled. \end{inparaenum} In contrast to the classical time-series models, the FDA is not prone to a high correlation between repeated measurements, as it views the whole curve as a single entry. Hence, functional time series (FTS) modeling has been increasingly attractive recently \citep{shang2021functional}.

Recent literature has shown that the FDA models have been successfully used within the field of hydrology. For example, \cite{ShHyd2011} proposed a nonparametric FTS model to forecast monthly sea surface temperatures. \cite{suhaila2011} converted rainfall time series data into smooth curves to describe the uni-modal rainfall patterns. \cite{chebana2012} introduced several FDA tools for flood frequency analysis. \cite{Adham2014} tested the FTS model to transform runoff data into smooth curves representing the surface runoff pattern. \cite{masselot2016} applied several FDA models to analyze the flow volume and the whole river flow curve during a given period using precipitations curves. \cite{ternynck2016} adapted and applied two functional data classification techniques to analyze flood hydrographs. \cite{curceac2019} used a nonparametric FDA model to forecast hourly air temperature up to one day in advance. \cite{Beyaztas2019} proposed a FTS model to construct a reliable predictive strategy of drought interval occurrences. \cite{Hael2020} used FTS model to provide a clear understanding of the rainfall patterns and predict future rainfall curves. The numerical results produced by the studies mentioned above have shown that the FDA-based models perform better than Box-Jenkins and AI-based models in forecasting hydrological time-series data.

\subsection{Research objectives}

The FTS models such as functional autoregressive of order one (FAR(1)) model of \cite{Bosq2000} or functional principal component regression (FPCR) of \cite{HydSh2009} were generally used only the lagged variable(s) as predictor in the modeling phase. However, hydrological time series data, such as river flow, is linked to other hydrometeorological variables, such as temperature, humidity, rainfall, wind speed, and evaporation. Therefore, such models may not produce reliable results in predicting the time series data future values since they ignore other variables that affect the response variable. For example, a recent study, \cite{Tyralis} has showed that exogenous predictors offer improvements in the long run in daily river flow forecasting by performing large-scale benchmark tests using data from 511 catchments, USA. In this study, a novel FTS model was proposed called functional autoregressive with exogenous input of order one (FARX(1)) to predict future river flow curve time series. The proposed model uses the past lags of the functional time series and other hydrometeorological variables. Hence, the proposed model investigates the effects of the exogenous variables on the predicted curve of the river flow.

The FARX(1) model was first proposed by \cite{Damon2002}, and they used a nonparametric procedure to estimate their proposed models. In doing so, the authors projected the functional observations into $q+1$ spaces, where $q$ denotes the number of exogenous variables. They chose the subspace generated by the eigenvectors of the appropriate covariance operator associated with the greatest eigenvalues for each space. In addition, \cite{Chen2018} proposed a FARX(1) model for forecasting natural gas demand and supply. The authors used the Fourier basis expansion method to transform the infinite-dimensional model into a finite-dimensional space and used the maximum likelihood method to estimate the model parameters. A detailed review of FAR and FARX models can be found in \cite{Chen2020}. The studies above have shown that the FARX model produces improved forecasting performance compared with FAR and traditional time series models. However, they have some drawbacks. For example, the method of \cite{Damon2002} requires choosing $q+1$ parameters in the model estimation phase and a complicated cross-validation procedure. Also, the maximum likelihood method in which used by \cite{Chen2018} to estimate the model parameters produces unstable estimates and encounter a singular matrix problem when a large number of exogenous variables are included in the model \citep{matsui2009, Beyaztas2020}. In this article, an extended version of FARX(1) model of \cite{Damon2002} and \cite{Chen2018}, was proposed. The proposed model differs from the FARX(1) models of \cite{Damon2002} and \cite{Chen2018} by using different parameter estimation strategy to improve the prediction performance of the model. 

In summary, the proposed method works as follows; first, it stacks the lagged and exogenous functional variables into a single function. It uses $B$-spline basis expansion to project the infinite-dimensional model into the finite-dimensional space. Second, it applies a partial least squares model on the vector-valued predictors and responses to estimate the model parameters. Partial least squares method allows bypassing the singular matrix problem and increases the prediction accuracy of the FARX(1) model, \citep[see, e.g.,][]{Beyaztas2020}. Finally, a recursion formula is used to calculate future curve time series.

In practice, the true time series model's form is unspecified, and the significant variables for the model are unknown. For this reason, there is a need for variable selection procedure to determine the significant exogenous variables. For this purpose, several variables selection procedures based on several criteria, such as the Akaike information and Bayesian information criteria, have been proposed, \citep[see, e.g.,][]{Damon2015}. However, information criterion-based variable selection procedures may be computationally intensive when there are many exogenous variables in the FARX(1) model. For this reason, a forward procedure to determine the significant exogenous variables was introduced (see Section~\ref{sec:vsp}). The numerical results, which will be discussed in detail in Sections~\ref{sec:MC} and~\ref{sec:results}, reveal that the proposed model produces improved prediction results compared with other existing models.

Point forecast may not provide an accurate inference for a future realization of a time series since the uncertainty associated with the point forecasts are unknown in practice. On the other hand, a prediction interval can provide better statistical inferences considering the uncertainty of each point forecasts \citep{Chatfield1993, Kim2001, HydSh2009}. Traditional prediction intervals computed based on the error variance require some distributional assumptions which are unknown in practice. Furthermore, the construction of a prediction interval may be affected due to any departure from the assumptions and may provide inaccurate results. An alternative way to construct a prediction interval without considering distributional assumptions is using the bootstrap method. In the FTS context, several bootstrap methods have been proposed to construct prediction intervals for the response functions. For example, \cite{HydSh2009} proposed a nonparametric bootstrap approach for the FTS models; \cite{Vilar2018} proposed two bootstrap approaches for the nonparametric autoregressive and partial linear semi-parametric models. In this paper, for further investigation of the prediction uncertainty, an extended version of the bootstrap approach of \cite{HydSh2009} was proposed to construct the pointwise prediction intervals for the response functions under the proposed FARX(1) model (see Section~\ref{sec:boot} for details). The proposed bootstrap approach differs from the method of \cite{HydSh2009} in two respects. First, while the bootstrap method of \cite{HydSh2009} uses the estimated future realizations of the principal component scores in the algorithm, the proposed approach uses a recursion formula. Second, the bootstrap method of \cite{HydSh2009} uses residuals obtained based on the estimated and forecasted principal component scores. The proposed algorithm uses residuals that are computed based on the observed and estimated response functions. This allows capturing the error functions more appropriately since the proposed estimation procedure is different from that of \cite{HydSh2009}. The proposed bootstrap approach also differs from the bootstrap methods of \cite{Vilar2018}. In the proposed method, the response, predictors, and errors are in the form of functions. On the other hand, the methods of \cite{Vilar2018} are based on scalar response and functional predictors.

\section{Methodology}\label{sec:methodology}

Let $\Y(s) = \left\lbrace \Y_t(s);~t=1, 2, \cdots, N \right\rbrace$ denote a functional time series where each functional element $\Y_t(s)$ is defined on a bounded interval $s \in \left[ 0, \mathcal{I} \right]$. Samples of functional time series are assumed to be elements of square-integrable functions residing in Hilbert space $\mathcal{L}_2$. In this study, it was assumed that $\Y_t(s)$ for $t = 1, \cdots, N$, are zero-mean stochastic processes; $\text{E} \left[ \Y_t(s) \right] = 0$. In addition, it was considered the prediction of unobservable future realization of $\Y(s)$, $\left\lbrace Y_h(s);~ h = N+1, N+2, \cdots\right\rbrace$, where $h$ denotes forecast horizon. Let $g(s) = \left\lbrace Y_2(s), \cdots, Y_N(s) \right\rbrace$ and $f(\nu) = \left\lbrace Y_1(\nu), \cdots, Y_{N-1}(\nu) \right\rbrace$ respectively denote the functional response and a functional predictor which is also a lagged variable. Then, the first-order autocorrelation of series $\Y(s)$ can be modeled by FAR(1) as follows:
\begin{equation}\label{eq:far}
g(s) = \int_{\nu} f(\nu) \beta(\nu, s) d \nu + \epsilon(s), \qquad \nu \in [0, \nu],
\end{equation}
where $\beta(\nu, s)$ and $\epsilon(s)$ are the regression coefficient function and the random error function that commonly follows a Gaussian process with mean zero and variance-covariance matrix $\mathbf{\Sigma}$, respectively. Let $\widehat{\beta}(\nu, s)$ denotes an estimate of $\beta(\nu, s)$. Then, the one-step-ahead prediction of $\Y_{N+1}(s)$ can be obtained as follows:
\begin{equation}\label{eq:osf}
\widehat{\Y}_{N+1}(s) = \int_{\mathcal{I}} \Y_N(\nu) \hat{\beta}(\nu, s) d \nu. 
\end{equation}
From~\eqref{eq:far} and~\eqref{eq:osf}, FAR(1) model uses only the lagged variable in the model. On the other hand, other factors may relate to time-series data. Thus, the FAR(1) model may not provide the best results in predicting the unobservable future realization of $\Y(s)$. Thus, the FARX(1) model, which uses the lagged information and other variables, is considered to obtain forecasts of FTS.

To start with, let $\X_m(s) = \left\lbrace \X_{m,t}(s);~t=1, 2, \cdots, N,~m = 1, 2, \cdots, M \right\rbrace$ denote $M$ functional time series which are assumed mean zero processes $\text{E} \left[ \X_m(s) \right] = 0$ for $m = 1, 2, \cdots, M$ have an effect on $\Y(s)$ and . Denote by $\gamma_m(\nu) = \left\lbrace \X_{m,1}(\nu), \cdots, \X_{m,N-1}(\nu) \right\rbrace$ and $\pmb{\gamma}(\nu) = \left[ \gamma_1(\nu), \cdots, \gamma_M(\nu) \right]$ the function for $m$\textsuperscript{th} exogenous variable and a function including $M$ sets of functional exogenous variables, respectively. Now let $\pmb{\X}(\nu) = \left[ f(\nu), \pmb{\gamma}(\nu) \right]$ be a function including $M+1$ functional lagged variables. Then, it was considered the following FARX(1) model:
\begin{equation}\label{eq:farx}
g(s) = \int_{\nu} \pmb{\X}(\nu) \theta(\nu, s) d \nu + \epsilon(s),
\end{equation}
where $\epsilon(s)$ denotes the error function and $\theta(\nu, s)$ is a bivariate coefficient function comprising the coefficient functions of $M+1$ functional lagged variables.

In formula~\eqref{eq:farx}, the relationship between the functional response and vector of functional lagged variables is characterized by the regression coefficient function $\theta(\nu, s)$. In the literature, several methods have been proposed to estimate $\theta(\nu, s)$, such as least squares (LS) \citep{yamanishi2003, ramsay2006}, maximum likelihood (ML), and maximum penalized likelihood (MPL) \citep{matsui2009}. Although these methods work well under certain circumstances, they have some drawbacks. For instance, the LS and ML-based methods produce unstable estimates for $\theta(\nu, s)$ \citep{matsui2009}. In contrast, the MPL method produces a consistent estimate for the regression coefficient surface; it is computationally inefficient. As an alternative, it was considered the functional partial least squares regression (FPLSR) approach of \cite{PredSap} and \cite{Beyaztas2020} to estimate the regression coefficient function. FPLSR is computationally more efficient than the LS, ML and, MPL methods; in turn, it increases the prediction accuracy of the FARX(1) in~\eqref{eq:farx}.

\subsection{Functional partial least squares regression}

The FPLSR model in its original form is considered as an iterative procedure. Each iteration produces orthogonal components by maximizing squared covariance between linear spans of the response and lagged variables. The FPLSR components associated with the FARX(1) in~\eqref{eq:farx} are obtained as solutions of Tucker's criterion extended to functional variables as follows:
\begin{equation}\label{eq:tucker}
\underset{\begin{subarray}{c}
  \kappa \in \mathcal{L}_2,~ \Vert \kappa \Vert_{\mathcal{L}_2} = 1 \\
  \zeta \in \mathcal{L}_2,~ \Vert \zeta \Vert_{\mathcal{L}_2} = 1
  \end{subarray}}{\max} \text{Cov}^2 \left( \int_{\nu} \pmb{\X}(\nu) \kappa(\nu) d \nu, ~ \int_{S} g(s) \zeta(s) d s \right).
\end{equation}
The FPLSR components are also equal to the eigenvectors of the Escoufier operators \citep{PredSap}. Let $Z$ be a random variable with values in $\mathcal{L}_2$ Hilbert space. Let $W^{g}$ and $W^{\pmb{\X}}$ denote the Escoufier's operators \citep{Escoufier} associated with $g(s)$ and $\pmb{\X}(\nu)$, respectively, as follows:
\begin{equation}
W^{g} = \int_{\mathcal{I}} \text{E} \left[ g(s) Z \right] g(s) d s, \qquad W^{\pmb{\X}} = \int_{\mathcal{I}} \text{E} \left[ \pmb{\X}(\nu) Z \right] \pmb{\X}(\nu) d \nu, \qquad \forall Z \in \mathcal{L}_2.
\end{equation}
Denote by $i = 1, 2, \cdots$ the iteration number. Then, at each step $i$, the $i$\textsuperscript{th} FPLSR component, denoted by $\eta_i$, is determined by an iterative stepwise procedure. $\eta_i$ is obtained using the residuals of the regressions of the response ($g_{i}(s)$) and lagged functions ($\pmb{\X}_i(\nu)$) on the FPLSR component computed at previous step. In more detail, let $g_{i}(s)$ and $\pmb{\X}_i(\nu)$ are defined as follows:
\begin{align}
g_{i}(s) &= g_{i-1}(s) \zeta_i(x) \eta_i, \\
\pmb{\X}_i(\nu) &= \pmb{\X}_i(\nu) - p_i(\nu) \eta_i, 
\end{align}
where $g_0(s) = g(s)$, $\pmb{\X}_0(\nu) = \pmb{\X}(\nu)$, $\zeta_i(x) = \frac{\text{E} \left[ g_i(s) \eta_i \right]}{\text{E} \left[ \eta_i^2 \right]}$, and $p_i(\nu) = \frac{\text{E} \left[ \pmb{\X}_{i-1}(\nu) \eta_i\right]}{\text{E} \left[ \eta_i^2 \right]}$. Then, $\eta_i$ equals to the eigenvector associated with the largest eigenvalue of $W_{i-1}^{\pmb{\X}} W_{i-1}^{g}$, ($\lambda_{\max}$) as follows:
\begin{equation}
W_{i-1}^{g} W_{i-1}^{\pmb{\X}} \eta_i = \lambda_{\max} \eta_i,
\end{equation}
where $W_{i-1}^{g}$ and $W_{i-1}^{\pmb{\X}}$ are the Escoufier's operators of $g_{i-1}(s)$ and $\pmb{\X}_{i-1}(\nu)$, respectively. In other words, $\eta_i$ is the random variable that maximizes the Tucker's criterion~\eqref{eq:tucker}:
\begin{equation}
\eta_i = \int_{\mathcal{I}} \kappa_i(\nu) \pmb{\X}_{i-1}(\nu) d \nu,
\end{equation}
where the weight function $\kappa_i(\nu)$ is given by:
\begin{equation}
\kappa_i(\nu) = \frac{\int_{\mathcal{I}} \text{E} \left[ g_{i-1}(s) \pmb{\X}_{i-1}(\nu)\right] d s}{\sqrt{\int_{\mathcal{I}} \left( \int_{\mathcal{I}} \text{E} \left[ g_{i-1}(s) \pmb{\X}_{i-1}(\nu)\right] d s \right)^2 d \nu}}.
\end{equation}

Originally, the FPLSR components of FARX(1) are computed using the infinite-dimensional sample curves of the functional response and functional lagged variables. On the other hand, these curves are observed in a set of discrete-time points. Thus, the Escoufier's operators are not observed directly and needed to be estimated from the discrete observations. An approximation for the functional response and lagged variables by the $B$-spline basis expansion method was conducted to overcome this problem. Then, the partial least squares regression using the basis expansion coefficients is considered to approximate the FPLSR of $g(s)$ on $\pmb{\X}(\nu)$. Briefly, a basis expansion method approximates a function $y(u)$ as the linear combinations of basis functions $\phi_k(u)$ and the corresponding coefficients $c_k$: $y(u) \approx \sum_{k=1}^K c_k \phi_k(u)$, where $K$ represents the number of basis functions. Now, let us consider the basis function expansions of the functional response and lagged variables as follows:
\begin{align}
g(s) &= \sum_{k=1}^{K_{g}} c_k \phi_k(s) = \pmb{C} \pmb{\Phi}(s), \\
\pmb{\X}(\nu) &= \sum_{j=1}^{K_{\pmb{\X}}} d_j \psi(\nu) = \pmb{D} \pmb{\Psi}(\nu),
\end{align}
where $K_{g}$ and $K_{\pmb{\X}}$ respectively are the numbers of basis functions used for approximating the response and lagged functions, $\pmb{\Phi}(s) = \left( \phi_1(s), \cdots, \phi_{K_{g}}(s) \right)^\top$ and $\pmb{\Psi}(\nu) = \left( \psi_1(\nu), \cdots, \psi_{K_{\pmb{\X}}}(\nu) \right)^\top$ are the basis functions, and $\pmb{C} = \left( c_1, \cdots, c_{K_{g}} \right)^\top$ and $\pmb{D} = \left( d_1, \cdots, d_{K_{\pmb{\X}}} \right)^\top$ are the corresponding coefficient matrices.

Denote by $\pmb{\Phi} = \int_{S} \pmb{\Phi}(s) \pmb{\Phi}^\top(s) d s$ and $\pmb{\Psi} = \int_{\nu} \pmb{\Psi}(\nu) \pmb{\Psi}^\top(\nu) d \nu$ the symmetric matrices of the inner products of the basis functions. Let $\pmb{\Phi}^{1/2}$ and $\pmb{\Psi}^{1/2}$ be the the square roots of $\pmb{\Phi}$ and $\pmb{\Psi}$, respectively. Then, the following regression model can be considered to approximate the FPLSR of $g(s)$ on $\pmb{\X}(\nu)$:
\begin{equation}\label{eq:PLS_basis}
\pmb{C} \pmb{\Phi}^{1/2} = \pmb{D} \pmb{\Psi}^{1/2} \pmb{\Omega} + \pmb{\epsilon},
\end{equation}
where $\pmb{\Omega}$ and $\pmb{\epsilon}$ are the coefficient and residual matrices, respectively. Accordingly, the coefficient function $\theta(\nu,s)$ in FARX(1) can be approximated from \eqref{eq:PLS_basis} as follows:
\begin{equation}\label{eq:est_FPLSR}
\hat{\theta}(\nu,s) = \left[ \left( \pmb{\Psi}^{1/2} \right)^{-1} \widehat{\pmb{\Omega}} \left( \pmb{\Phi}^{1/2} \right)^{-1} \right] \pmb{\Psi}(\nu) \pmb{\Phi}(s),
\end{equation}
where $\widehat{\pmb{\Omega}}$ is the estimated coefficient matrix of $\pmb{\Omega}$. Finally, the one-step-ahead prediction of $\Y_{N+1}(t)$ using the proposed FARX(1) model can be obtained as follows:
\begin{equation}\label{eq:frc}
\hat{\Y}_{N+1}(s) = \int_{\mathcal{I}} \pmb{\X}_N(\nu) \hat{\theta}(\nu, s) d \nu.
\end{equation} 

Note that it was assumed that all the FTS objects are mean zero processes under the proposed FARX(1) model. On the other hand, when considering seasonal functional time series (i.e., $\text{E} \left[ \Y_t(s) \right] \neq 0$ and $\text{E} \left[ \X_m(s) \right] \neq 0$ for $m = 1, \cdots, M$), the centered FTS $\Y^*_t(s) = \Y_t(s) - \overline{\Y}(s)$ and $\X^*_m(s) = \X_m(s) - \overline{\X}_m(s)$ where $\overline{\Y}(s) = N^{-1} \sum_{t=1}^N \Y_t(s)$ and $\overline{\X}_m(s) = N^{-1} \sum_{t=1}^N \X_{tm}$ may be used to construct model and obtain one-step-ahead prediction as in~\eqref{eq:frc}. Then, the seasonal version of the predictions can be obtained by adding the mean of the response to the predictions i.e., $\hat{\Y}_{N+1}(s) = \hat{\Y}^*_{N+1}(s) + \overline{\Y}(s)$ where $\hat{\Y}^*_{N+1}(s)$ denotes the prediction calculated using the centered functional time series objects.

For the sake of clarity, a flow chart is provided in Figure~\ref{fig:flowchart} to demonstrate how the proposed FARX(1) model works for obtaining the one-step-ahead prediction of the response variable.

\tikzstyle{arrow} = [thick,->,>=stealth]
\tikzstyle{process} = [rectangle, minimum width=5cm, minimum height=1cm, text centered, text width=10cm, draw=black, fill=orange!30]

\begin{figure}
\begin{center}
\begin{tikzpicture}[node distance=2.4cm]
\node (pro1) [process] {Input the data observed at discrete time points};
\node (pro2) [process, below of=pro1, yshift=0.5cm] {Approximate the functional forms of discretely observed data using $B$-spline basis expansion model};
\node (pro3) [process, below of=pro2, yshift=0.5cm] {Stack all the exogenous lagged variables into single function $\pmb{\X}(\nu)$};
\node (pro4) [process, below of=pro3, yshift=0.5cm] {Construct the FARX(1) model as in~\eqref{eq:farx}};
\node (pro5) [process, below of=pro4, yshift=0.5cm] {Estimate the model parameter $\theta(\nu,s)$ using FPLSR as in~\eqref{eq:est_FPLSR}};
\node (pro6) [process, below of=pro5, yshift=0.5cm] {Obtain one-step-ahead prediction of response variable using the estimated parameter as in~\eqref{eq:frc}};
\draw [arrow] (pro1) -- (pro2);
\draw [arrow] (pro2) -- (pro3);
\draw [arrow] (pro3) -- (pro4);
\draw [arrow] (pro4) -- (pro5);
\draw [arrow] (pro5) -- (pro6);
\end{tikzpicture}
\end{center}
\caption{Flow chart of the proposed model}
\label{fig:flowchart}
\end{figure}
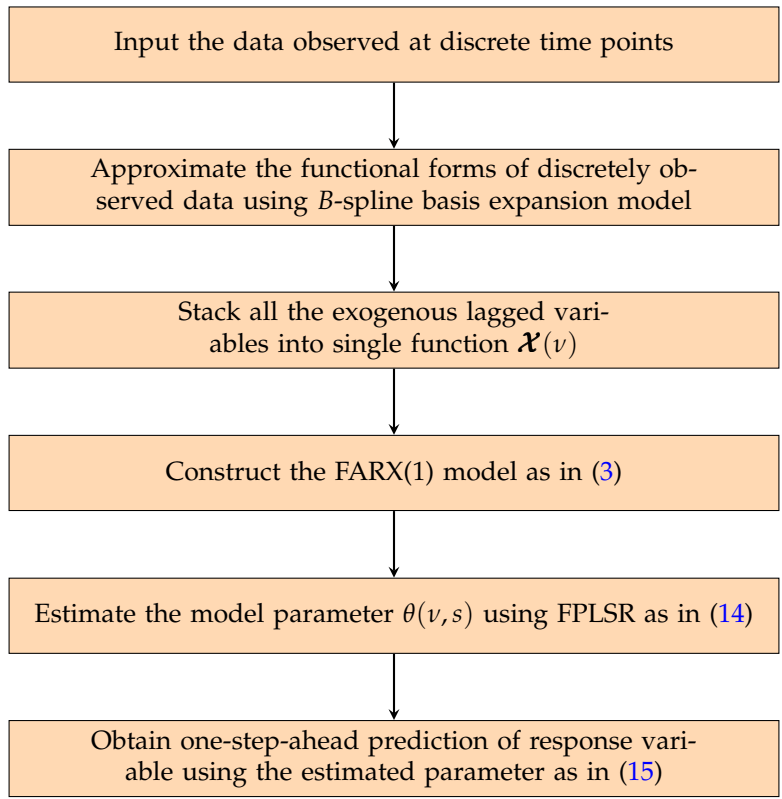

\subsection{Variable selection procedure}\label{sec:vsp}

The proposed FARX(1) model may include many functional lagged variables. In this case, not all the functional lagged variables might be significant for predicting the functional response variable's future realizations. Thus, the following forward selection procedure was proposed to determine the significant functional lagged variables.
\begin{itemize}
\item[Step 1.] Model construction: First, $(M+1)$ - FARX(1) model, each of which includes the common response and a functional lagged variable, are constructed as follows:
\begin{equation}
g(s) = \int_{\nu} \pmb{\X}_m(\nu) \theta_m(\nu, s) d \nu + \epsilon_m(s),
\end{equation}
where $\pmb{\X}_m(\nu) = [f(\nu), \gamma_1(\nu), \cdots, \gamma_M(\nu)]$ for $m = 1, \cdots, M+1$. Among these models, the one having the smallest mean squared error (MSE),
\begin{equation}
\text{MSE} = \frac{1}{N-1}\sum_{j=2}^{N} \left\Vert \Y_j(s) - \widehat{\Y}_j(s) \right\Vert^2_{\mathcal{L}_2},
\end{equation}
where $\widehat{\Y}_j(t)$ is the prediction of $\Y_j(t)$ for $j = 2, \cdots, N$, is chosen as an initial model. In MSE, $\left\Vert \cdot \right\Vert_{\mathcal{L}_2}$ denotes $\mathcal{L}_2$ norm, which is approximated by the Riemann sum \citep{LuoQi}.
\item[Step 2.] Model selection: Let $\pmb{\X}^{(1)}(\nu)$ and $\text{MSE}^{(1)}$ denote the functional lagged variable in the initial model and the MSE obtained from this model, respectively. Then, $M$ different FARX(1) models, including the common response and a functional lagged variable, are constructed as follows:
\begin{equation}
g(s) = \int_{\nu} \pmb{\X}_m(\nu) \theta_m(\nu, s) d \nu + \epsilon_m(s),
\end{equation}
where $\pmb{\X}_m(\nu) = \left[ \pmb{\X}^{(1)}(\nu), \X_m(\nu) \right] $, $\X_m(\nu) \neq \pmb{\X}^{(1)}(\nu)$, for $m = 1, \cdots, M$, and the MSE is calculated for each of these models. The functional lagged variable producing the smallest MSE, $\pmb{\X}^{(2)}(\nu)$, is chosen as the lagged variable for the current model if $\text{MSE}^{(2)} < \text{MSE}^{(1)}$ where $\text{MSE}^{(2)}$ is the calculated MSE when $\pmb{\X}^{(2)}(\nu)$ is used to estimate the current model. This process is repeated until all the significant variables are included in the model.
\end{itemize}

\subsection{Bootstrap prediction interval}\label{sec:boot}

A residual-based bootstrap model was proposed to construct prediction intervals for the future realizations of functional time series. In the proposed model, two error sources were taken into account: fitted model errors $\hat{\epsilon}^f(s) = g(s) - \hat{g}(s)$ and the residual component $\hat{\epsilon}^s(s) = \hat{g}(s) - \pmb{C} \pmb{\Phi}(s)$ where $\pmb{\Phi}(s)$ and $\pmb{C}$ denote the vectors of basis functions and associated coefficients, respectively. Then, the following summarizes the proposed bootstrap algorithm.

\begin{itemize}
\item[Step 1.] Construct bootstrap observations $\epsilon^*(s) = \left\lbrace \varepsilon^*_t(s);~t = 1, \cdots, N-1 \right\rbrace$ by random drawn from $\hat{\epsilon}^f(s)$.
\item[Step 2.] Compute the bootstrap observations $g^*(s)$ as follows:
\begin{equation}
g^*(s) = \int_{\nu} \pmb{\X}(\nu) \hat{\theta}(\nu, s) d \nu + \epsilon^*(s).
\end{equation}
\item[Step 3.] Compute the FPLSR estimate of the coefficient function, $\hat{\theta}^*(\nu, s)$ based on the bootstrapped data obtained in the previous step.
\item[Step 4.] Obtain the bootstrap future realization of $\Y_{N+1}(s)$ as follows:
\begin{equation}
\hat{\Y}^*_{N+1}(s) = \int_{\mathcal{I}} \pmb{\X}_N(\nu) \hat{\theta}^*(\nu, s) d \nu + \epsilon^*_N(s) + \varepsilon^*_N(s),
\end{equation}
where $\epsilon^*_N(s)$ and $\varepsilon^*_N(s)$ are random drawn from $\hat{\varepsilon}^f(s)$ and $\hat{\varepsilon}^s(s)$, respectively.
\item[Step 5.] Repeat Steps 1-4 $B$ times, where $B$ denotes the number of bootstrap replications, to obtain $B$ sets of bootstrap future values $\left\lbrace \hat{\Y}^{*,1}_{N+1}(s), \cdots, \hat{\Y}^{*,B}_{N+1}(s) \right\rbrace$.
\end{itemize}
Let $Q_{\alpha}(s)$ denote the $\alpha^{th}$ quantile of the generated $B$ sets of bootstrap future values. Then, the $100(1-\alpha)$ bootstrap prediction interval for $\Y_{N+1}(s)$ is computed as $\left[ Q_{\alpha/2}(s), ~ Q_{1-\alpha/2}(s)\right]$.

\subsection{Alternative models to the proposed FARX(1) model}\label{sec:ext}

In this study, the prediction performance of the proposed model was compared with some commonly used models. These models include the FARX(1) model of \cite{Damon2002}, the FAR(1) model of \cite{Chen2018}, FPCR, the traditional autoregressive integrated moving average (ARIMA), exponential smoothing (ETS), artificial neural network (ANN), and quantile regression (QR).

The FARX(1) model of \cite{Damon2002} works similarly to the proposed model, but it decomposes the functional observations into $q+1$ spaces and uses a nonparametric approach to estimate the model parameters. In the numerical analyses, the \texttt{R} package ``\texttt{far}'' of \cite{Damon2015} was used to evaluate the performance of FARX(1) model of \cite{Damon2002}. In addition, it was combined with the proposed bootstrap algorithm to construct pointwise prediction intervals. It is assumed that this method's residual component follows a normal distribution with mean zero and variance $\sigma^2$ (since the method of \cite{Damon2002} does not have a smoothing phase) where $\sigma^2$ was estimated from the estimated model.

The FAR(1) model of \cite{Chen2018} given by~\eqref{eq:far}. Unlike the parameter estimation strategy given by \cite{Chen2018}, the model parameter of FAR(1) was estimated using the FPLSR method instead of the maximum likelihood method. The aim of doing this, was to demonstrate how the use of FPLSR improves the predictive performance of the model. The numerical results showed that the FAR(1) model based on FPLSR produces similar or even better performance compared with the FARX(1) model of \cite{Damon2002}. The proposed bootstrap algorithm was also employed to FAR(1) model to construct prediction intervals.

The FPCR decomposes a functional time series into orthonormal principal components and associated uncorrelated principal component scores. Then, it generates the future realizations of the principal component scores using a univariate time series model. Finally, the functional time series's future realizations are obtained by multiplying principal components with the predicted principal components. Also, it uses a nonparametric bootstrap approach, which is similar to the proposed model, to construct pointwise prediction intervals for the response function. For more information about the FPCR, consult \cite{HydSh2009}. Also, refer to the \texttt{R} package ``\texttt{ftsa}'' \citep{ftsa}, which was used to obtain both point forecasts and bootstrap prediction intervals for the FPCR method in this study, and for the practical implementation of the FPCR method.

In the traditional ARIMA model, the time series data is first transformed into a stationary form. Then, this stationary form is used to construct the time series model. Based on the constructed model, future realizations are obtained using a recursion formula. On the other hand, in the traditional ETS model, the weights are first assigned to all observations, decreasing exponentially for the distant past data points. Then, the future realizations are calculated using a recursion formula. The bootstrap prediction intervals are obtained using the resampled errors (called ordinary bootstrap). The \texttt{R} package ``\texttt{forecast}'' \citep{HydKhan2008} was used in the numerical analyses for performing ARIMA and ETS (see \cite{HydKhan2008}.

ANN, which is a soft computing technique, is one of the most commonly used prediction models. The neural network is fitted using past lags of the time series as input variables and hidden layers. Then, the inputs are multiplied by the connection weights. Finally, these calculations are averaged to obtain predictions. The ANN model's prediction intervals are obtained iteratively by simulating the future sample path of the model. The error values are generated randomly from the distribution of errors, and the sample path is obtained by adding the generated error values to the estimated model. For more information about the ANN, see \cite{Zhang1998} and \cite{chetan2006hybrid}. In the analyses, \texttt{nnetar} function in the \texttt{R} package ``\texttt{forecast}'' \citep{HydKhan2008} was used to obtain ANN forecasts and bootstrap prediction intervals.

The QR, which was proposed by \cite{Koenker}, is used to investigate the effects of exogenous variables at different quantile levels of the response variable. Compared with traditional regression models, it provides a more general picture of the response variable's conditional distribution function and can model heteroscedasticity by perception and construction. Let $Y$ and $\bm{X}$ denote the response and the vector of exogenous variables that are assumed to affect the response variable, respectively. Then, for a given $\tau \in (0,1)$, the $\tau$\textsuperscript{th} conditional quantile of $Y$ given $\bm{X}$ can be expressed as follows:
\begin{equation}
Q_{\tau}(Y | \bm{X}) = \bm{X} \beta_{\tau},
\end{equation}
where $\beta_{\tau}$ denotes the vector of model parameters and is estimated by minimizing the check loss function $\rho_{\tau}(u) = u\{\tau - I (u < 0) \}$ as follows:
\begin{equation}
\widehat{\beta}_{\tau} = \underset{\beta_{\tau}}{\arg\min} \left[ \sum_{i=1}^N \rho_{\tau} \left( Y_i - \beta_{\tau} \right) \right].
\end{equation}

Consult \cite{Koenker} for the theoretical details of the QR model. Also, interested readers are referred to \cite{Papacharalampous} for its practical implementation in hydrology. In the numerical analyses, the \texttt{rq} function in the \texttt{R} package ``\texttt{quantreg}'' \citep{quantreg} was used to obtain the point forecasts of QR model, where the quantile level was chosen as $\tau = 0.5$. In addition, the prediction intervals for the QR were calculated by fitting the same model on the data for two quantile levels, $\tau_1 = 0.025$ and $\tau_2 = 0.975$.

\section{A Monte Carlo experiment}\label{sec:MC}

\begin{figure}[!htbp]
  \centering
  \includegraphics[width=4.8cm,height=5cm]{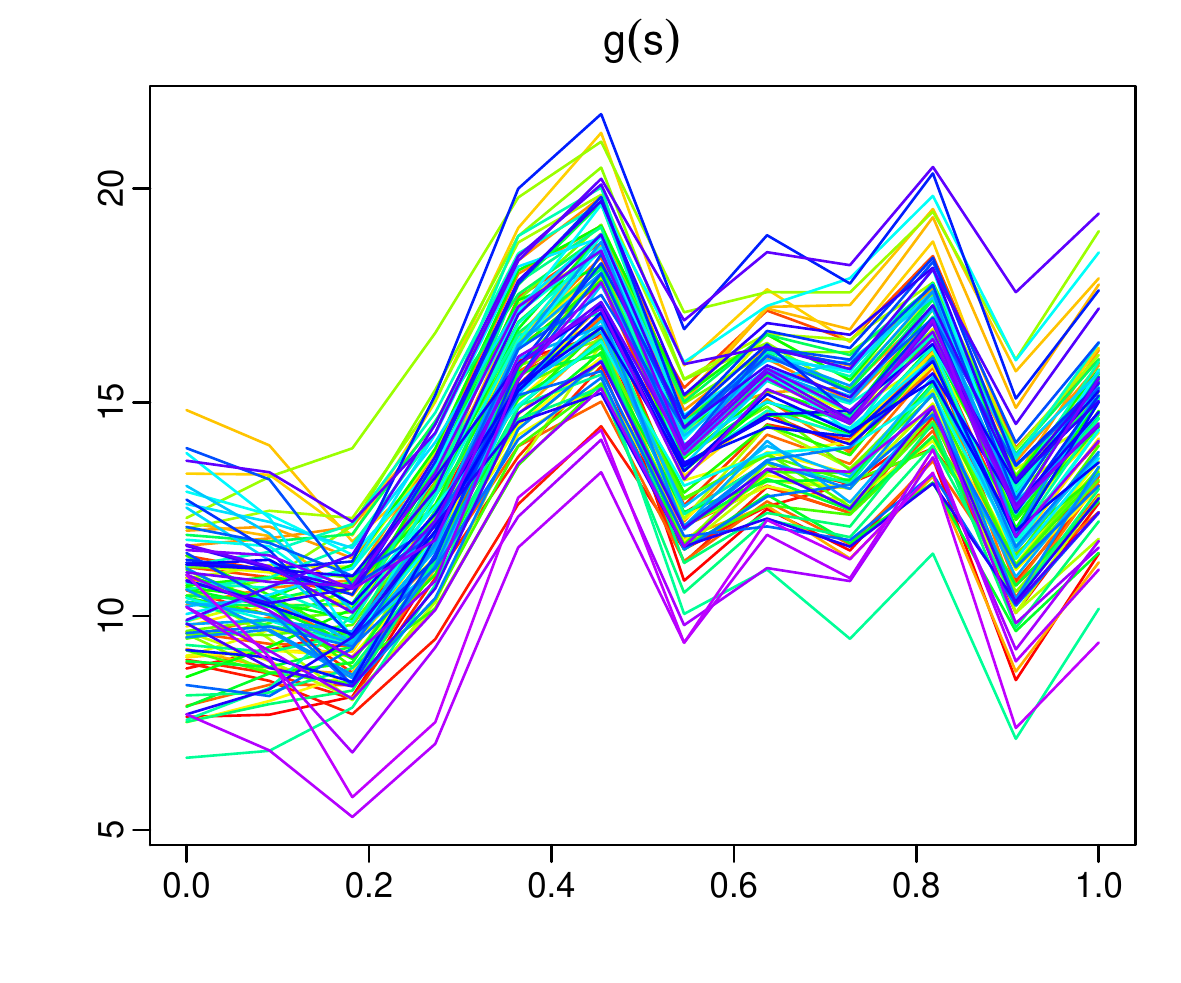}
  \includegraphics[width=4.8cm,height=5cm]{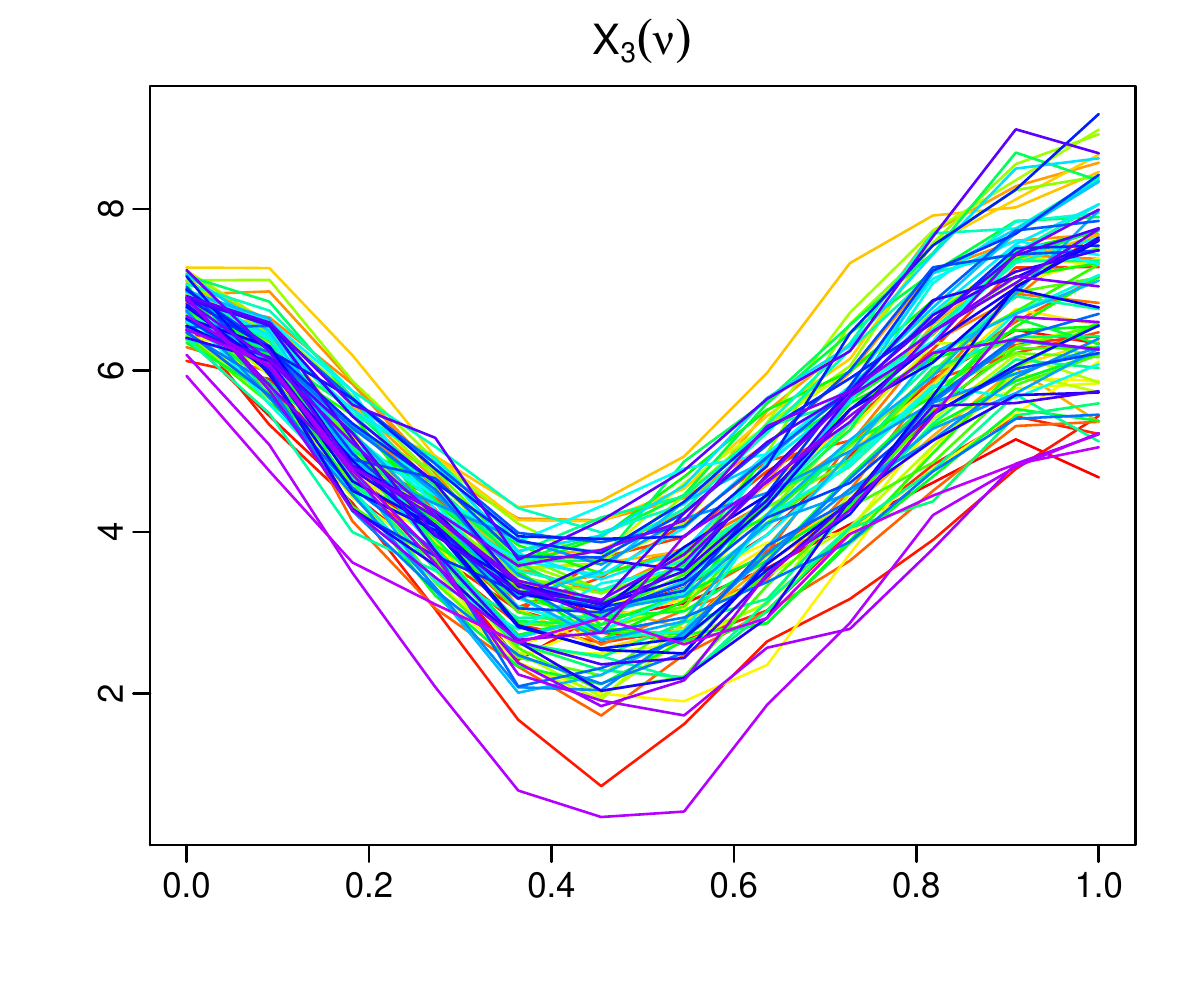}
  \includegraphics[width=4.8cm,height=5cm]{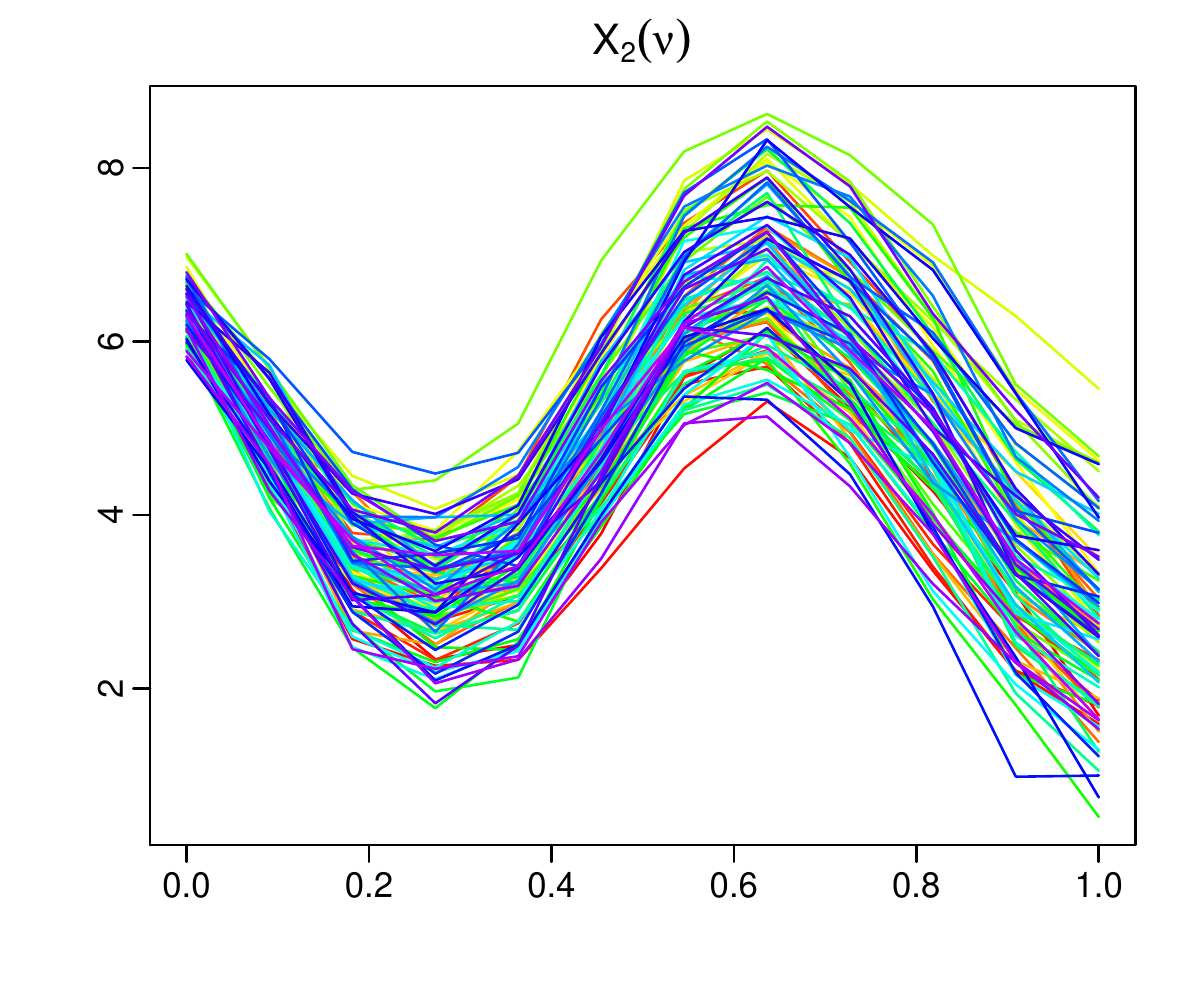}
  \\
  \includegraphics[width=4.8cm,height=5cm]{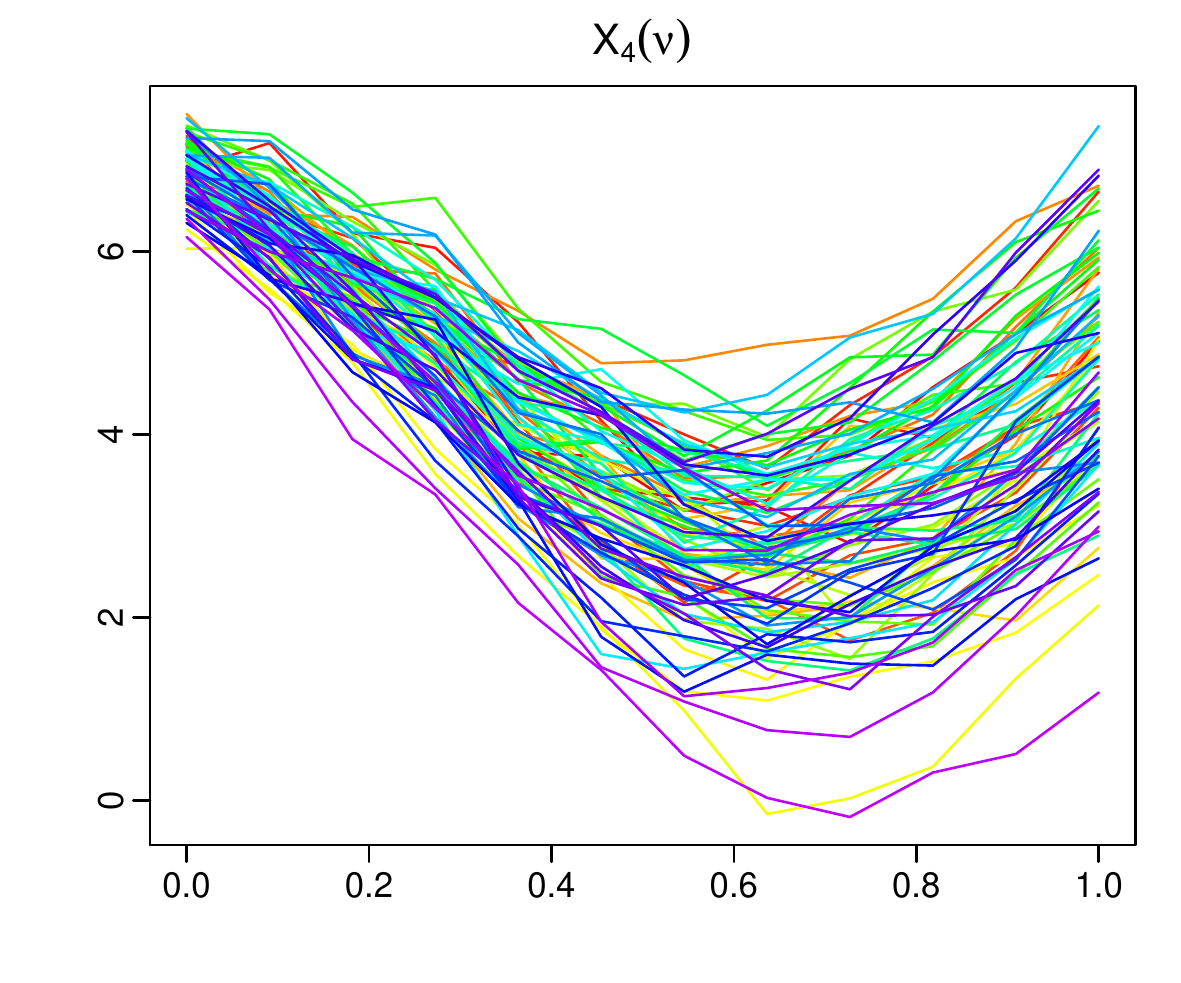}
  \includegraphics[width=4.8cm,height=5cm]{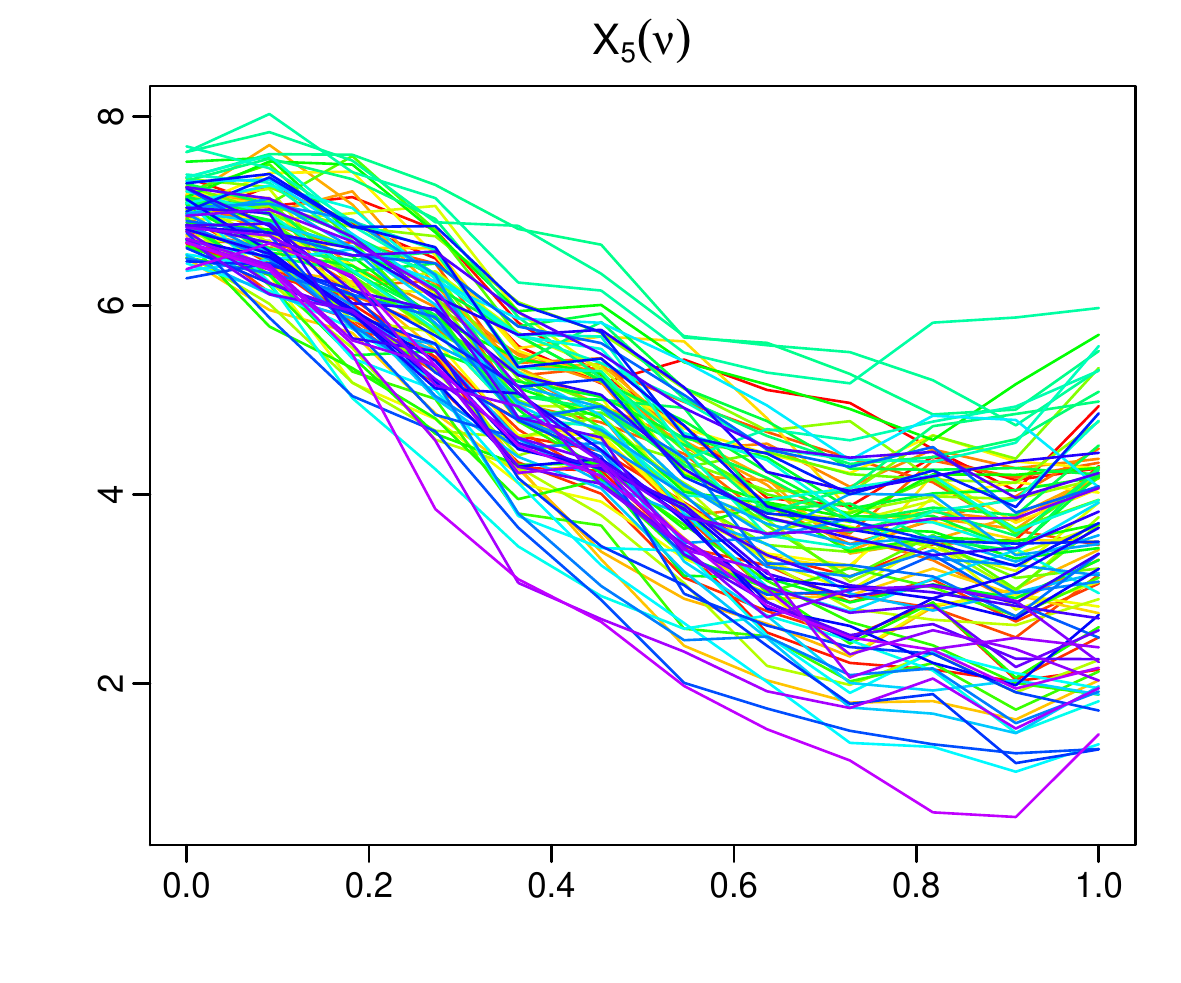}
  \includegraphics[width=4.8cm,height=5cm]{Fig_2e.pdf}
  \caption{Plots of the generated functional response and predictor variables.}
  \label{fig:Fig_2}
\end{figure}

A Monte Carlo experiment was performed to explore the prediction performance of the proposed model, and the results were compared with the models mentioned in Section~\ref{sec:ext}. Throughout the experiment, $m = 5$ functional predictors were generated using the following process:
\begin{equation}
\X_{m,t}(\nu) = 5 + V_{m,t}(\nu) + U_{m,t}(\nu),
\end{equation} 
where $V_{m,t}(\nu)$ and $U_{m,t}(\nu)$, for $m = 1, \cdots, 5$, were generated from the following FAR(1) and cosine processes as follow:
\begin{align}
V_{m,t}(\nu) &= \int_0^1 \psi(\nu, u) V_{m,t-1}(\nu) ds + \varrho_t(\nu), \\
U_{m,t}(u) &= 2 \cos(\pi t_j / m) + N(0, 0.1^2),
\end{align}
where $\varrho_t(\nu)$ is a realization of an independently and identically distributed Brownian motion,  $\psi(\nu, u) = 0.34 e^{\frac{1}{2} \nu^2 + u^2}$, and $j = 1, \cdots, 12$. Then, the response function, $g(s)$ was generated as follows:
\begin{equation}
g_t(s) = g_{t-1}(s) + \int_0^1 \X_{1,t}(\nu) \beta_1(\nu, s) d \nu + \int_0^1 \X_{3,t}(\nu) \beta_3(\nu, s) d \nu + \int_0^1 \X_{5,t}(\nu) \beta_5(\nu, s) d \nu + \epsilon(s),
\end{equation}
where $\epsilon(s) \sim N(0, 0.1^2)$, and the coefficient functions were considered as follows:
\begin{align}
\beta_1(\nu, s) &= 2 \sin \left( (1-\nu)^2 (s-0.5)^2\right), \\
\beta_3(\nu, s) &= 2 \cos \left( e^{-5 (\nu-0.5) - 5 (s - 0.5)} + 8 e^{5 ( \nu - 1.5) -5 (s-0.5)}\right),\\
\beta_5(\nu, s) &= 2 \cos \left( \sqrt{\nu s} \right).
\end{align}
An example of the generated functional response and predictor variables are presented in Figure~\ref{fig:Fig_2}.

Throughout the experiment, $N = 100$ functions were generated at 12 equally spaced points in the interval $s,~\nu \in [0,1]$. The time series models were constructed based on the first 99 functions to obtain one-step-ahead prediction (100\textsuperscript{th} function). This procedure was repeated for $MC = 500$ times. For each predicted curve, five performance metrics: root mean squared prediction error (RMSPE), mean absolute prediction error (MAPE), root median squared prediction error (RMESPE), percent bias (PBIAS), and relative error (RE) were calculated to evaluate the prediction performance of the models \citep{tung2020survey, khosravi2018quantifying}:
\begin{align}
\text{RMSPE} &= \frac{1}{12} \sum_{j=1}^{12} \sqrt{\left\Vert \frac{g^{\text{pred}}(s_j) - g(s_j)}{g(s_j)} \right\Vert^2_{\mathcal{L}_2}}, \\
\text{MAPE} &= \frac{1}{12} \sum_{j=1}^{12} \left\Vert \frac{\left\vert g^{\text{pred}}(s_j) - g(s_j) \right\vert}{g(s_j)} \right\Vert_{\mathcal{L}_2}, \\
\text{RMESPE} &= \mathcal{M} \left\lbrace  \sqrt{\left\Vert \frac{g^{\text{pred}}(s_j) - g(s_j)}{g(s_j)} \right\Vert^2_{\mathcal{L}_2}} \right\rbrace, \qquad j = 1, \cdots, 12, \\
\text{PBIAS} &= \sum_{j=1}^{12} \left\Vert \frac{g^{\text{pred}}(s_j) - g(s_j)}{g(s_j)} \right\Vert_{\mathcal{L}_2}, \\
\text{RE} &= \frac{1}{12} \sum_{j=1}^{12} \left\Vert \frac{g^{\text{pred}}(s_j) - g(s_j)}{g(s_j)} \right\Vert_{\mathcal{L}_2} \times 100,
\end{align}
where $g(s_j)$ and $g^{\text{pred}}(s_j)$ respectively are the observed and predicted value at $j^{\text{th}}$ time point and $\mathcal{M}$ is the median operator. Also, two bootstrap performance metrics: the coverage probability deviance (CPD), which is the absolute difference between the nominal and empirical coverage probabilities, and the interval score (score) were calculated as follows:
\begin{align}
\text{CPD} &= \left\vert (1 - \alpha) - \frac{1}{12} \sum_{i=1}^{12} \mathbb{1} \left\{  Q_{\alpha/2}(s) \leq g(s) \leq Q_{1 - \alpha/2}(s) \right\} \right\vert, \\
\text{score} &= \frac{1}{12} \sum_{i=1}^{12} \left[ \left\{ Q_{1 - \alpha/2}(s) - Q_{\alpha/2}(s) \right\} \right. \\
&+ \frac{2}{\alpha} \left( Q_{\alpha/2}(s) - g(s) \right) \mathbb{1} \left\{ g(s) < Q_{\alpha/2}(s) \right\} \\
&+ \left. \frac{2}{\alpha} \left( g(s) - Q_{1 - \alpha/2}(s) \right) \mathbb{1} \left\{ g(s) > Q_{1 - \alpha/2}(s) \right\} \right],
\end{align}
where $\mathbb{1}\{\cdot\}$ denotes the binary indicator function. In the numerical analyses, $B = 100$ and $\alpha = 0.05$ were chosen to construct 95\% pointwise prediction intervals.

The performance of the proposed model is based on the number of basis functions ($K_{\Y}$ and $K_{\pmb{\X}}$) and the number of PLS components used to estimate the model parameter. Thus, the optimum number of basis functions and PLS components were chosen from two small sets of basis functions and PLS components according to its prediction performance for the proposed model. For this purpose, first, $K_{\Y} = K_{\pmb{\X}} = [4, \cdots, 10]$ number of basis functions and $i = [1, \cdots, 10]$ number of PLS components were used to estimate the model parameter. The number of basis functions and PLS components, which provide the smallest performance metrics, were chosen to be used in the analyses.

\begin{figure}[!htbp]
  \centering
  \includegraphics[width=4.8cm,height=5cm]{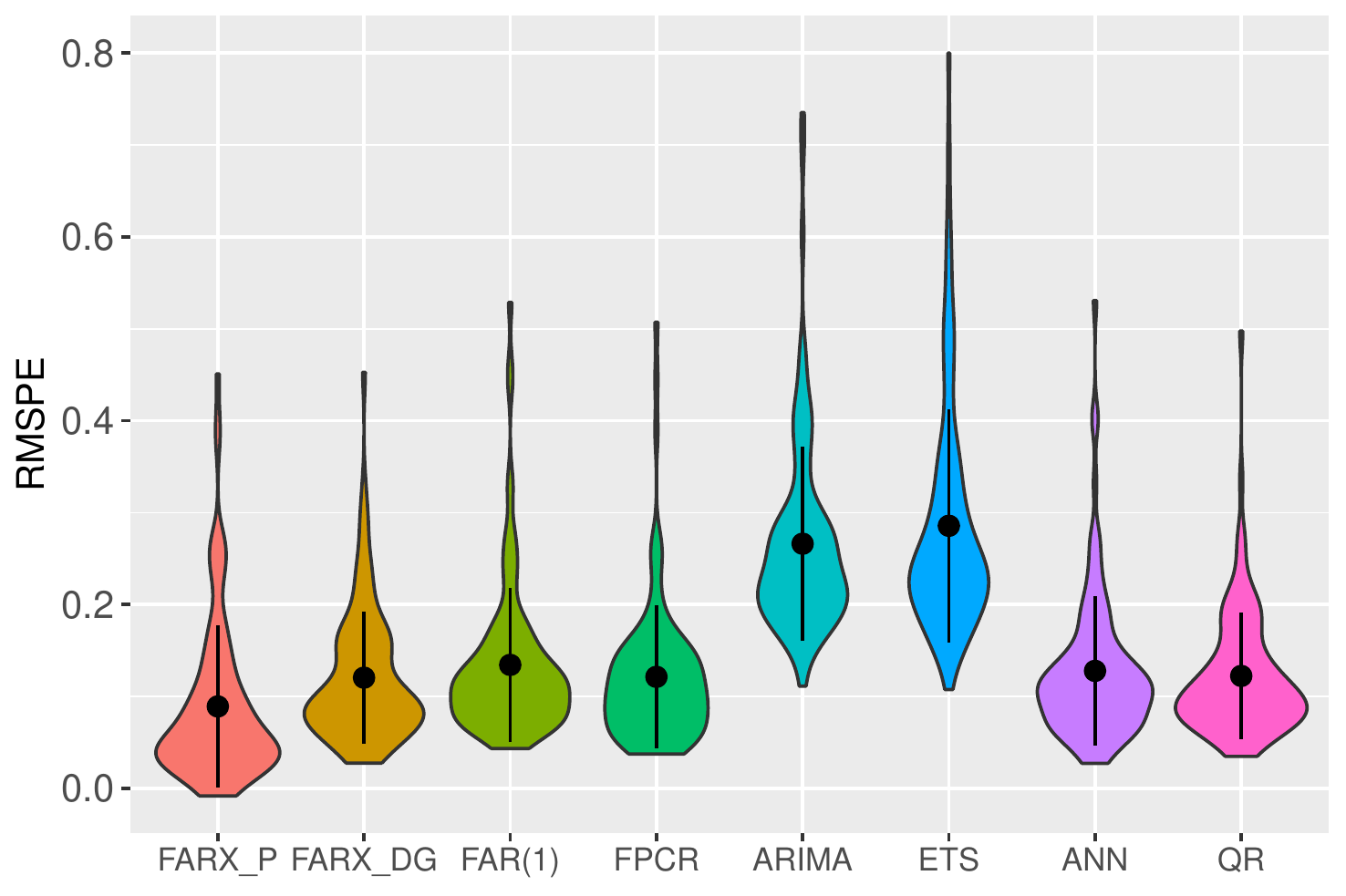}
  \includegraphics[width=4.8cm,height=5cm]{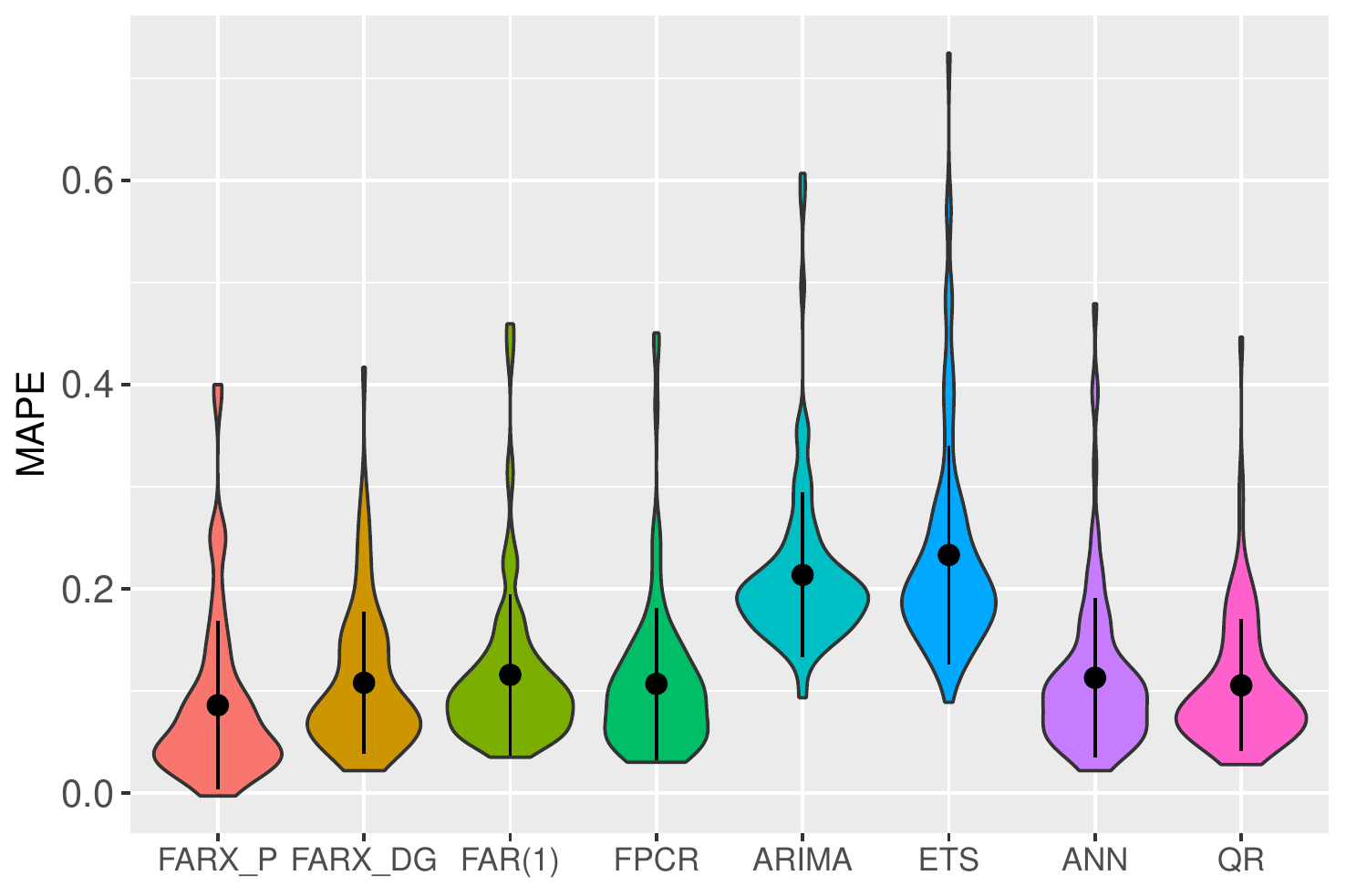}
  \includegraphics[width=4.8cm,height=5cm]{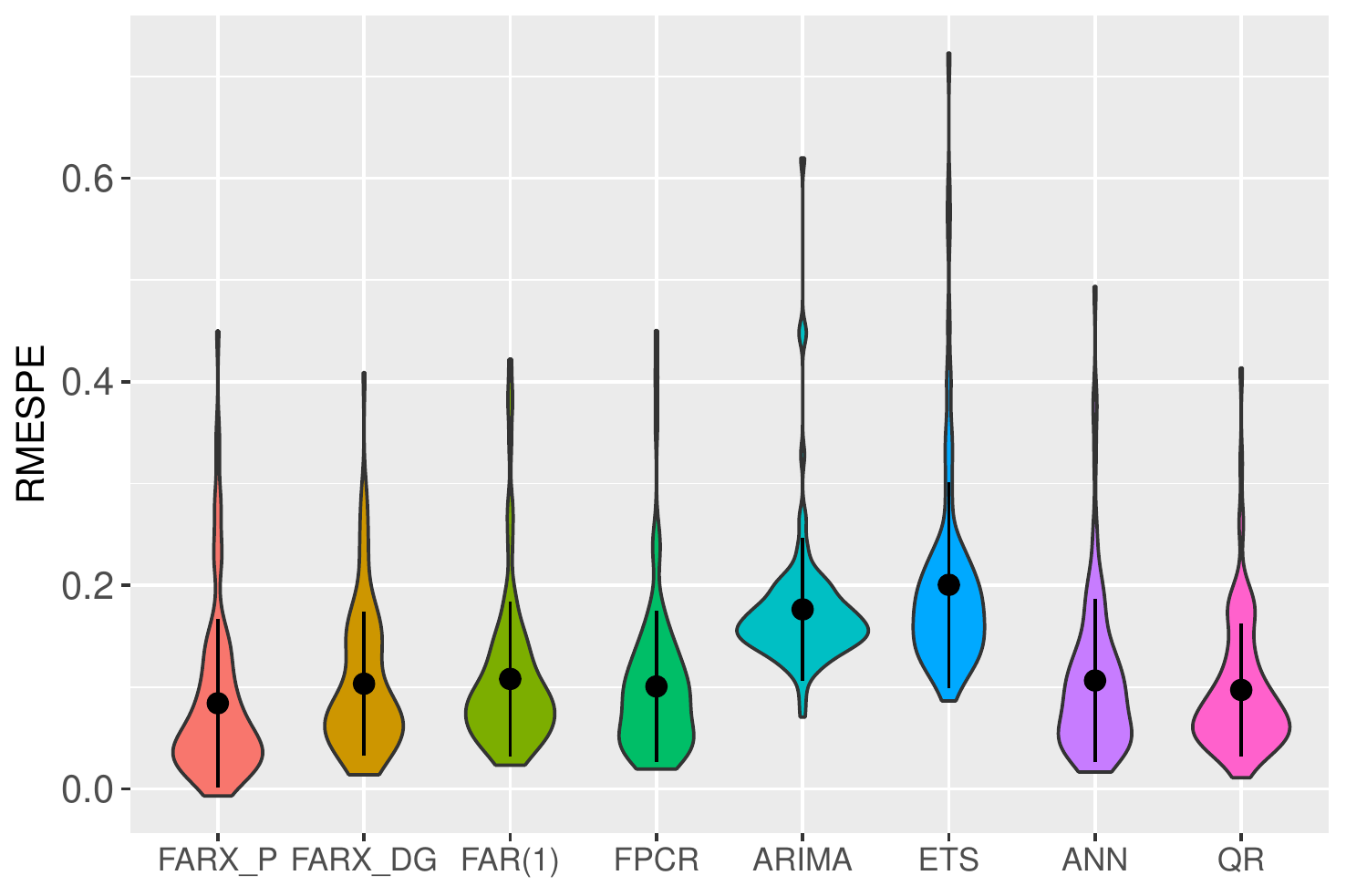}
  \\
  \includegraphics[width=4.8cm,height=5cm]{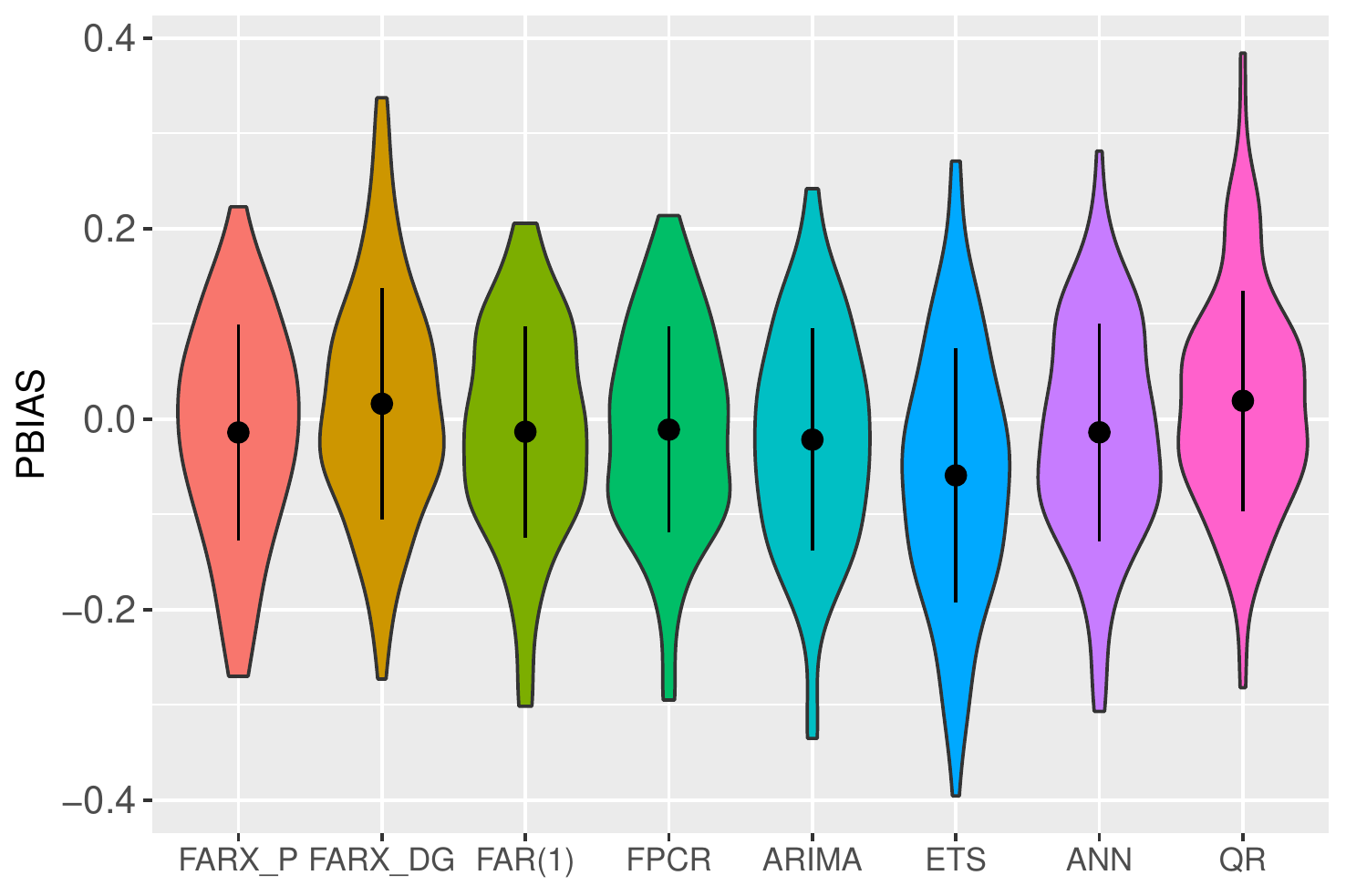}
  \includegraphics[width=4.8cm,height=5cm]{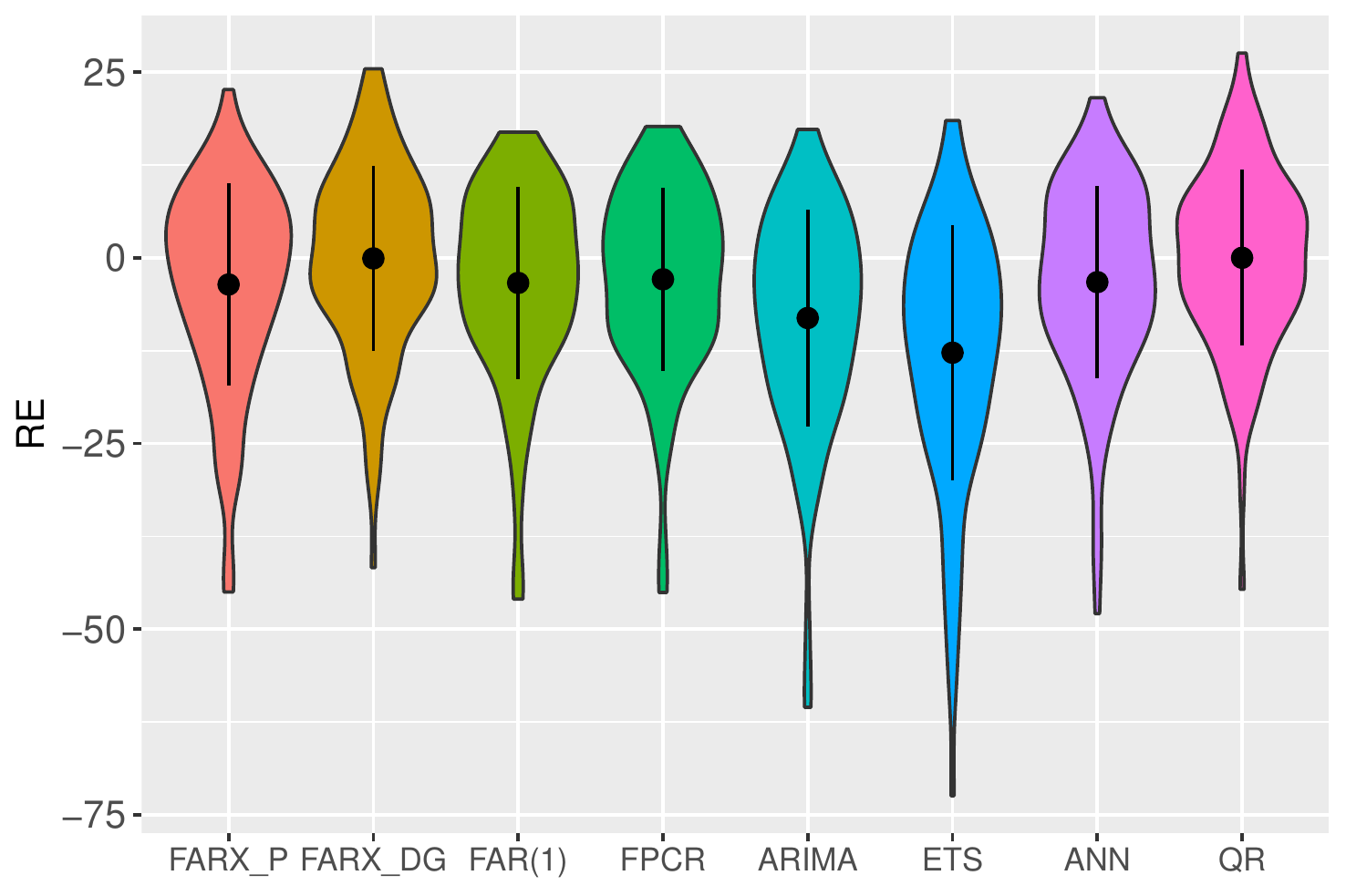}
  \caption{\textbf{Monte Carlo results}: Violin plots of the calculated RMSPE, MAPE, RMESPE, PBIAS, and RE values for the models. In the plots, FARX\underline{ }P and FARX\underline{ }DG denote the proposed FARX(1) model and the FARX(1) model of \cite{Damon2002}, respectively.}
  \label{fig:Fig_3}
\end{figure}

\begin{figure}[!htbp]
  \centering
  \includegraphics[width=6.9cm]{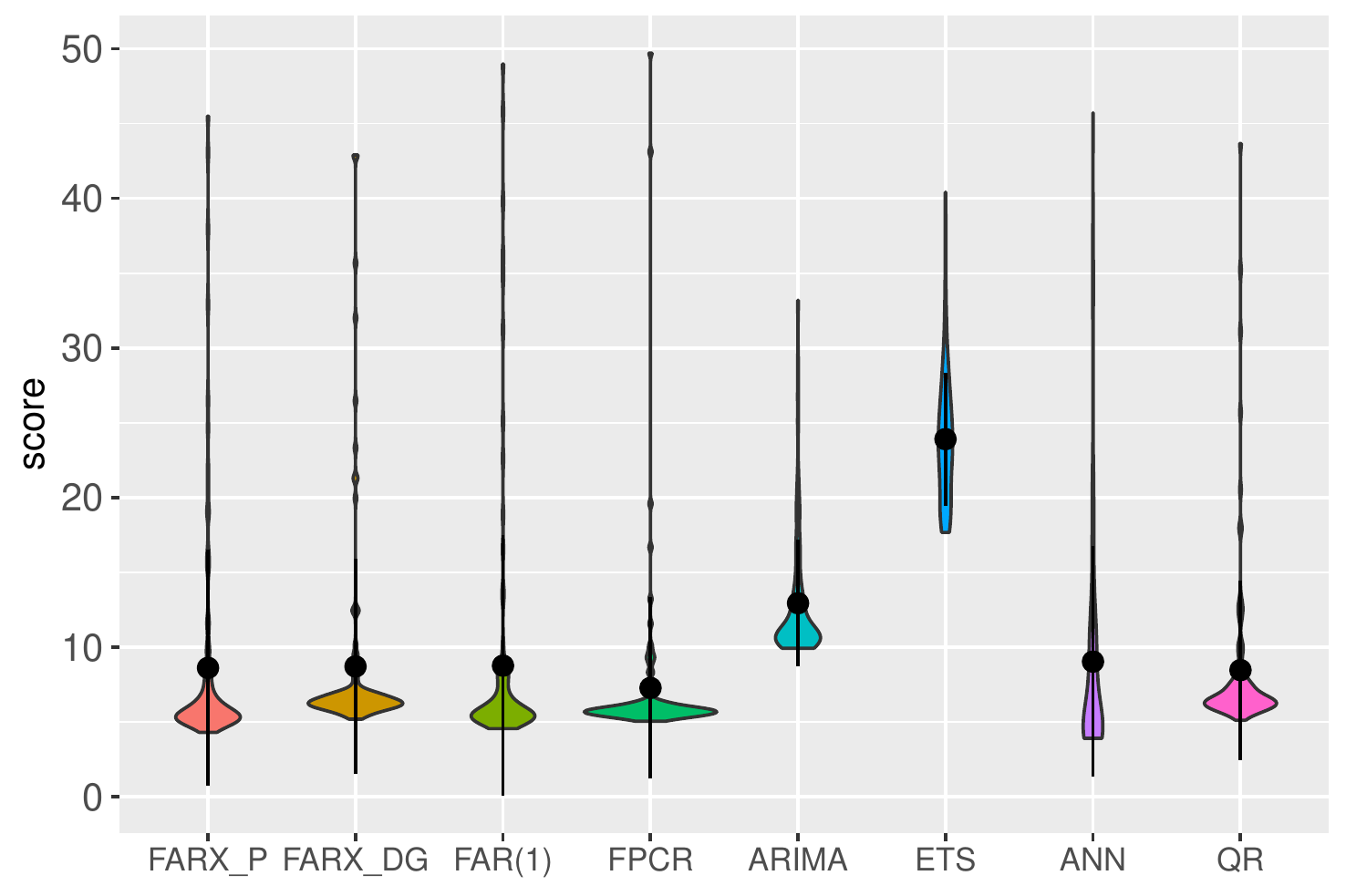}
  \includegraphics[width=6.9cm]{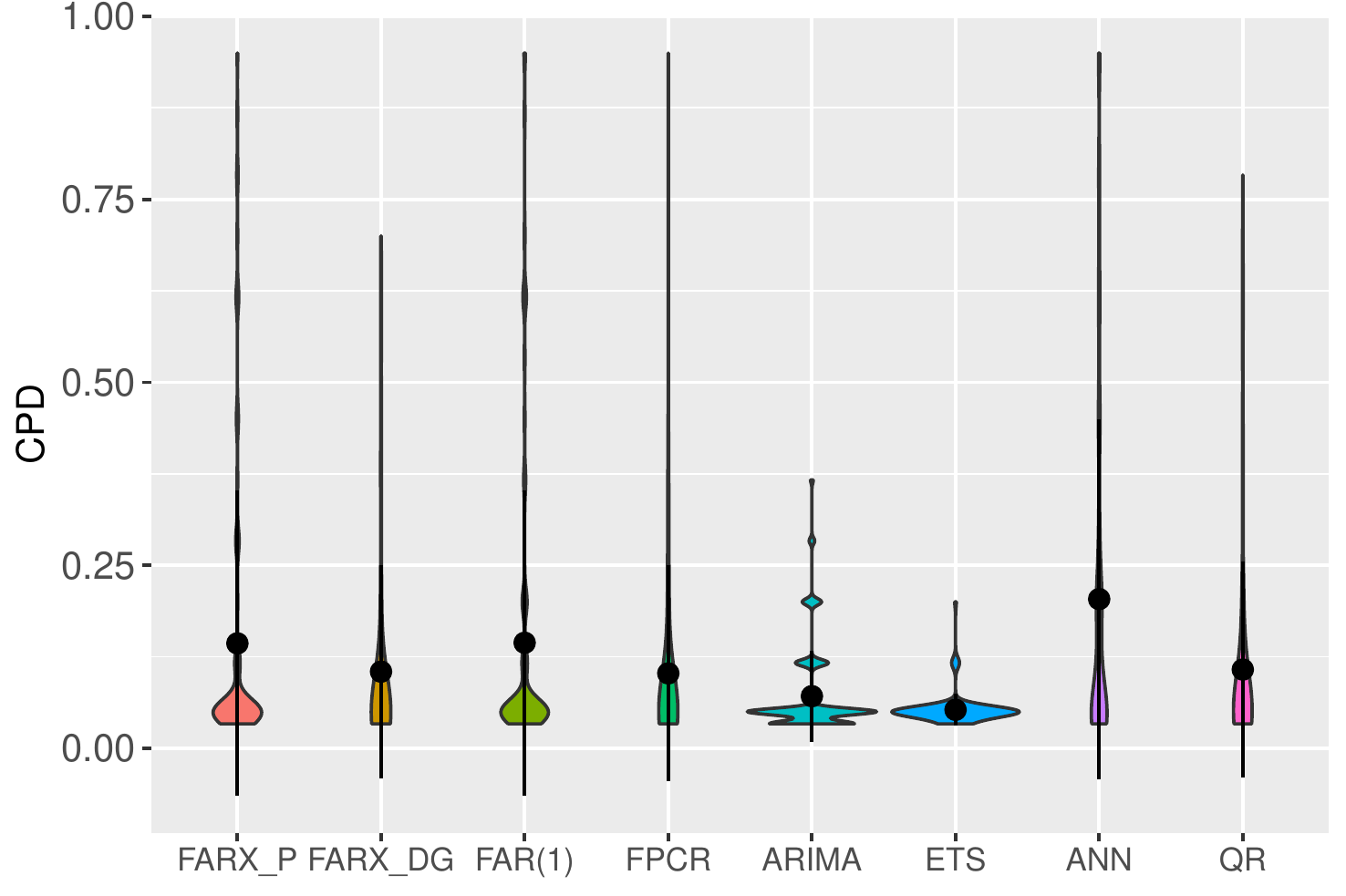}
  \caption{\textbf{Monte Carlo results}: Violin plots of the calculated score and CPD values for the models. In the plots, FARX\underline{ }P and FARX\underline{ }DG denote the proposed FARX(1) model and the FARX(1) model of \cite{Damon2002}, respectively.}
  \label{fig:Fig_4}
\end{figure}

The results produced by the Monte Carlo experiments are presented in Figures~\ref{fig:Fig_3} and~\ref{fig:Fig_4}. The obtained records demonstrated that the functional models performed better than the traditional ARIMA and ETS models. Only the ANN and QR produced a competitive performance for point forecasts versus functional models. The results further showed that the proposed model had superior performance, among others. It has smaller RMSPE, MAPE, RMESPE, PBIAS, and RE values than the FARX(1) model of \cite{Damon2002}, FAR(1), FPCR, ARIMA, ETS, ANN, and QR. Also, for the score and CPD metrics (Figure \ref{fig:Fig_4}), the functional models produced smaller score values than the traditional models (except QR), while all the models tended to produce similar CPD values (except ANN). While it seems that the traditional ARIMA and ETS models have slightly better CPD values than those of the proposed model, this is because these two models have larger prediction intervals. Among the traditional methods, only the QR produced the competitive score and CPD values to the functional models.

\section{Case study and data set description}

The second largest river in Western Asia is the Tigris River; the river has its basin in Iran, Iraq, Turkey, and Syria. The river is mainly fed by water from precipitation from the Armenian Highlands and various tributaries from the Zagros Mountains in Turkey, Iran, and Iraq \citep{kibaroglu2014analysis}. The water yield of the Tigris River is higher than that of the Euphrates River. At the Iraqi-Syrian-Turkish border, it has been estimated that the annual natural flow of the Tigris River is around 21 BCM. However, the water development projects going on in Iraq and Turkey have affected the flow volume of the Tigris in recent years as the recorded flow volume for Baghdad has exhibited a significant negative trend. Over the past 40 years, the Mesopotamian Marshlands have also experienced dwindling water supplies \citep{al2019development}. Turkey depends on the Tigris River for its Southeastern Gulf Anatolia Project (GAP). At the same time, Iraq has constructed several dams and embarked on various diversion projects on the Tigris River, focusing more on the Tharthar Canal between the Tigris and Euphrates \citep{unver1997southeastern}. All the riparian countries depend mainly on the Tigris for agriculture as its water for irrigation projects. The water quality in the basin is primarily affected by the increasing salinity rate due to sustained dependence on the river for irrigation and high evaporation rates. Besides the historic agreements that co-addressed the Tigris and Euphrates Rivers, much attention has not been given to the Tigris Basin's water resources. There is currently no basin-wide agreement, and the Tigris River has attracted only one bilateral agreement \citep{al2011toward}. The largest surface water resource within the study area is the Euphrates-Tigris-Shatt Al-Arab river system with a total topographic catchment of $>$900,000 km$^2$ from its head in the Taurus-Zagros Mountain Range to the Mesopotamian lowlands, with the Shatt Al Arab being the only outlet to the Persian Gulf \citep{isaev2009hydrography}.

The Tigris Basin has a semi-humid climate in the headwaters to the north and a semi-arid environment in southern Iraq near the Euphrates' confluence. The river basin has mean annual basin precipitation of around 400 to 600 mm, but 800 mm in the upper part and 150 mm in the lower part. The Tigris Basin records higher mean precipitation (about 300 mm/year) than the Euphrates Basin. This could be due to the high rates of rainfall in the Zagros Mountains that border the Tigris Basin on the east \citep{al2014climate}. Precipitation is experienced in the basin from November to April, while snowfall is experienced in the mountains from January to March. Because of the semi-arid to the arid climate of the Iraqi and Syrian lowlands, a significant amount of water is lost to evapotranspiration in the Mesopotamian region. The Tigris River records air temperatures at -35°C in winter to 40°C in the summer \citep{bozkurt2015projected}. Baghdad, the capital city of Iraq, records an average rainfall of about 216 mm with seasonal variabilities from December to February. The Tigris river has an estimated mean flow of about 235 $m^3$/s; the region experiences a maximum temperature of about 45°C in the summer and about 10°C during winter. Historical data throughout (1977-2013) for river flow, evaporation, temperature, rainfall with monthly scale were obtained from "The Meteorological Organization of Seismology (IMOS), Ministry of Agriculture and Water Resources, Iraq," belong to Tigris river.

\begin{figure}[!htbp]
\centering
\includegraphics[width=12cm]{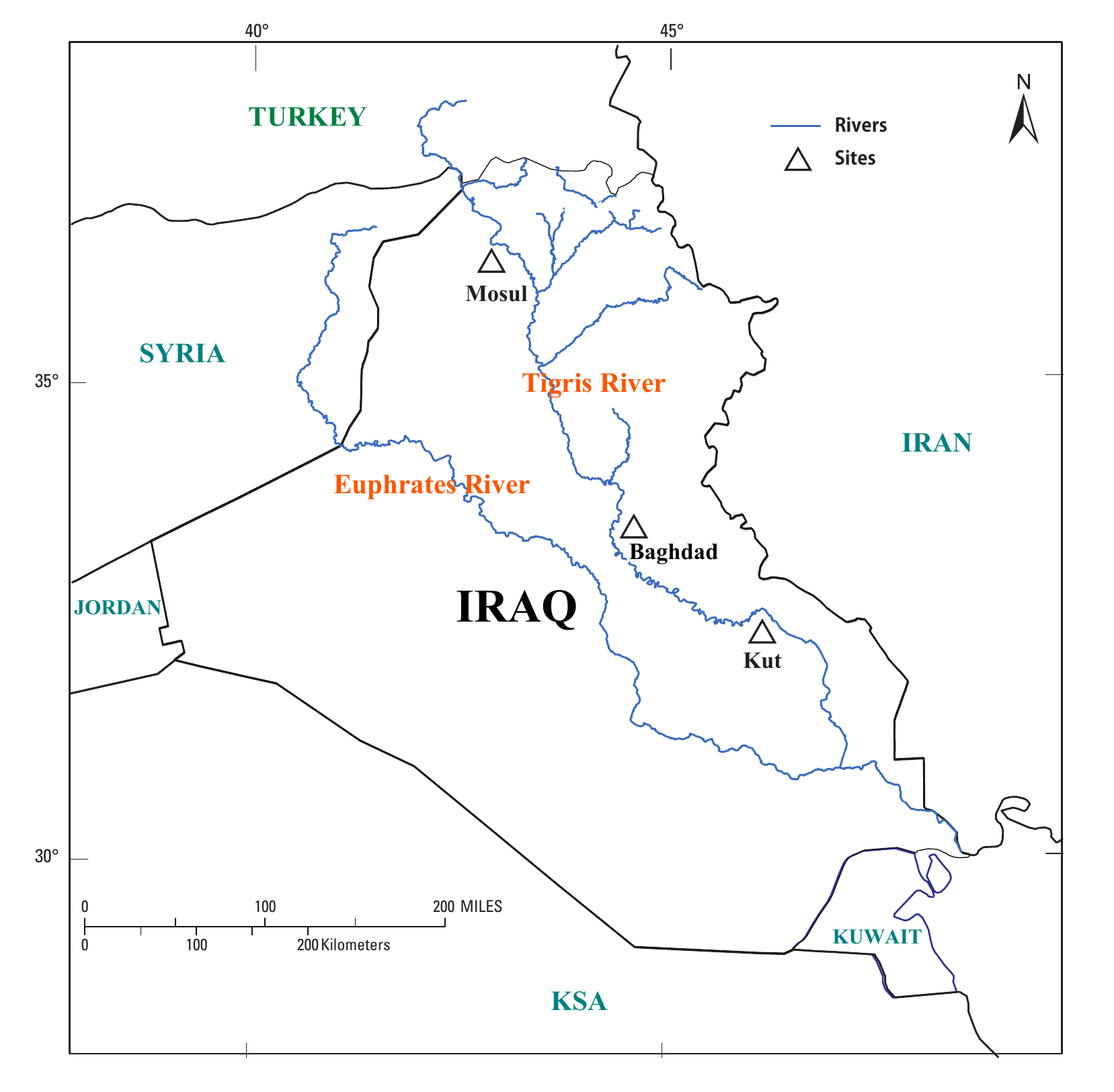}
\caption{The locations of the studied meteorological stations within Iraq region.}\label{fig:Fig_5}
\end{figure}

\begin{figure}[!htbp]
  \centering
  \includegraphics[width=5.9cm]{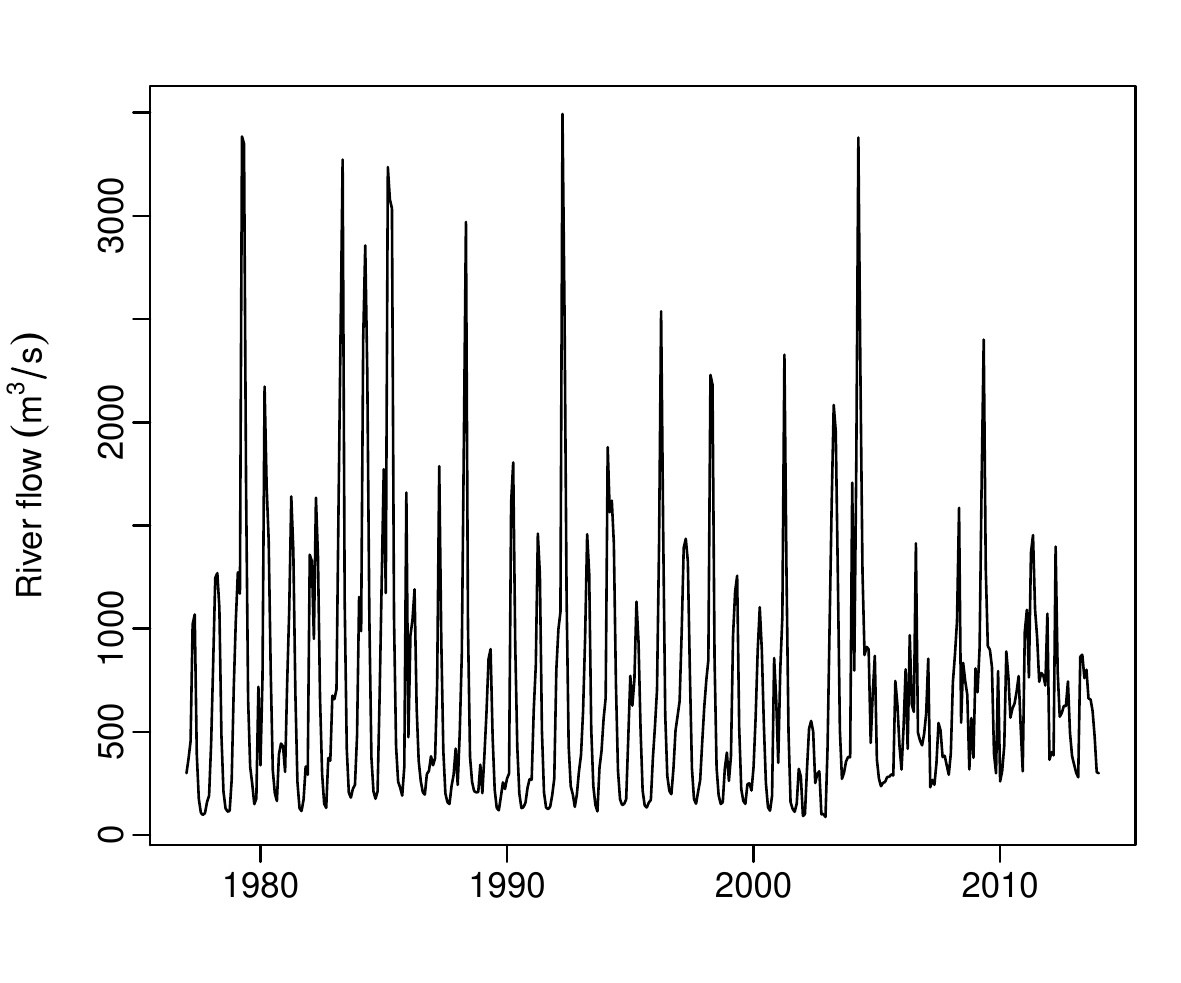}
\qquad
  \includegraphics[width=5.9cm]{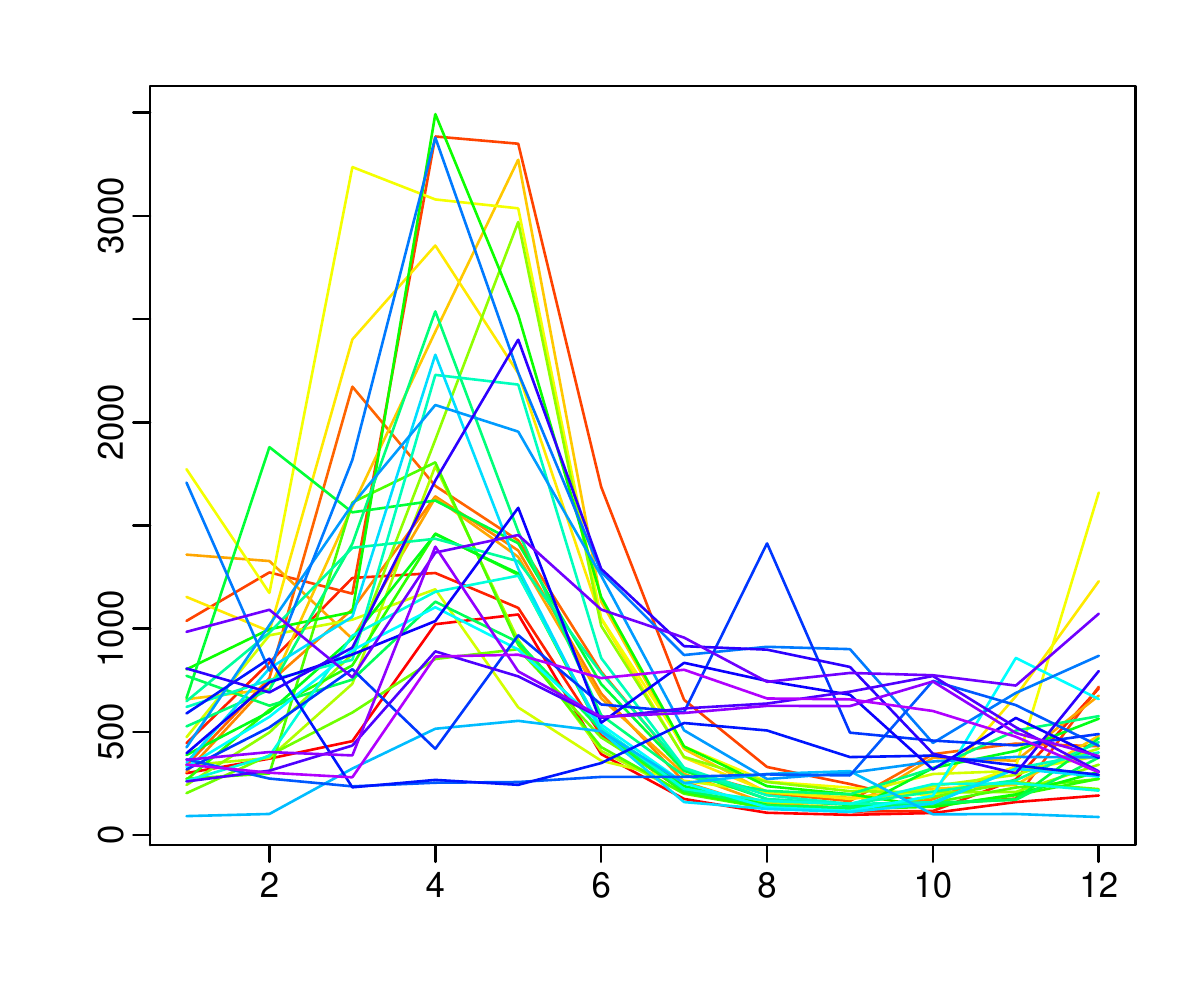}
  \\
  \includegraphics[width=5.9cm]{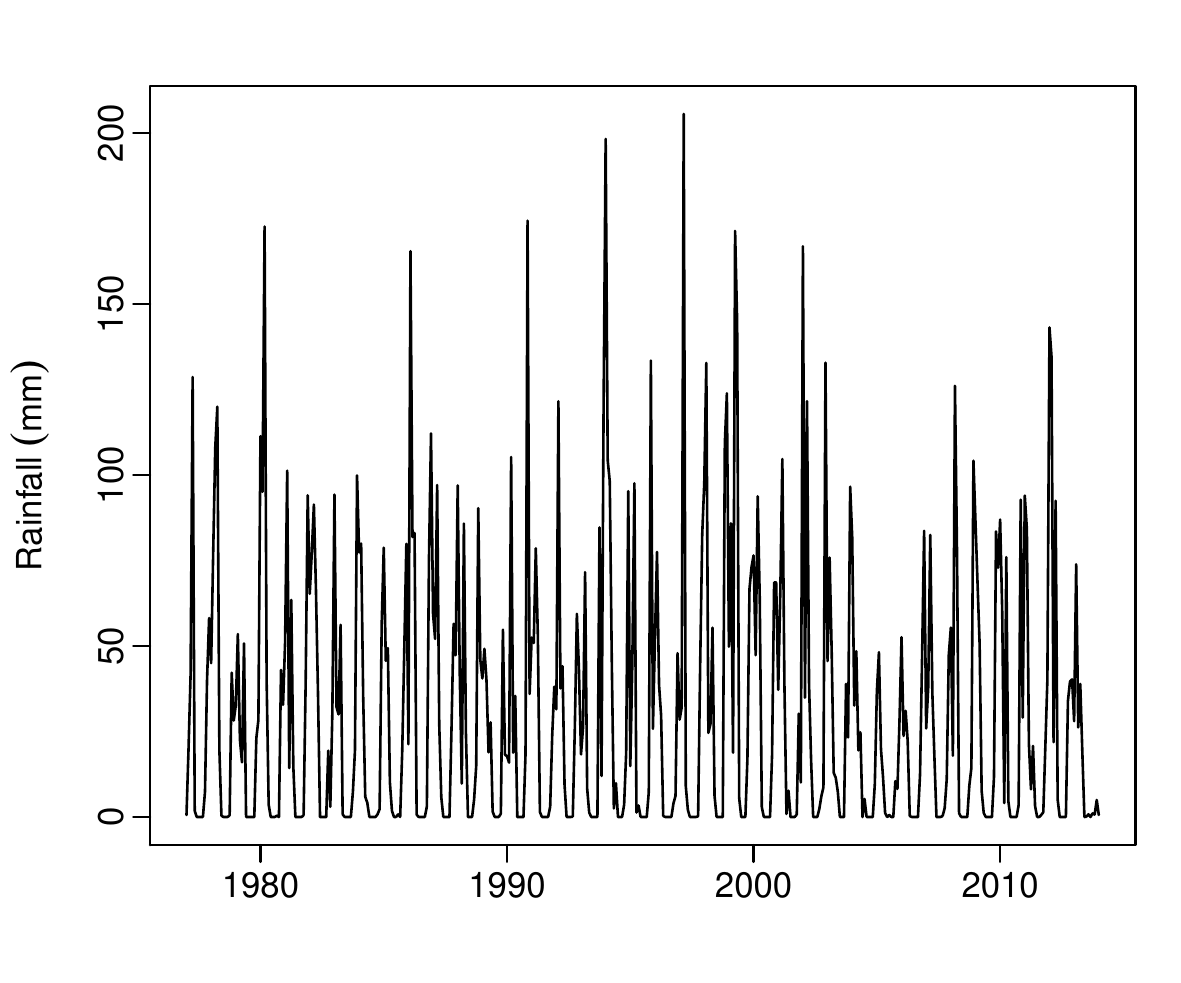}
\qquad
  \includegraphics[width=5.9cm]{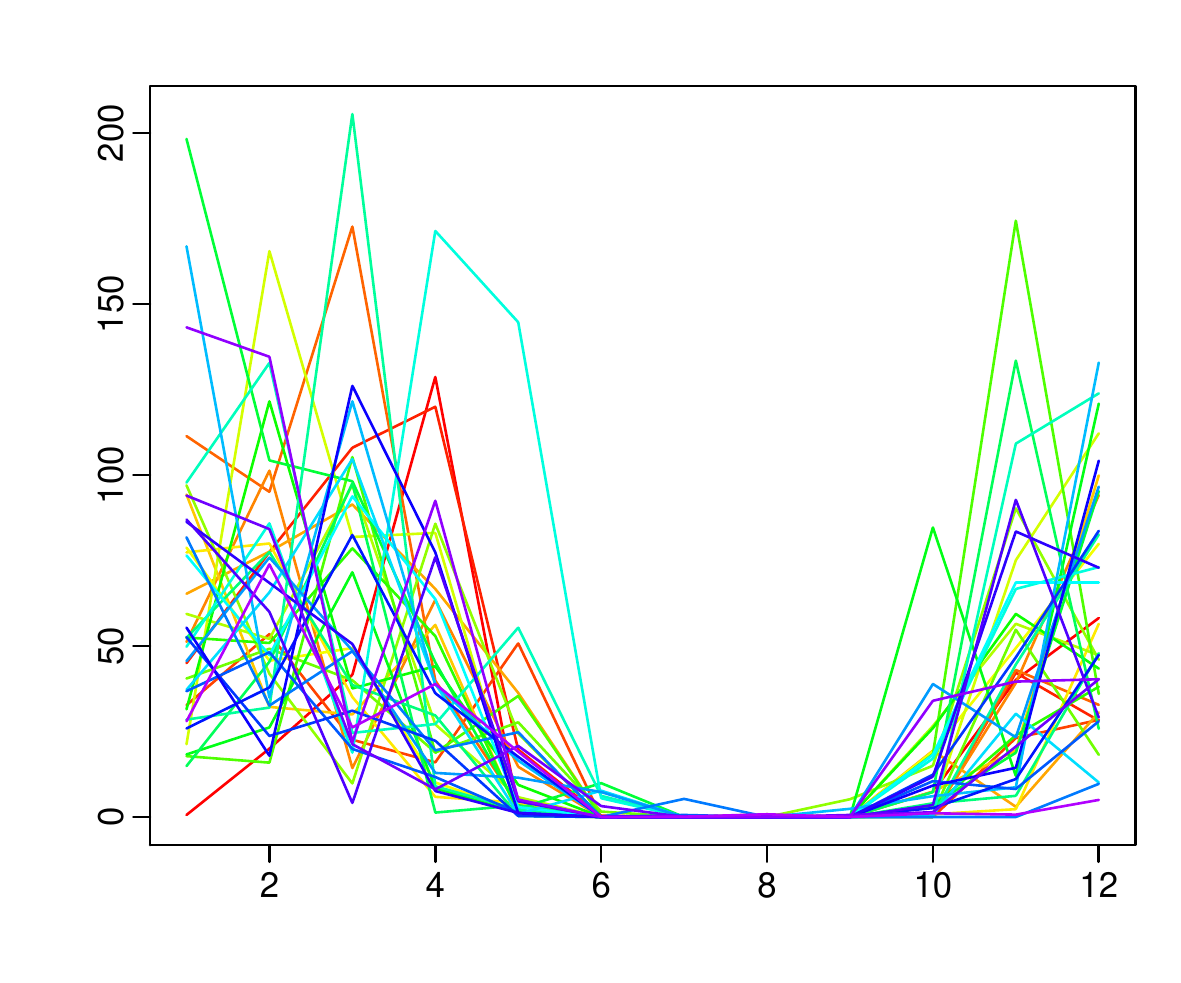}
  \\
  \includegraphics[width=5.9cm]{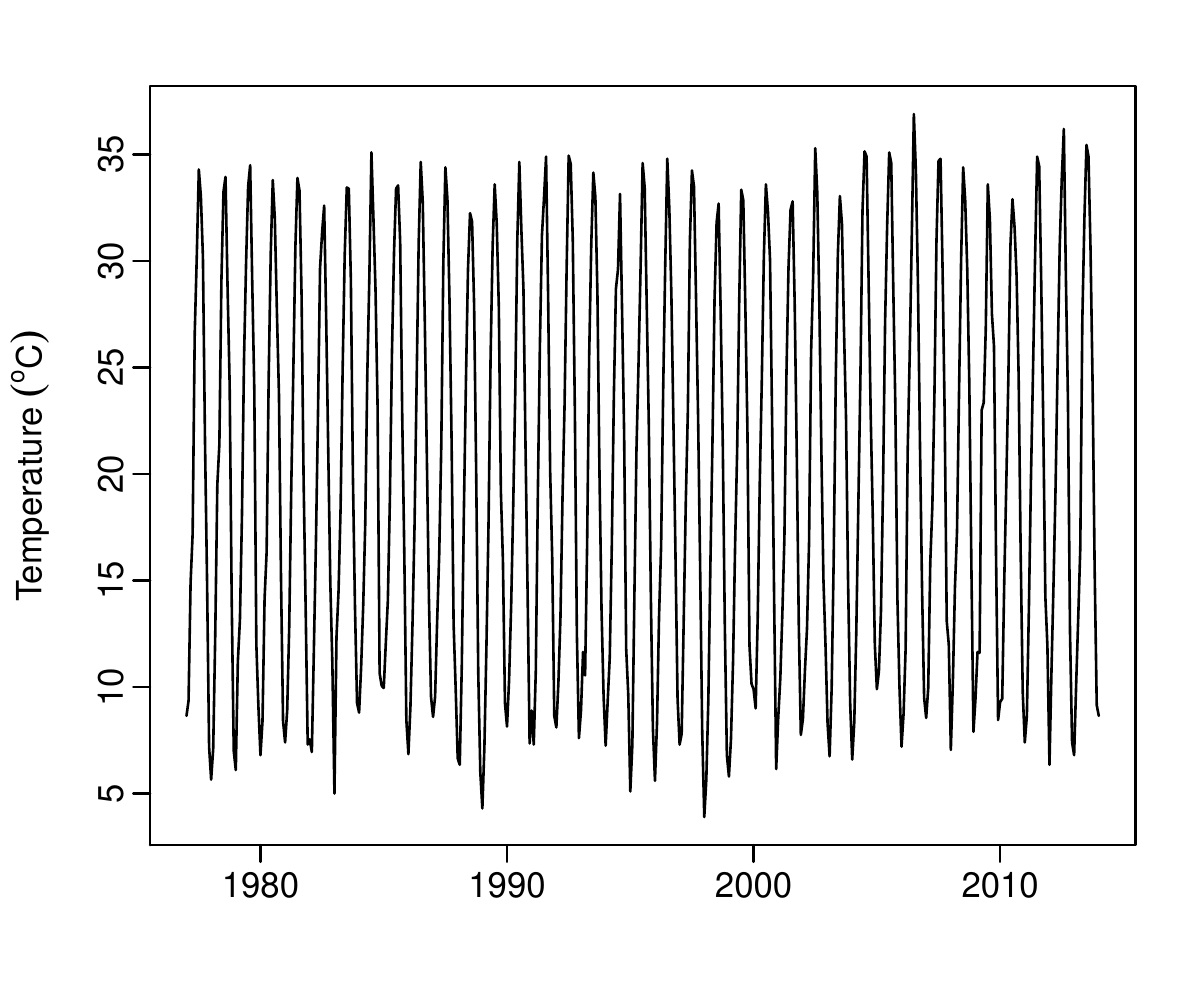}
\qquad
  \includegraphics[width=5.9cm]{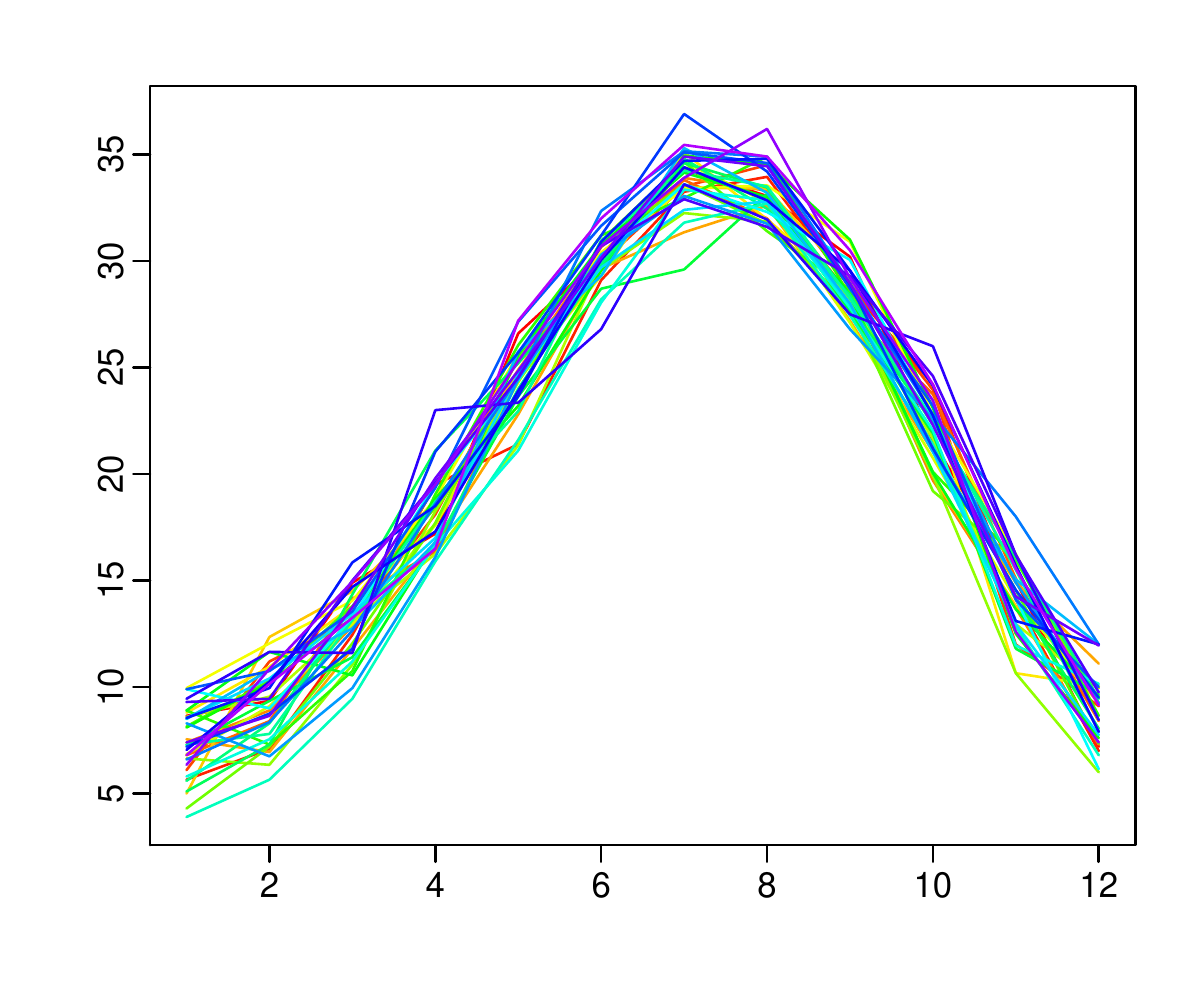}
  \\
  \includegraphics[width=5.9cm]{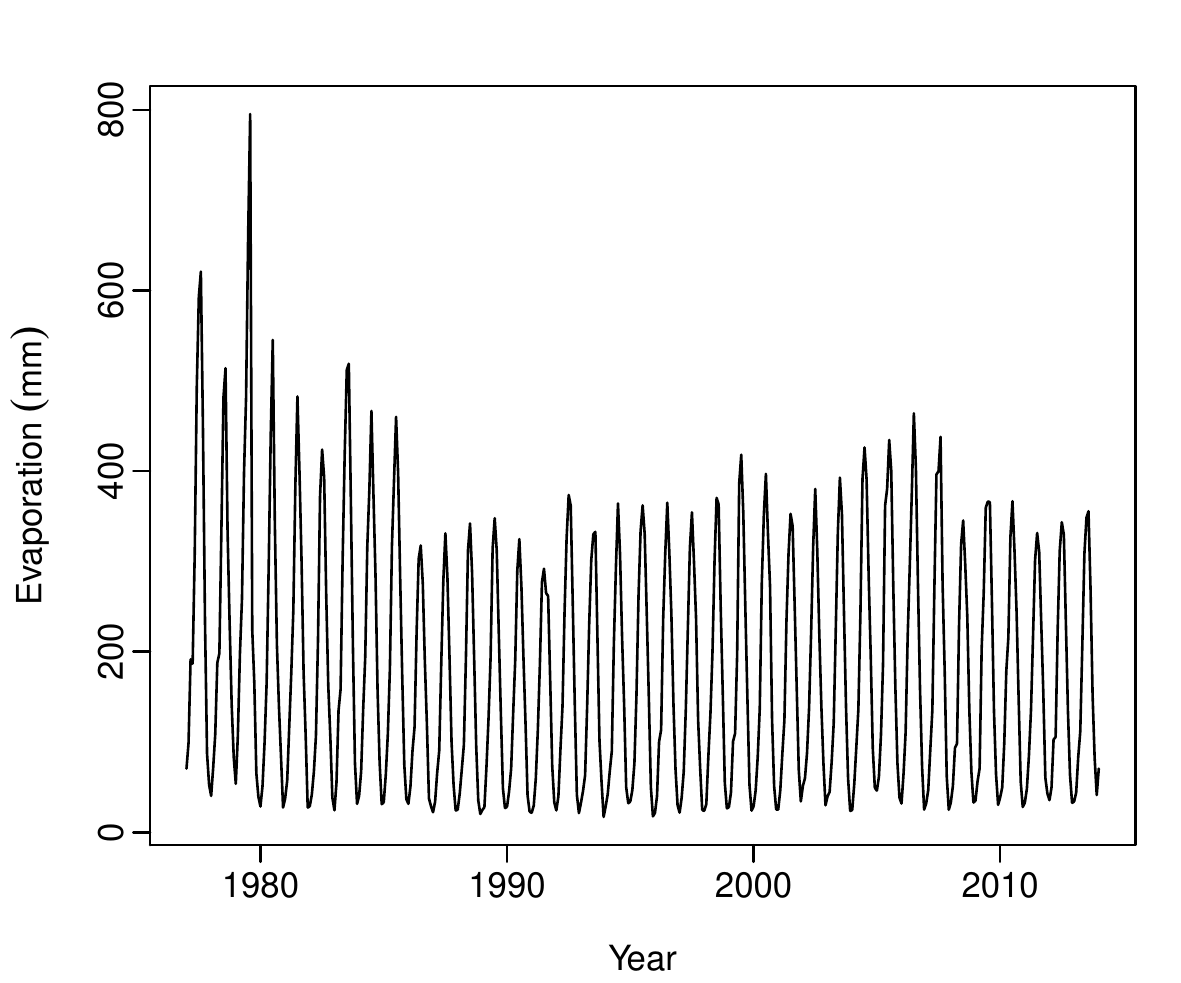}
\qquad
  \includegraphics[width=5.9cm]{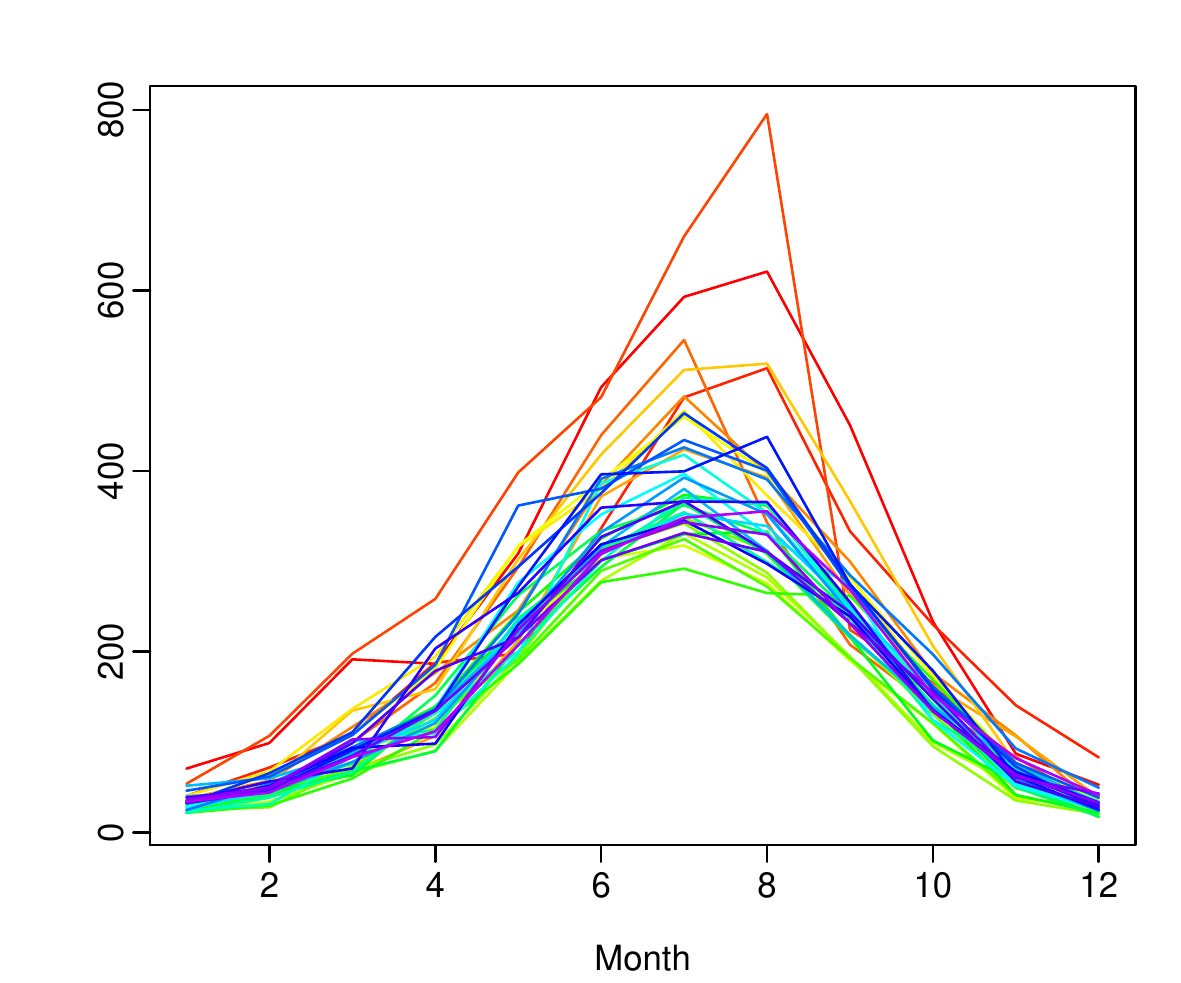}
  \caption{Traditional time series (first column) and functional time series (second column) plots of the variables: river flow (first row), rainfall (second row), temperature (third row), and evaporation (fourth row) for Mosul station.}
  \label{fig:Fig_6}
\end{figure}

\begin{figure}[!htbp]
  \centering
  \includegraphics[width=5.9cm]{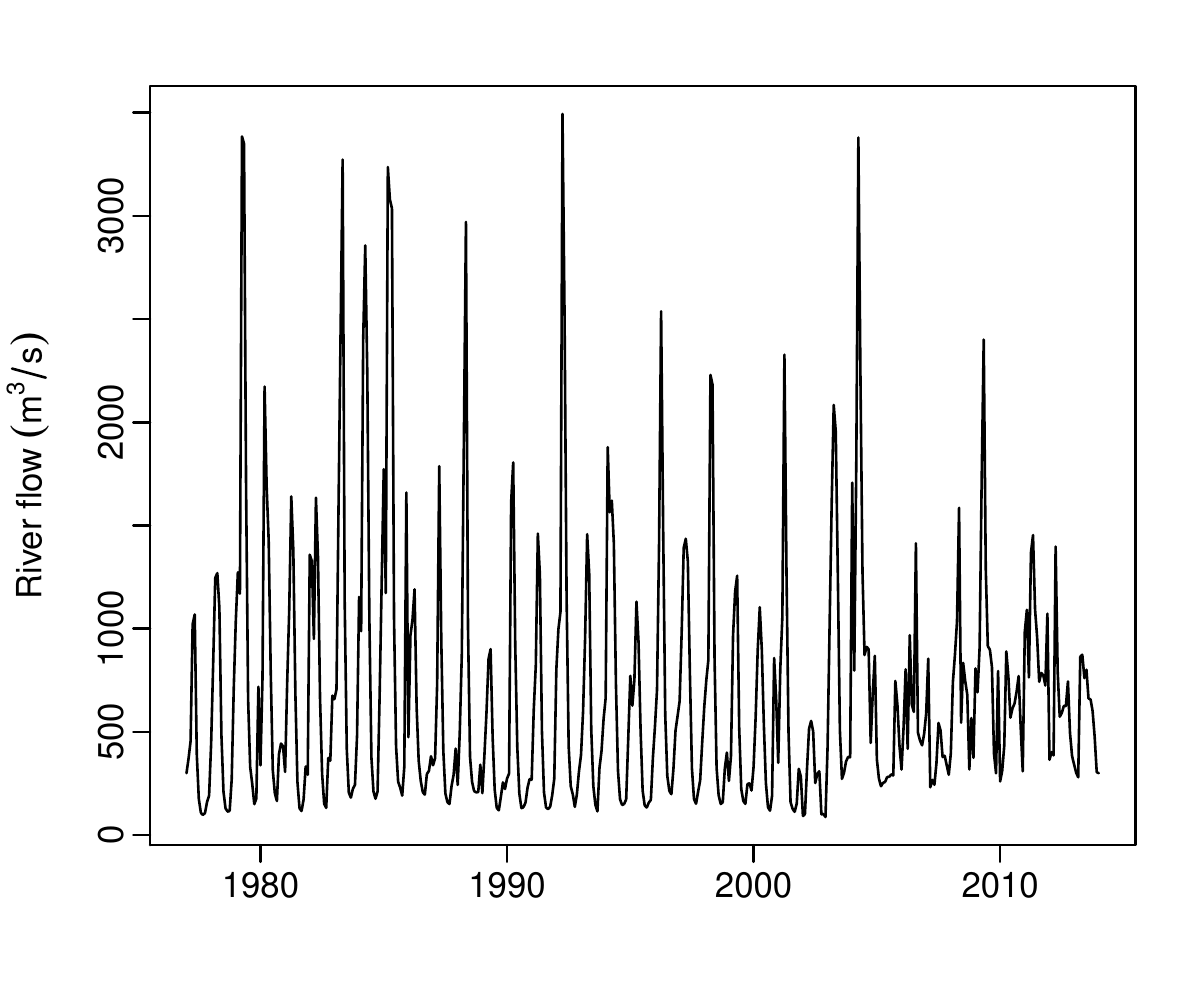}
\qquad
  \includegraphics[width=5.9cm]{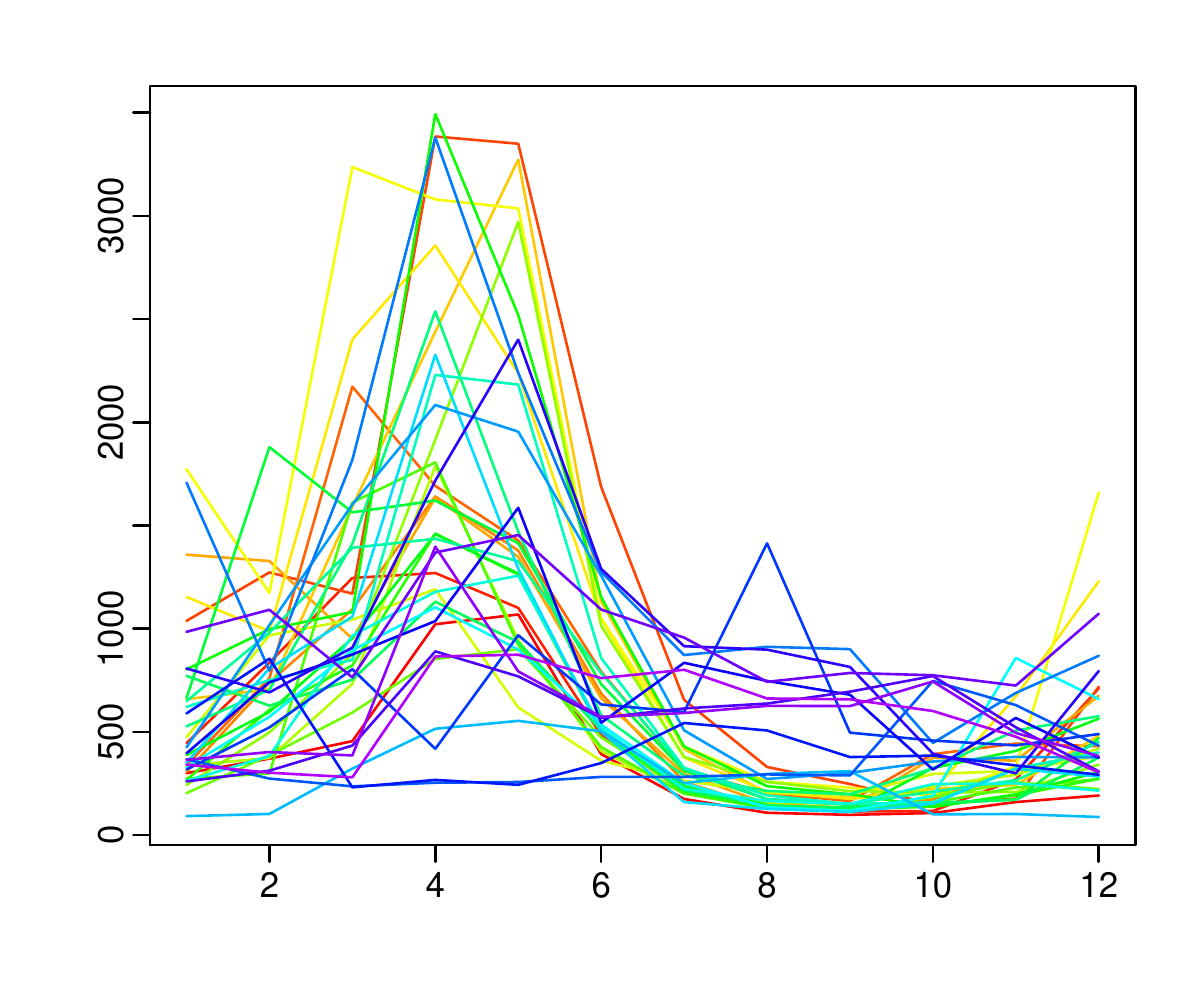}
  \\
  \includegraphics[width=5.9cm]{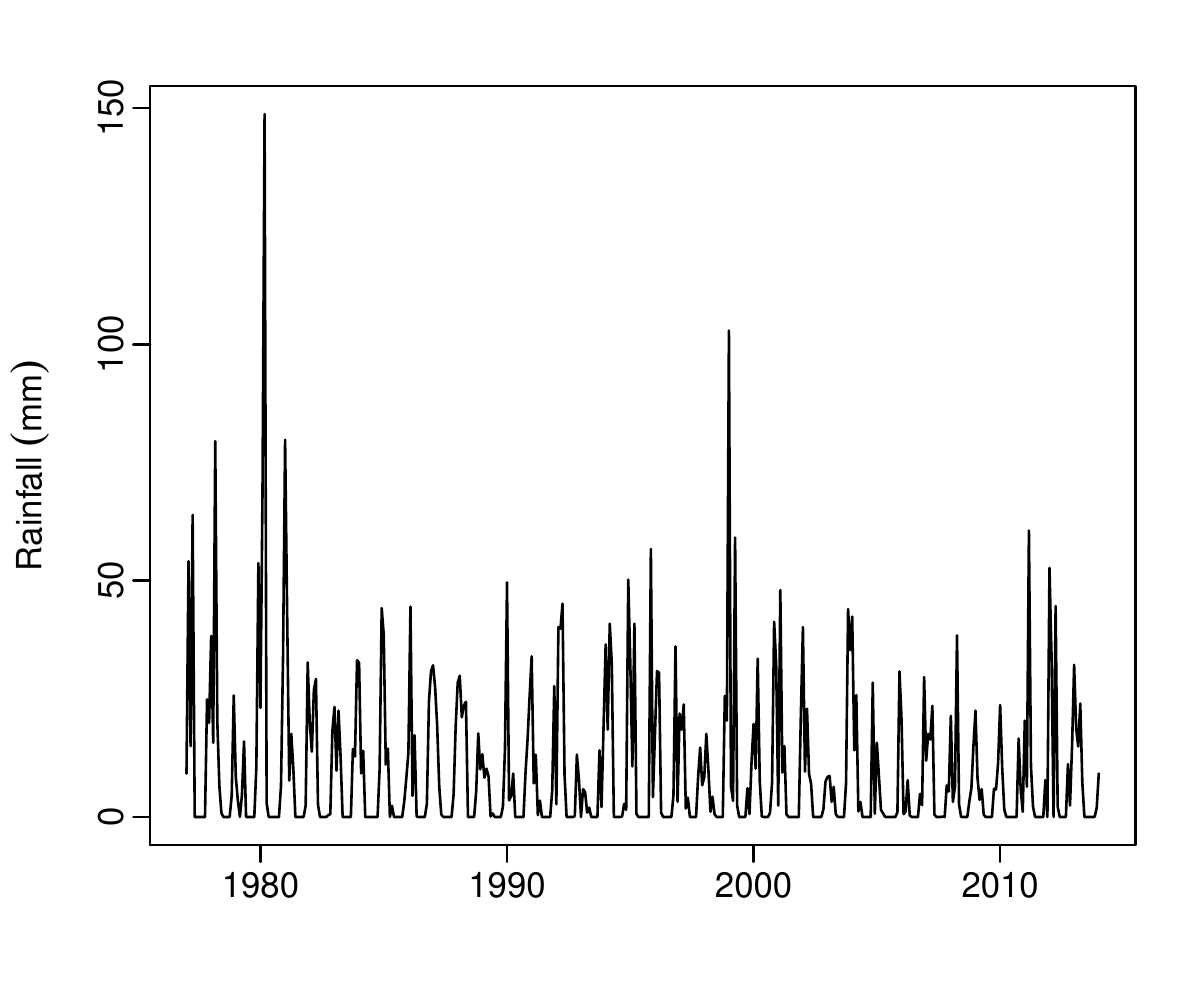}
\qquad
  \includegraphics[width=5.9cm]{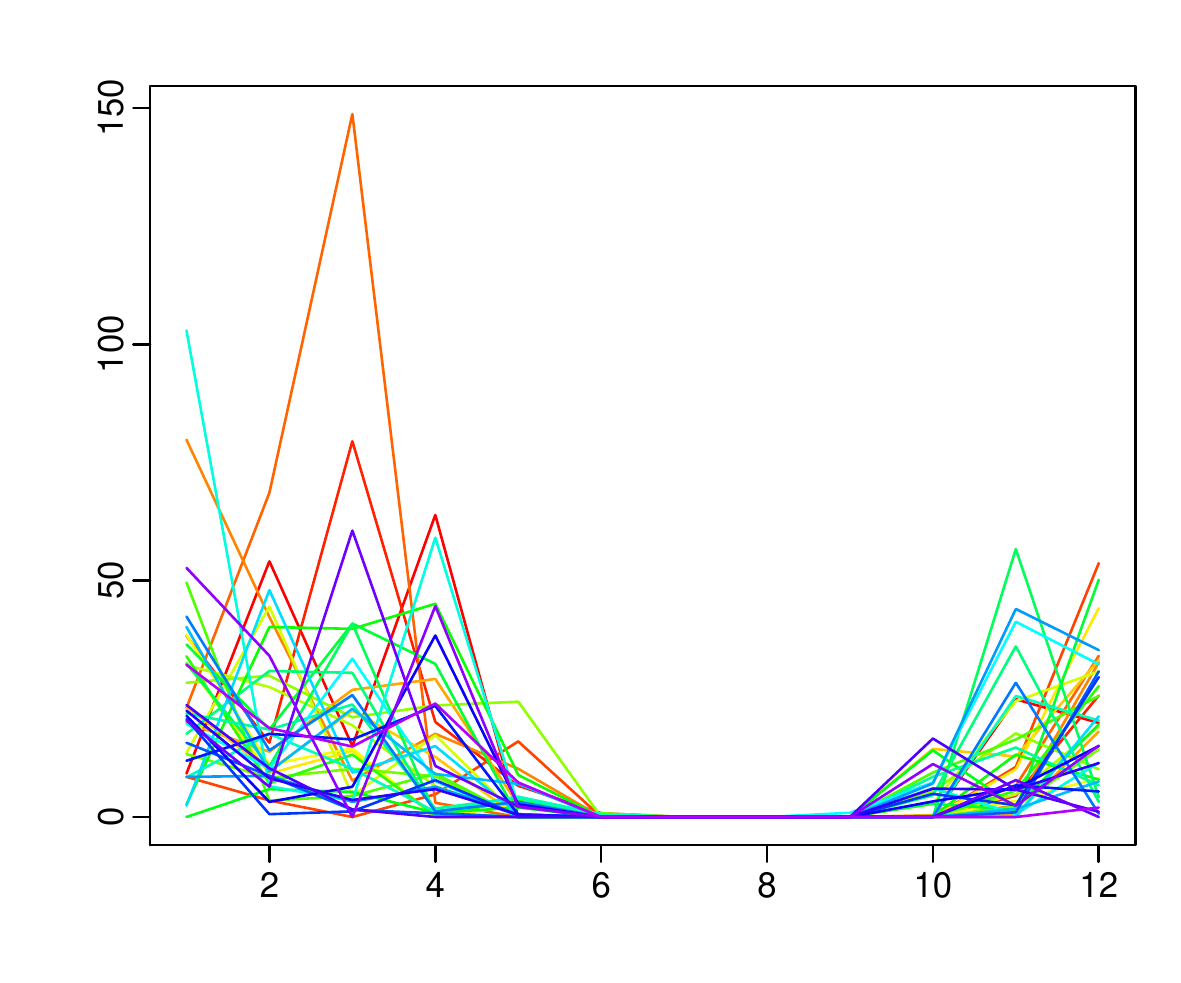}
  \\
  \includegraphics[width=5.9cm]{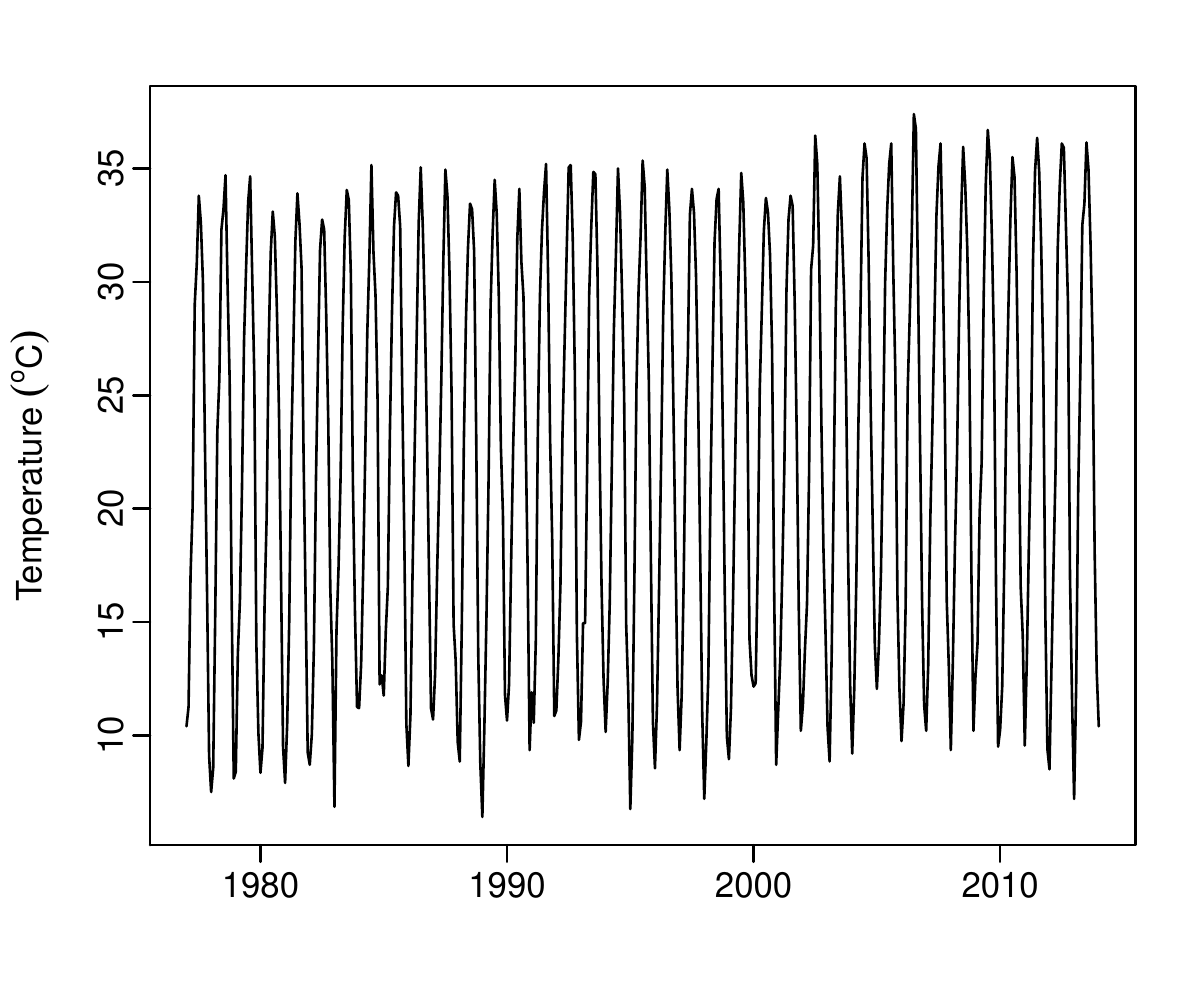}
\qquad
  \includegraphics[width=5.9cm]{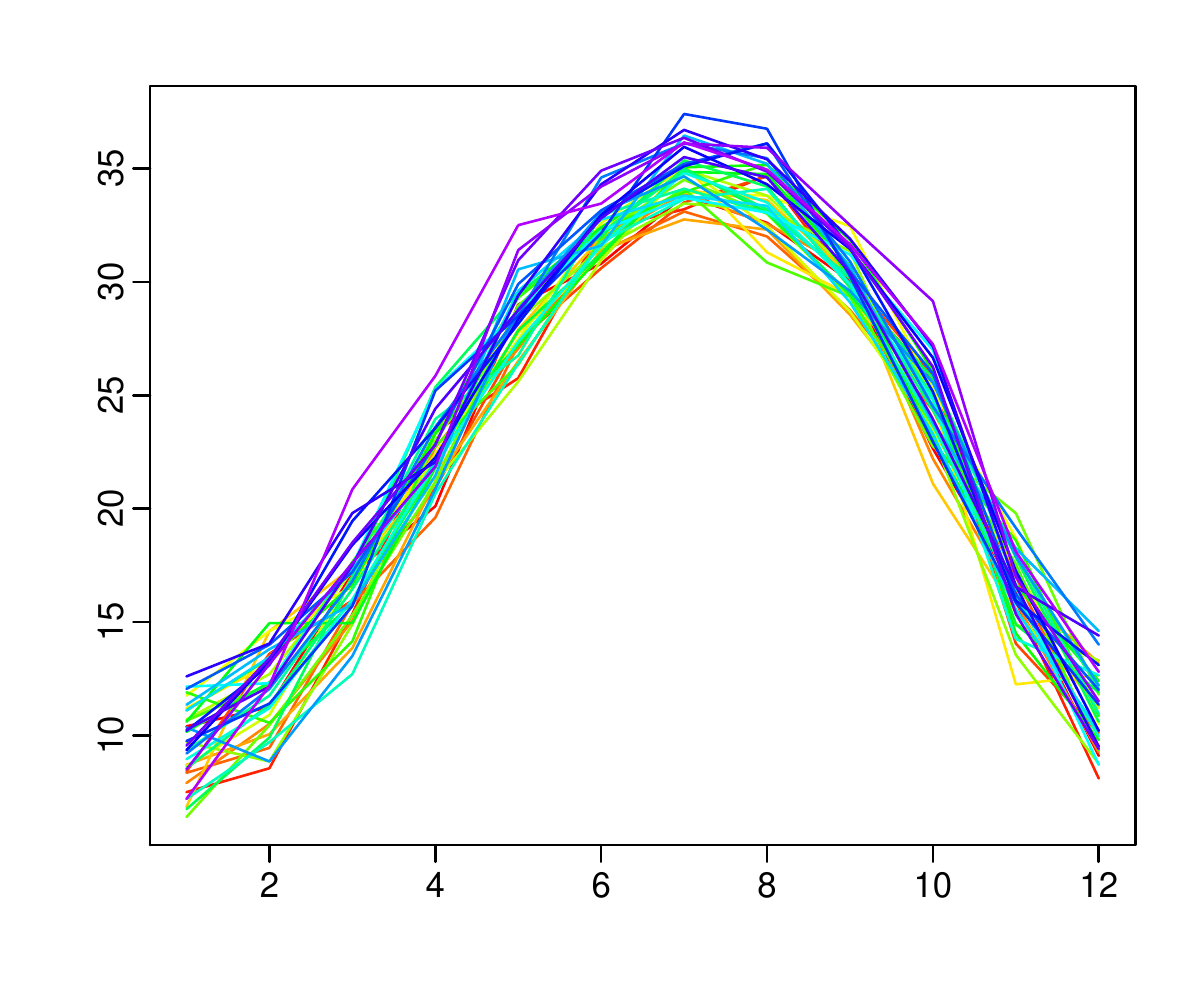}
  \\
  \includegraphics[width=5.9cm]{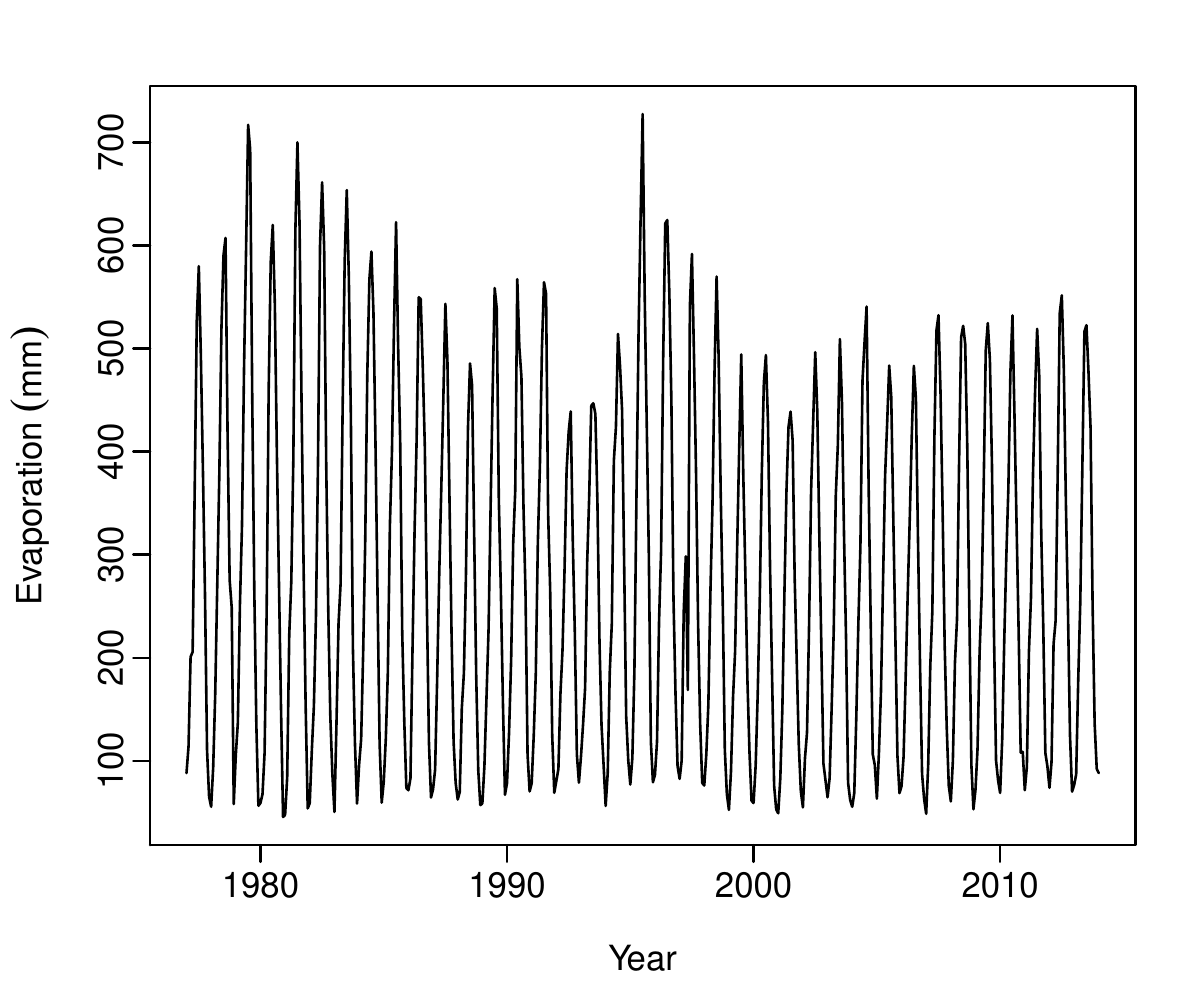}
\qquad
  \includegraphics[width=5.9cm]{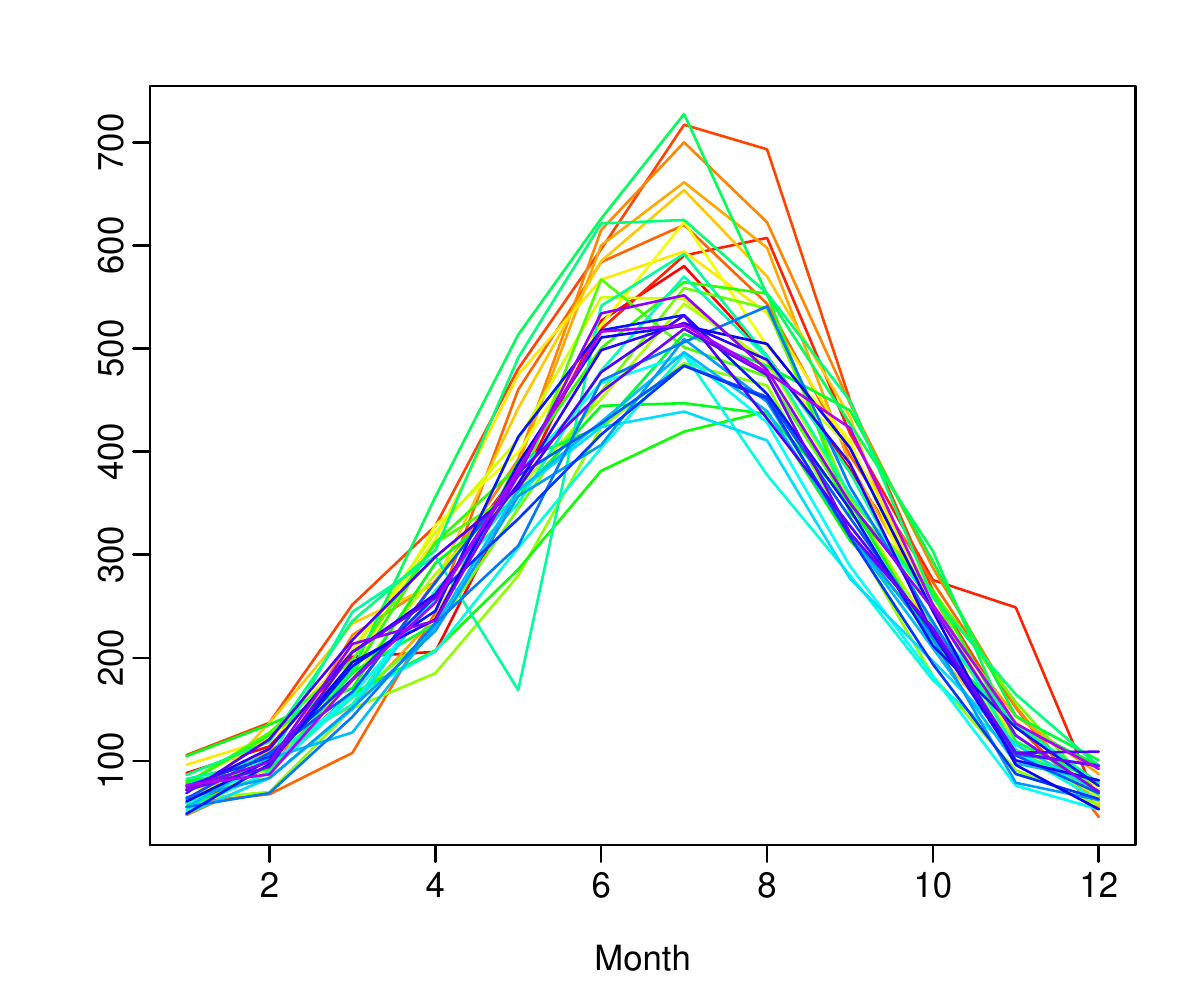}
  \caption{Traditional time series (first column) and functional time series (second column) plots of the variables: river flow (first row), rainfall (second row), temperature (third row), and evaporation (fourth row) for Baghdad station.}
  \label{fig:Fig_7}
\end{figure}

\begin{figure}[!htbp]
  \centering
  \includegraphics[width=5.9cm]{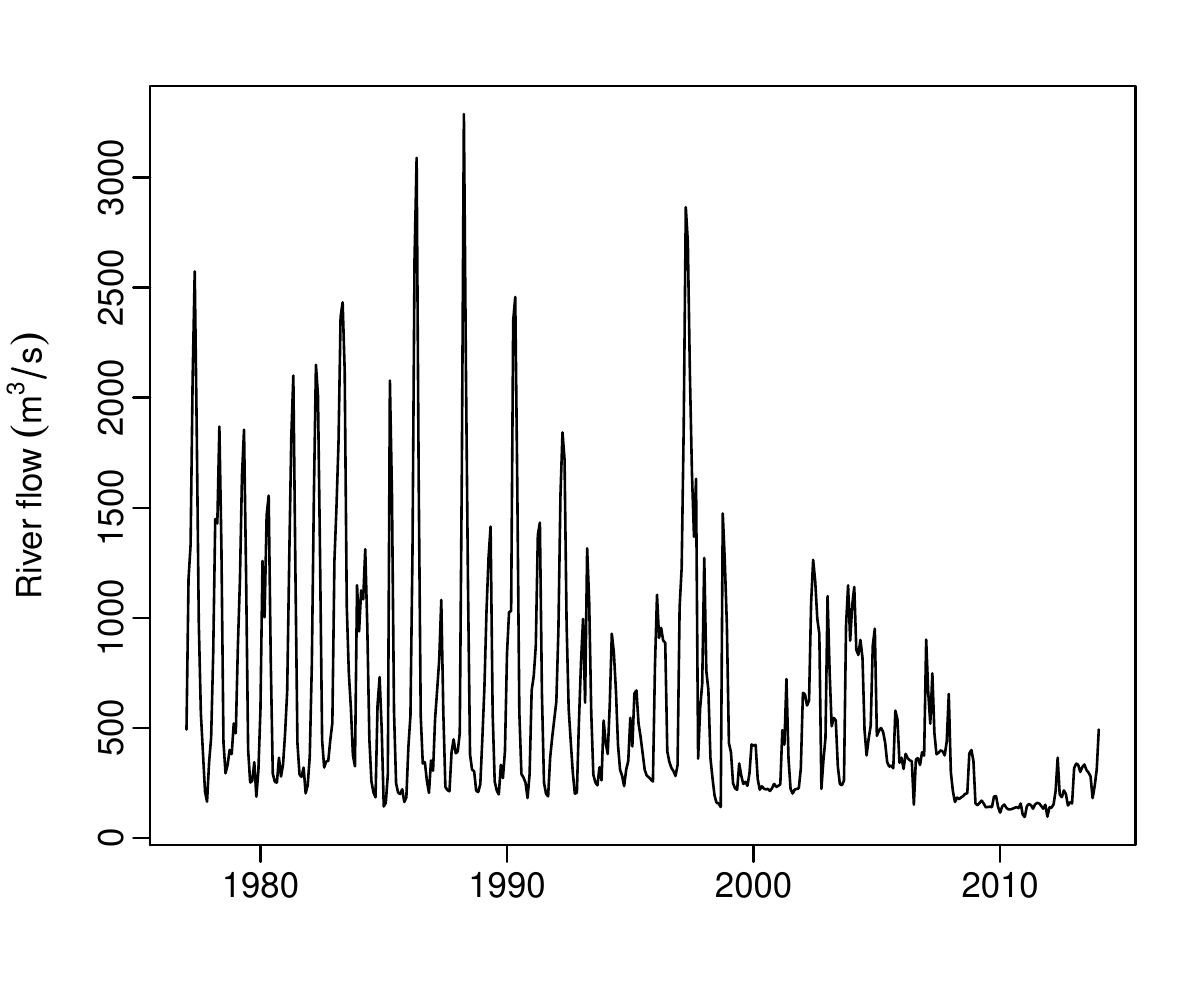}
\qquad
  \includegraphics[width=5.9cm]{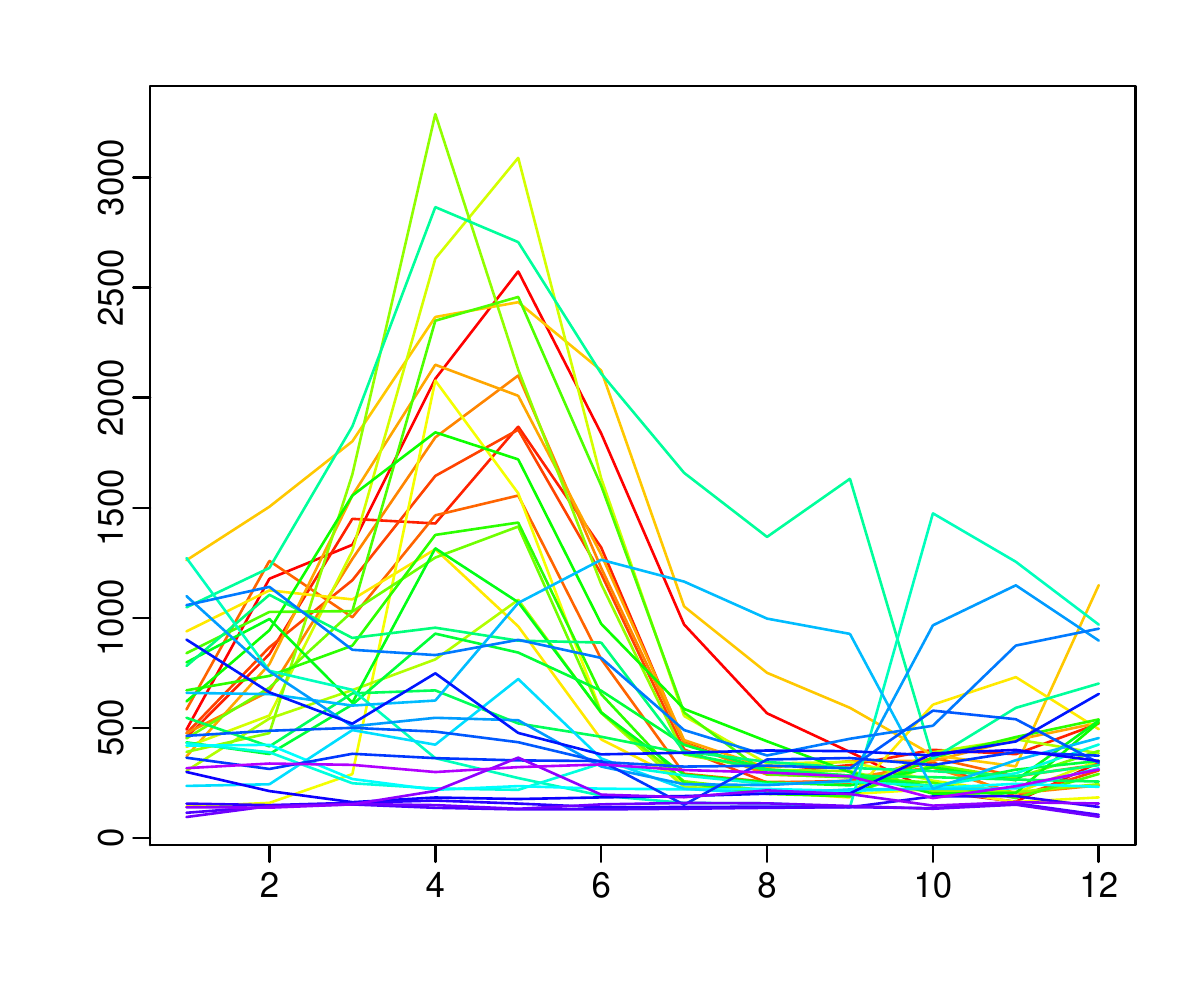}
  \\
  \includegraphics[width=5.9cm]{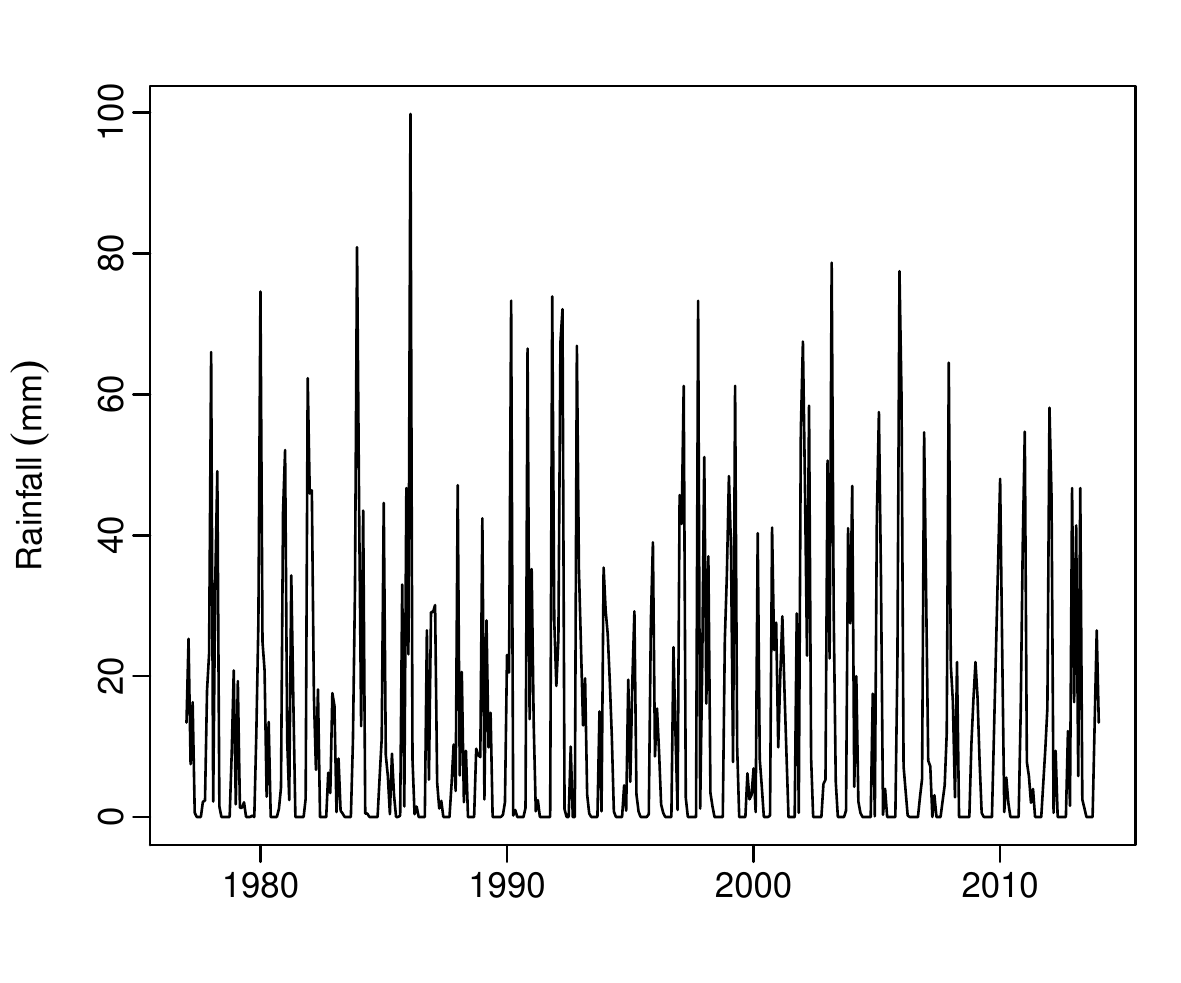}
\qquad
  \includegraphics[width=5.9cm]{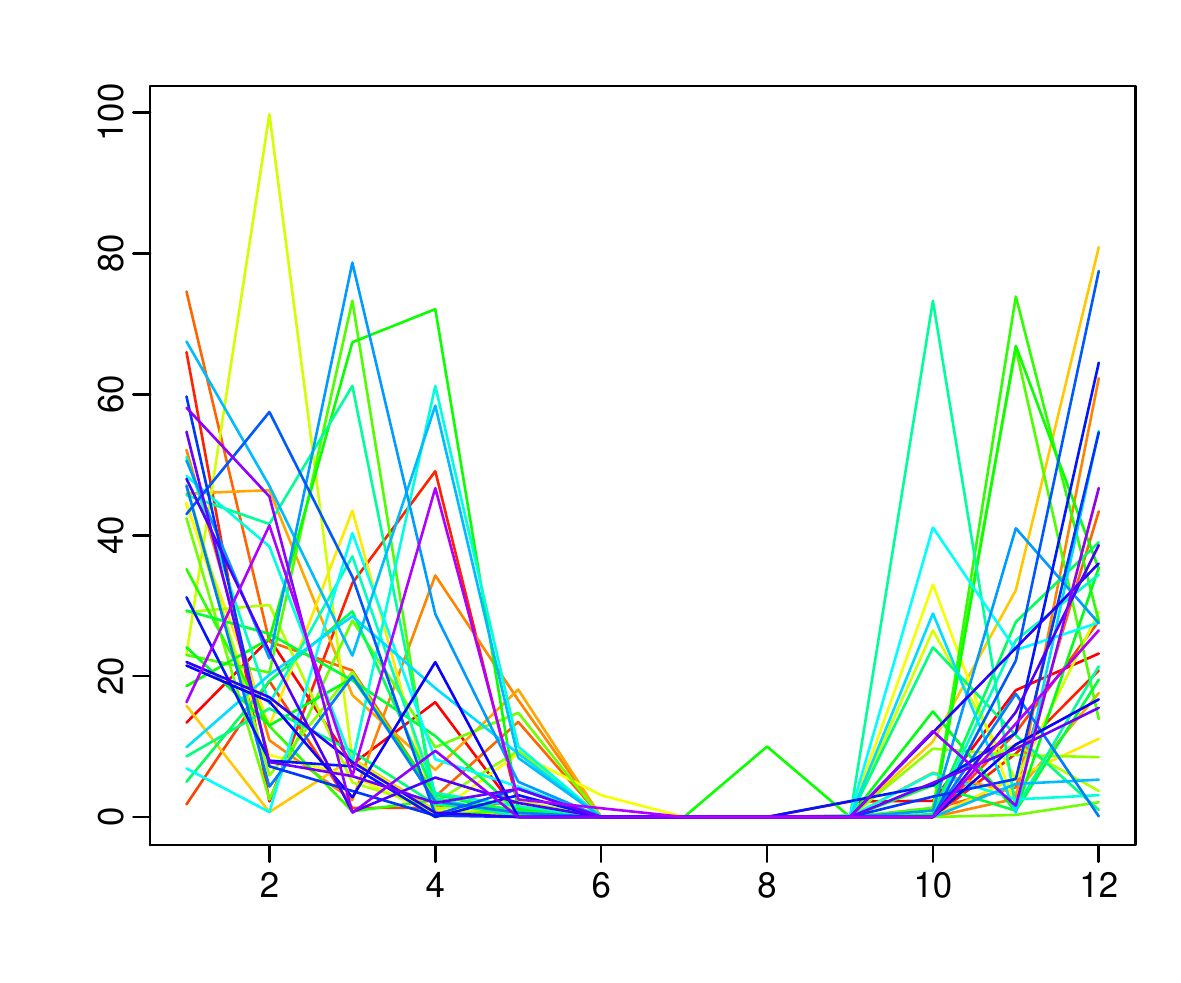}
  \\
  \includegraphics[width=5.9cm]{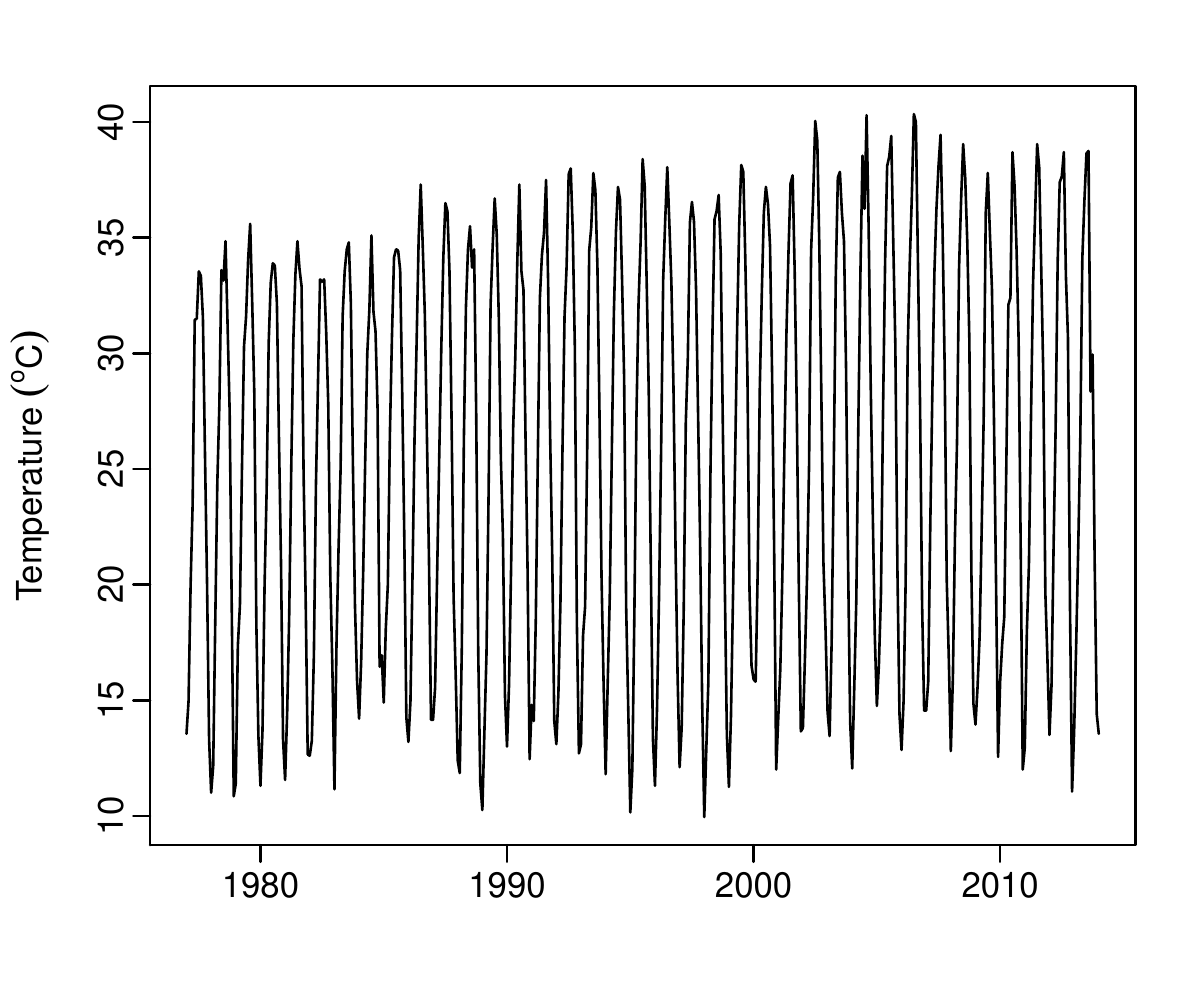}
\qquad
  \includegraphics[width=5.9cm]{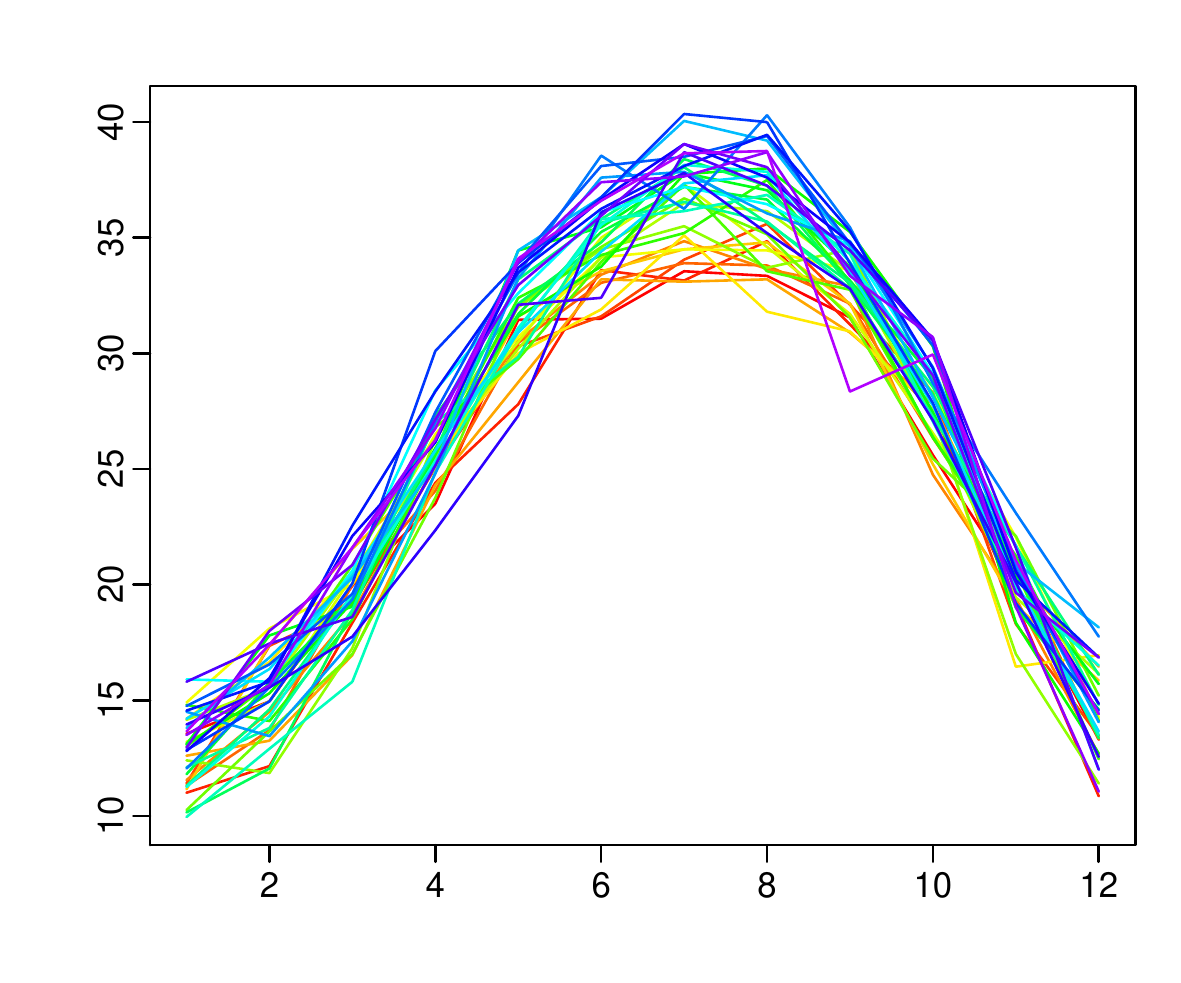}
  \\
  \includegraphics[width=5.9cm]{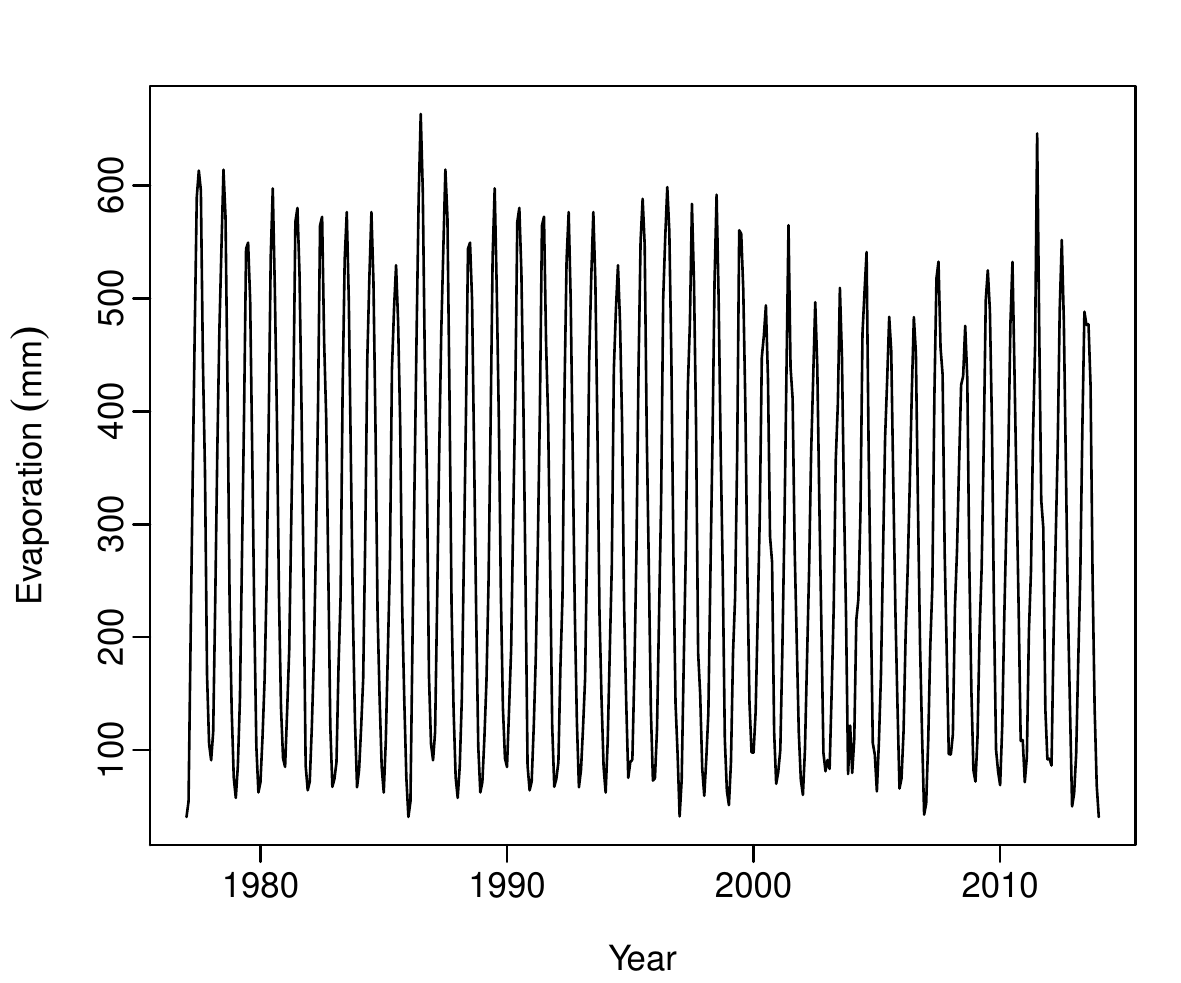}
\qquad
  \includegraphics[width=5.9cm]{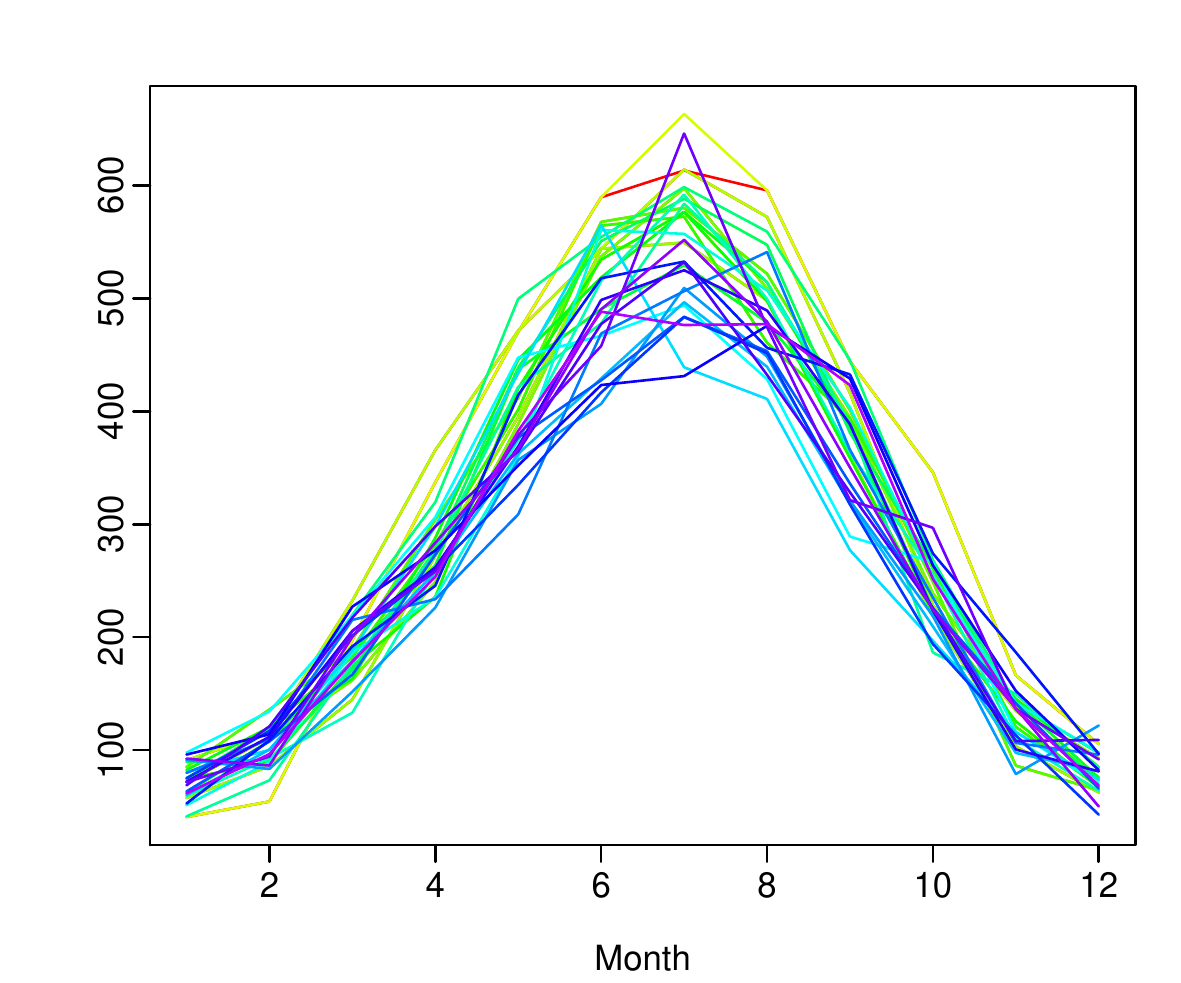}
  \caption{Traditional time series (first column) and functional time series (second column) plots of the variables: river flow (first row), rainfall (second row), temperature (third row), and evaporation (fourth row) for Kut station.}
  \label{fig:Fig_8}
\end{figure}

The studied station locations are presented in Figure~\ref{fig:Fig_5}. For each station, the dataset includes four monthly variables: river flow, rainfall, temperature, and evaporation from January 1977 to December 2013 (37 years in total). The observations for all variables were considered the functions of months ($1 \leq s \leq 12$). See Figures~\ref{fig:Fig_6}-\ref{fig:Fig_8} for the plots of variables of all three stations. Also, the pointwise summary statistics of the functional variables are presented in Table \ref{tab:summary_stat}.

\begin{center}
\tabcolsep 0.15in
\begin{longtable}{@{}llcccccc@{}}
\caption{Pointwise summary statistics of the functional time series.}\label{tab:summary_stat}\\
\toprule
Station & Variable & Min & Mean & Max & STD & Skewness & Kurtosis \\
\endfirsthead
Station & Variable & Min & Mean & Max & STD & Skewness & Kurtosis \\
\toprule
\endhead
\multicolumn{8}{r}{{Continued on next page}}\\
\endfoot
\endlastfoot
\midrule
\multirow{4}{*}{Mosul} & River flow & 87.70 & 701.50 & 3494.00 & 633.02 & 1.97 & 7.49 \\
& Rainfall & 0.00 & 30.77 & 205.60 & 39.84 & 1.56 & 5.27 \\
& Temperature & 3.90 & 20.28 & 36.90 & 9.47 & 0.05 & 1.58 \\
& Evaporation & 17.00 & 182.56 & 795.50 & 139.45 & 0.80 & 3.21 \\
\midrule
\multirow{4}{*}{Baghdad} & River flow & 292.70 & 833.30 & 2865.00 & 503.36 & 1.84 & 6.16 \\
& Rainfall & 0.00 & 10.07 & 148.70 & 16.43 & 3.00 & 17.71 \\
& Temperature & 6.40 & 22.73 & 37.40 & 9.05 & -0.11 & 1.57 \\
& Evaporation & 45.60 & 297.40 & 727.60 & 139.45 & 0.40 & 1.96 \\
\midrule
\multirow{4}{*}{Kut} & River flow & 95.50 & 610.20 & 3289.00 & 555.43 & 1.92 & 6.85 \\
& Rainfall & 0.00 & 12.67 & 99.80 & 18.96 & 1.77 & 5.64 \\
& Temperature & 9.95 & 25.77 & 40.35 & 8.78 & -0.14 & 1.60 \\
& Evaporation & 40.80 & 286.3 & 663.10 & 174.4 & 0.53 & 2.06 \\
\bottomrule
\end{longtable}
\end{center}

\section{Application results and assessment}\label{sec:results}

In this section, the modeling results of the prediction process are presented. The following procedure was used for evaluating the prediction performance of the models. The dataset was split into a training sample for each station, consisting of years from 1977 to 2006 (30 years in total) and a testing sample, consisting of years from 2007 to 2013 (7 years in total). Herein, with the proposed functional model, it was aimed to predict the whole trajectory for one year after the training sample (i.e., 12-months prediction) using the entire dataset in the training sample. Therefore, with the historical data from 1977 to 2006, using the functional models, the one-step-ahead predictions of river flows were obtained for 2007 (i.e., the whole trajectory for the year 2007). Then, the training dataset was extended with one year and the one-step-ahead prediction of river flows was obtained for the year 2008. The prediction process was progressed through this expanding-window approach until the training samples cover the entire dataset. For non-functional models, two scenarios were considered to obtain predictions. In the first scenario, using the historical data from 1977 to 2007, twelve-steps-ahead monthly forecasts of river flows were obtained for 2008. Similar to the functional models, the twelve-step-ahead predictions for the year 2009 were obtained by increasing the training dataset by one year. The prediction process was repeated until the training samples cover the entire dataset. On the other hand, in the second scenario, the non-functional models were used to obtain a one-step-ahead monthly forecast of river flows for the first month of 2008. Then, the training dataset was extended by one month to obtain the predictions for the second month of 2008. Again, this expanding window approach was repeated until all the predictions are obtained. For each predicted curve, twelve-steps-ahead, and one-step ahead discrete values, five performance metrics; RMSPE, MAPE, RMESPE, PBIAS, RE, and two bootstrap performance metrics; score, and CPD were computed to evaluate the point and interval prediction accuracy of the proposed method against the benchmark models. It is worthwhile to note that, the predictive performance of the models may be affected by the presence of heteroscedasticity. Thus, for all the methods used in this study (except QR), all the numerical analyses were performed using the seasonally differenced time series. The predicted series and the point and interval forecasts are then converted to the original scale. On the other hand, the QR results were obtained using original data because the QR can model heteroscedasticity by perception and construction.

\begin{figure}[!htbp]
  \centering
  \includegraphics[width=4.8cm,height=4.2cm]{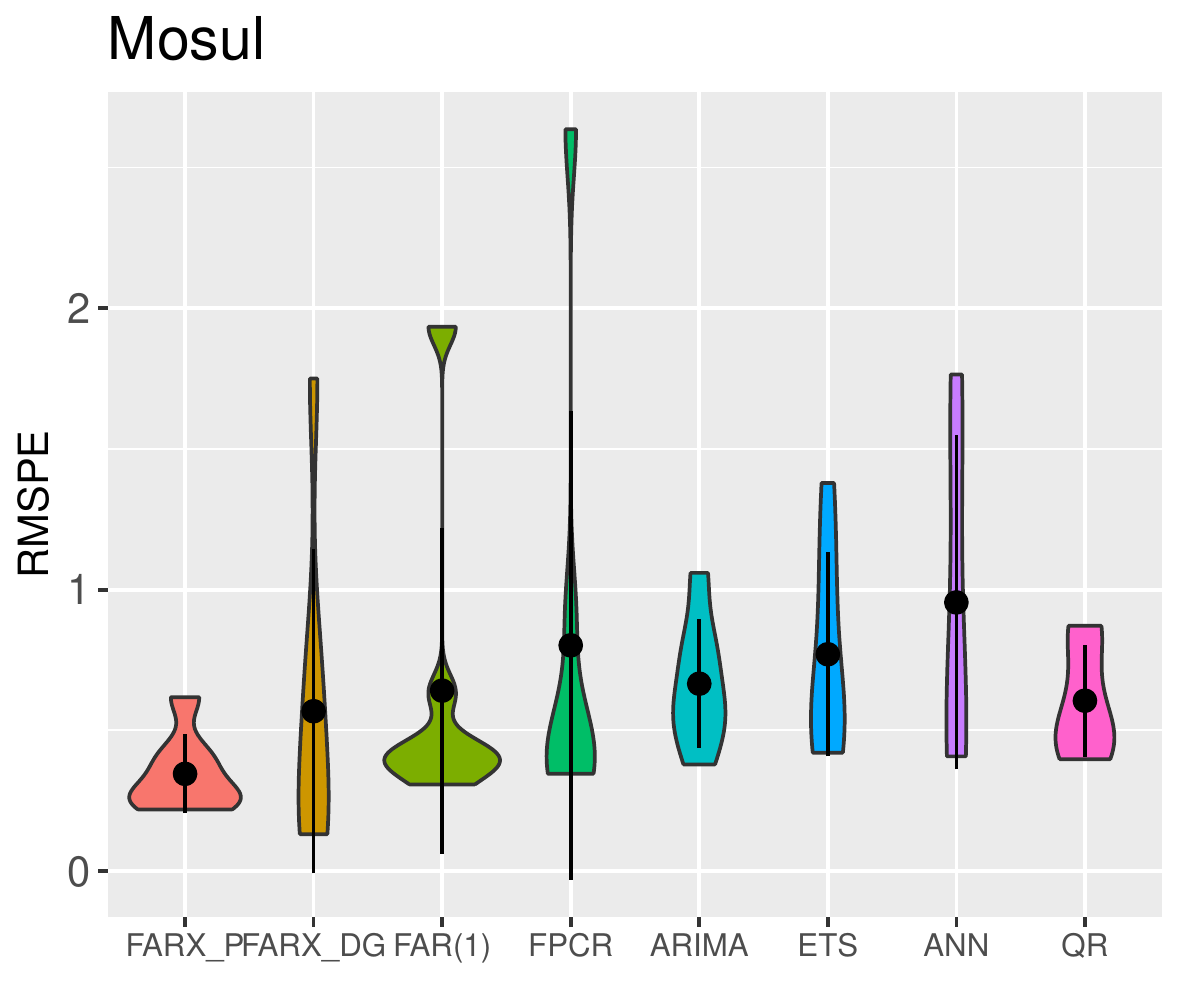}
  \includegraphics[width=4.8cm,height=4.2cm]{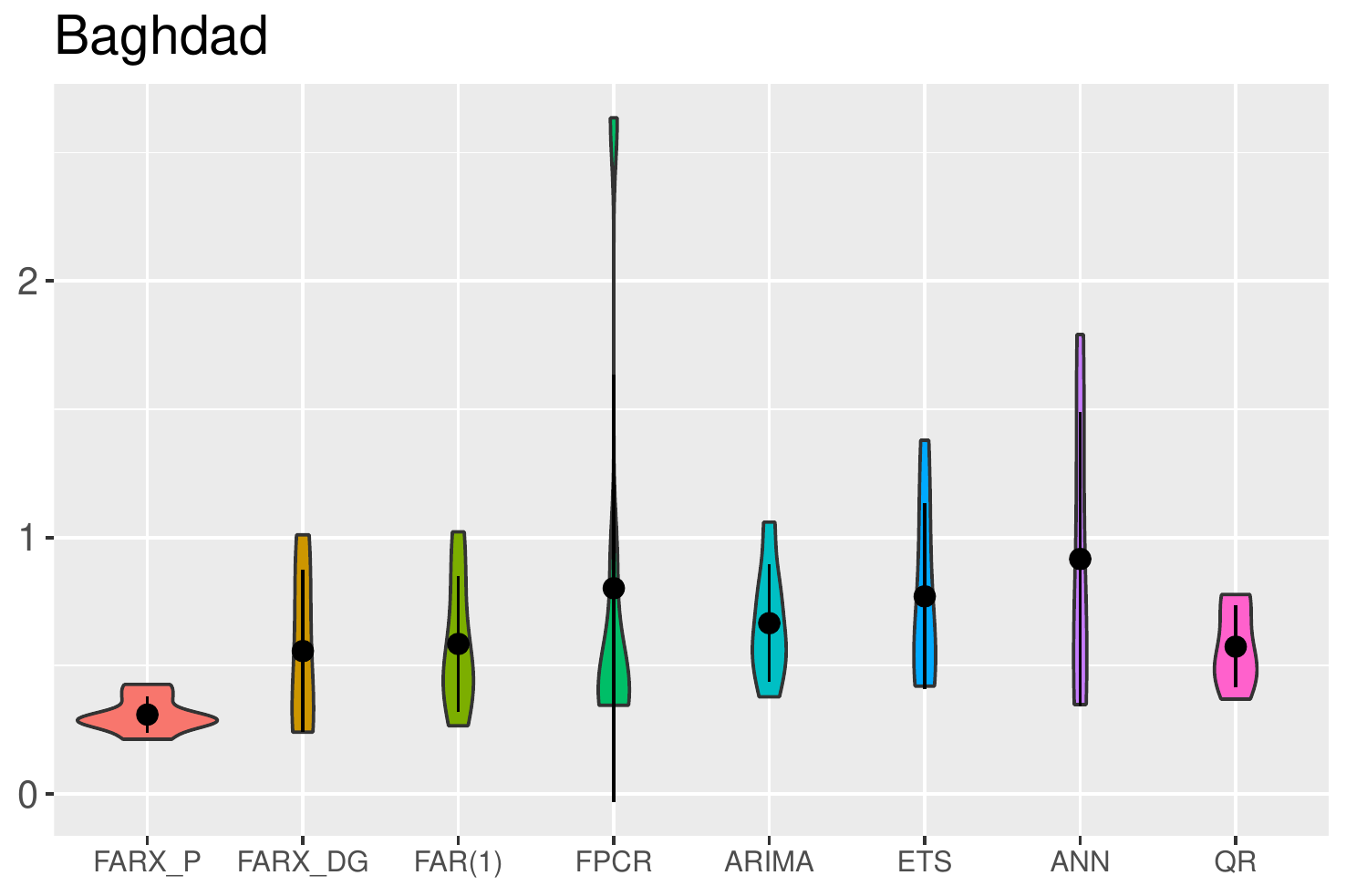}
  \includegraphics[width=4.8cm,height=4.2cm]{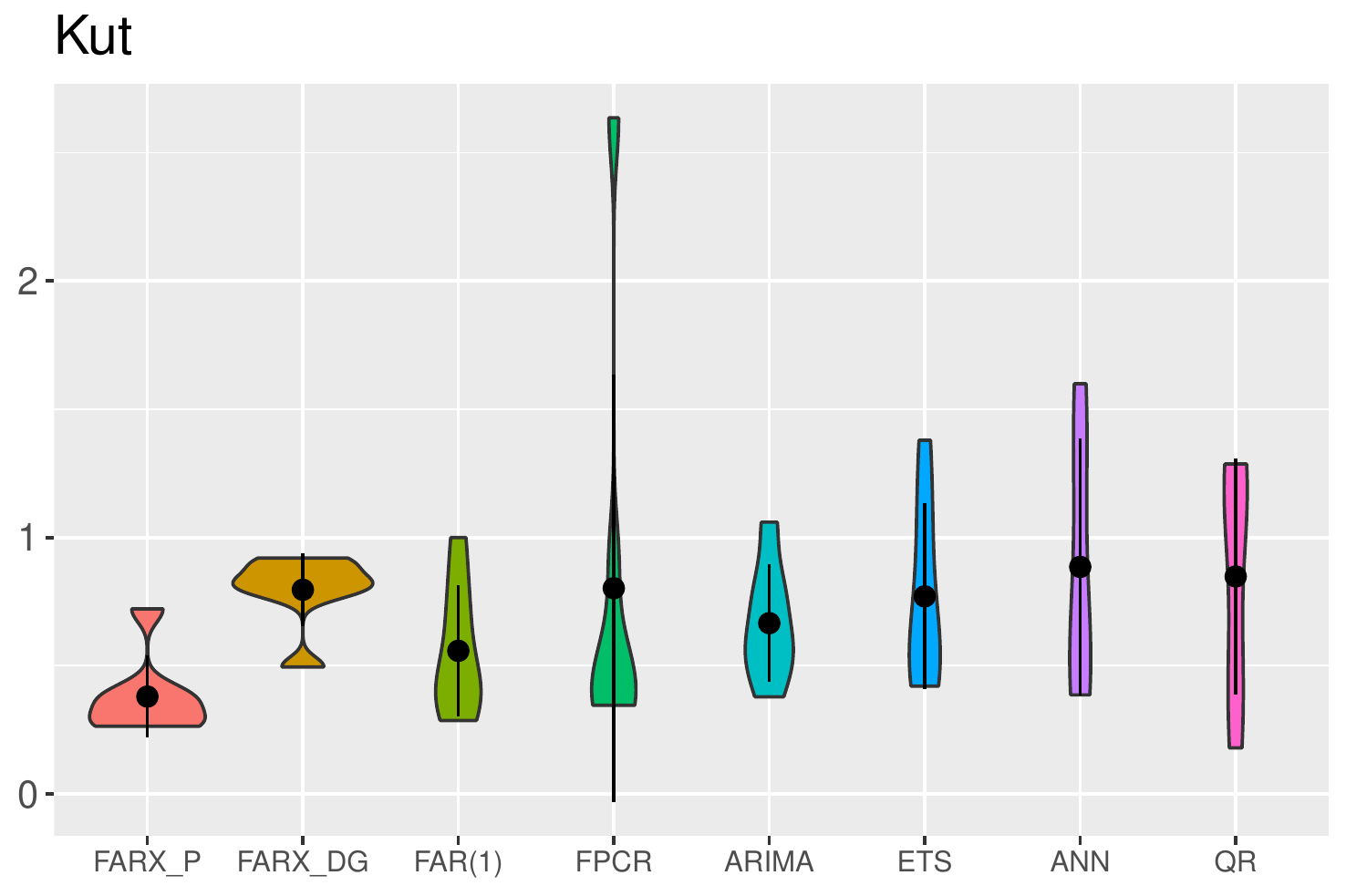}
  \\
  \includegraphics[width=4.8cm,height=4.2cm]{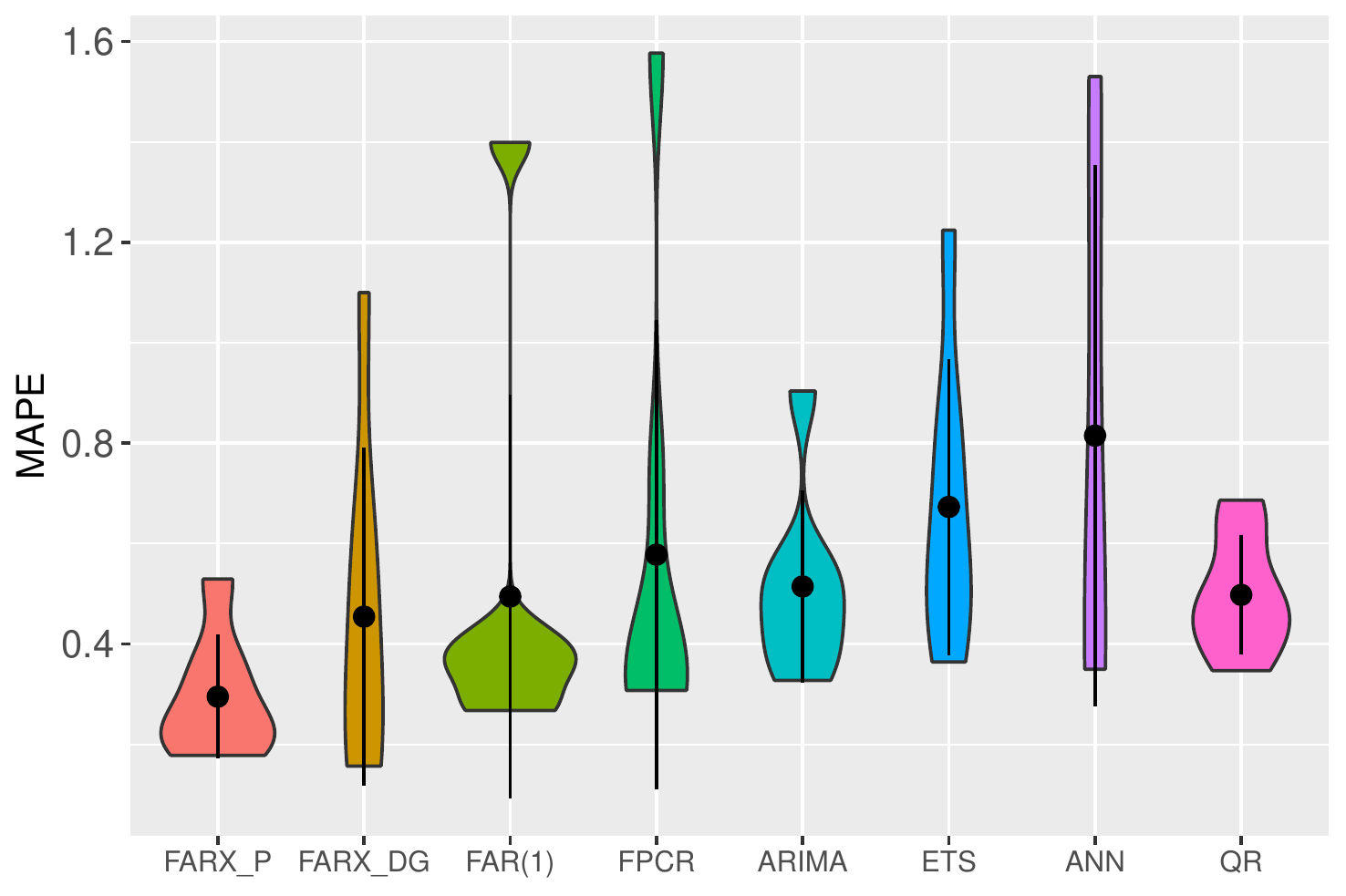}
  \includegraphics[width=4.8cm,height=4.2cm]{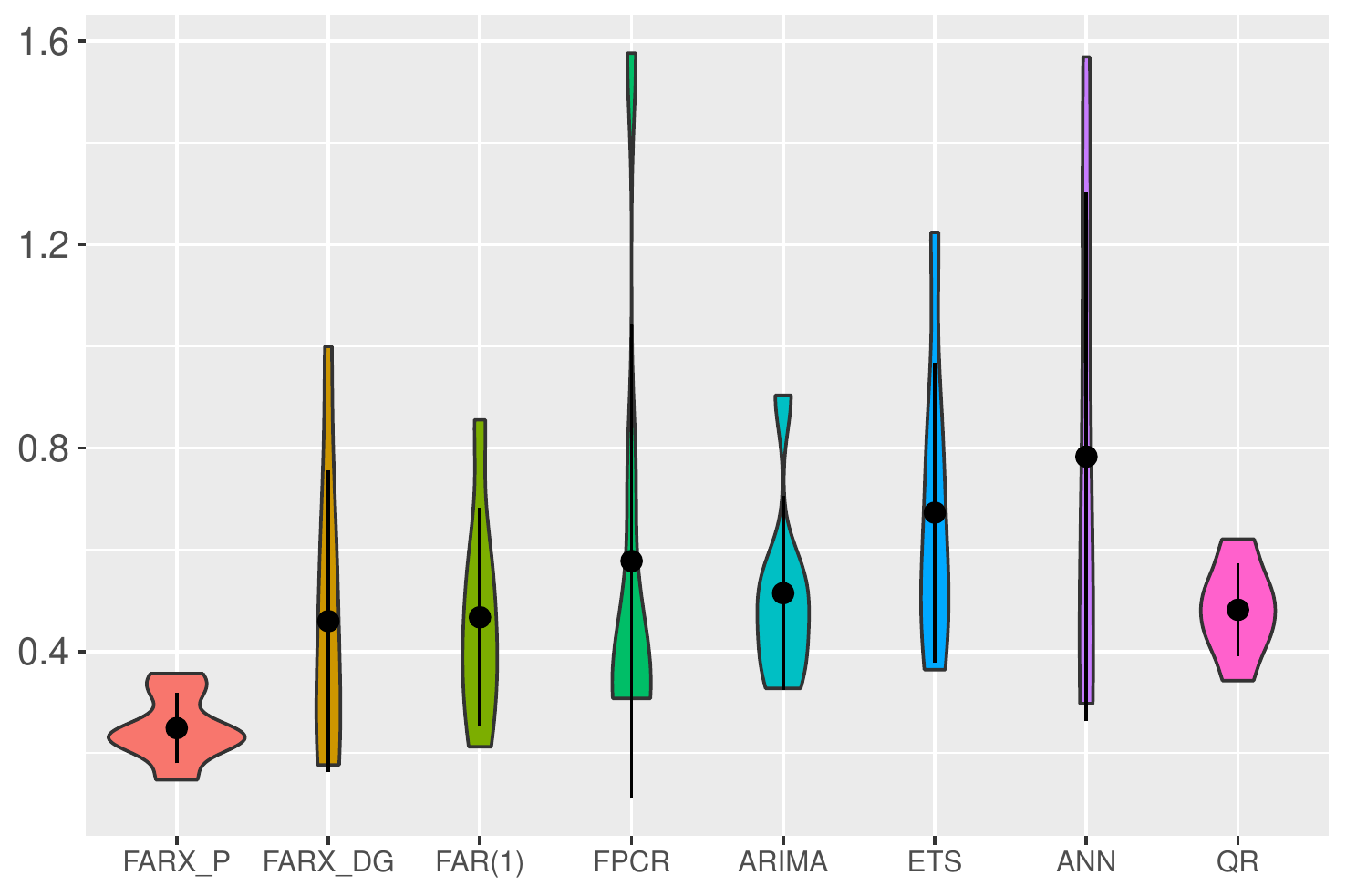}
  \includegraphics[width=4.8cm,height=4.2cm]{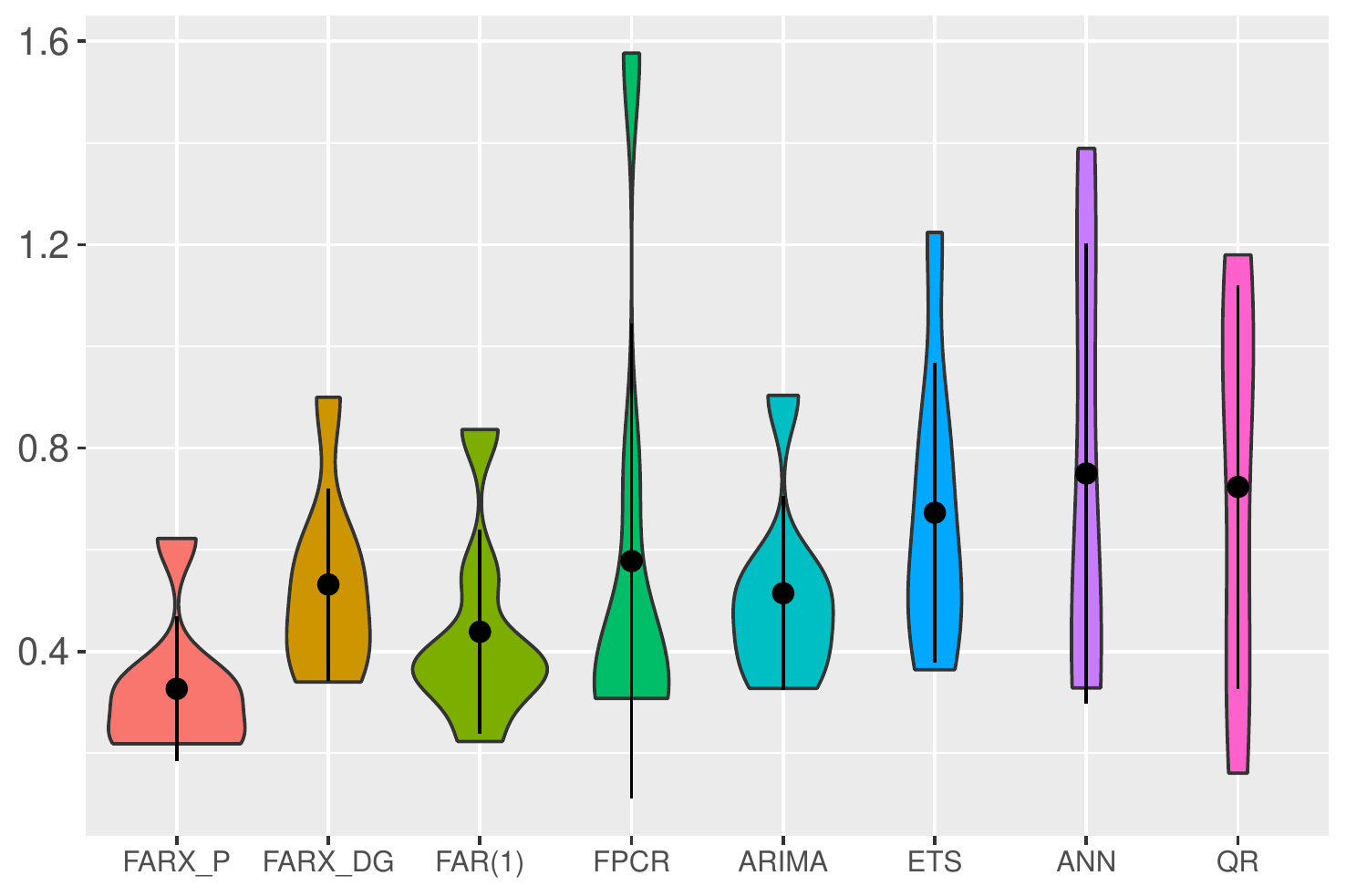}
  \\
  \includegraphics[width=4.8cm,height=4.2cm]{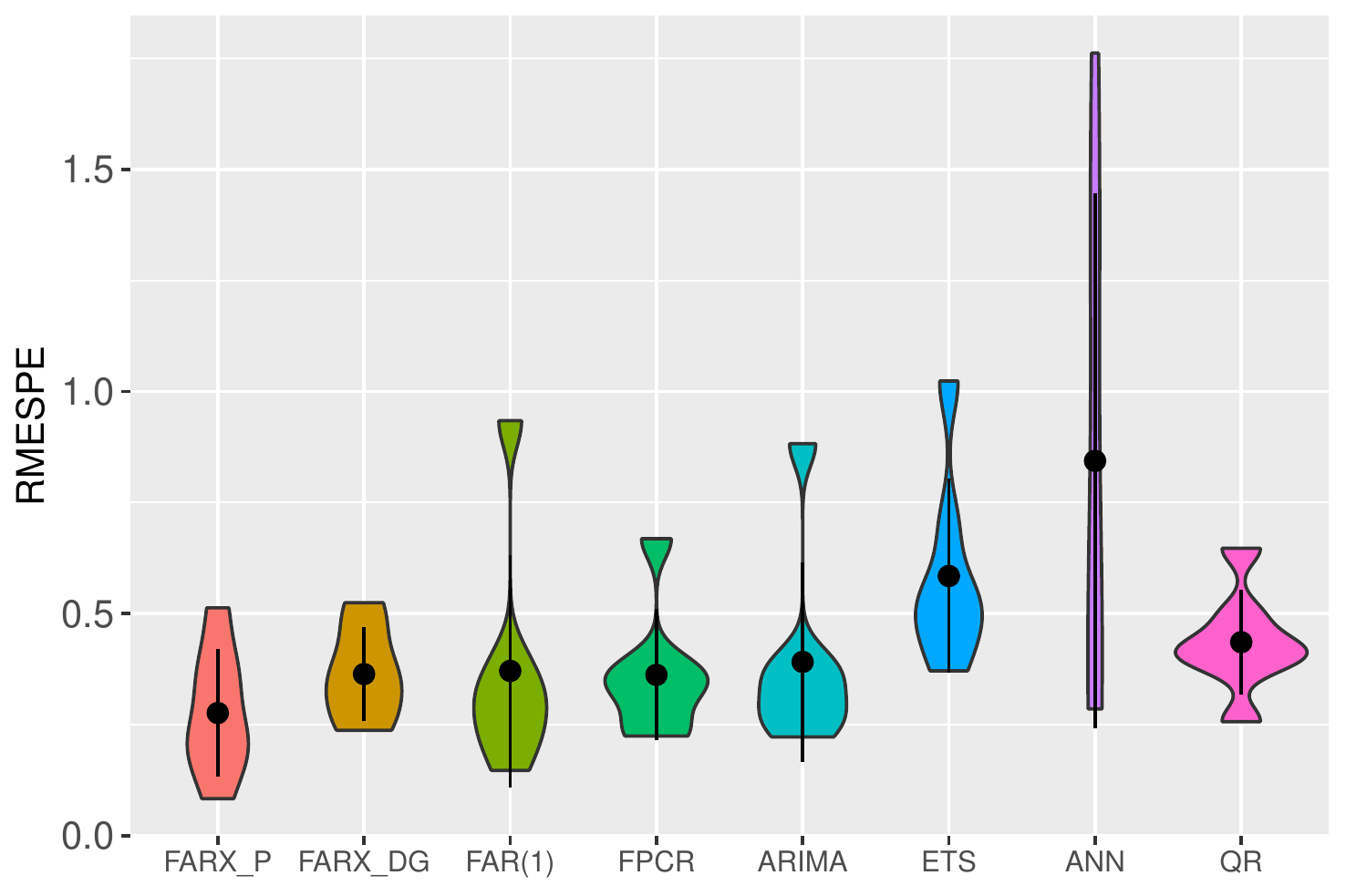}
  \includegraphics[width=4.8cm,height=4.2cm]{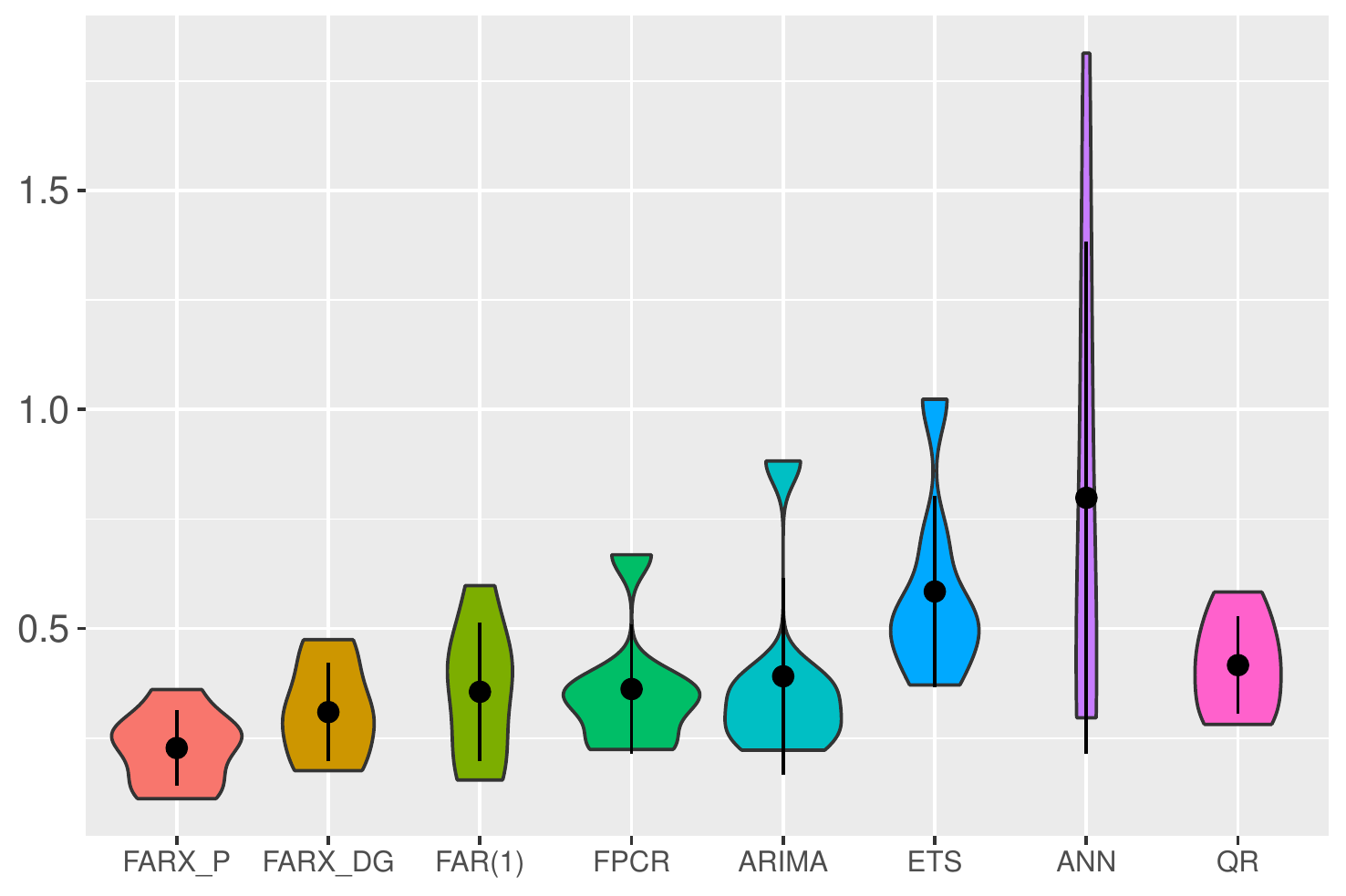}
  \includegraphics[width=4.8cm,height=4.2cm]{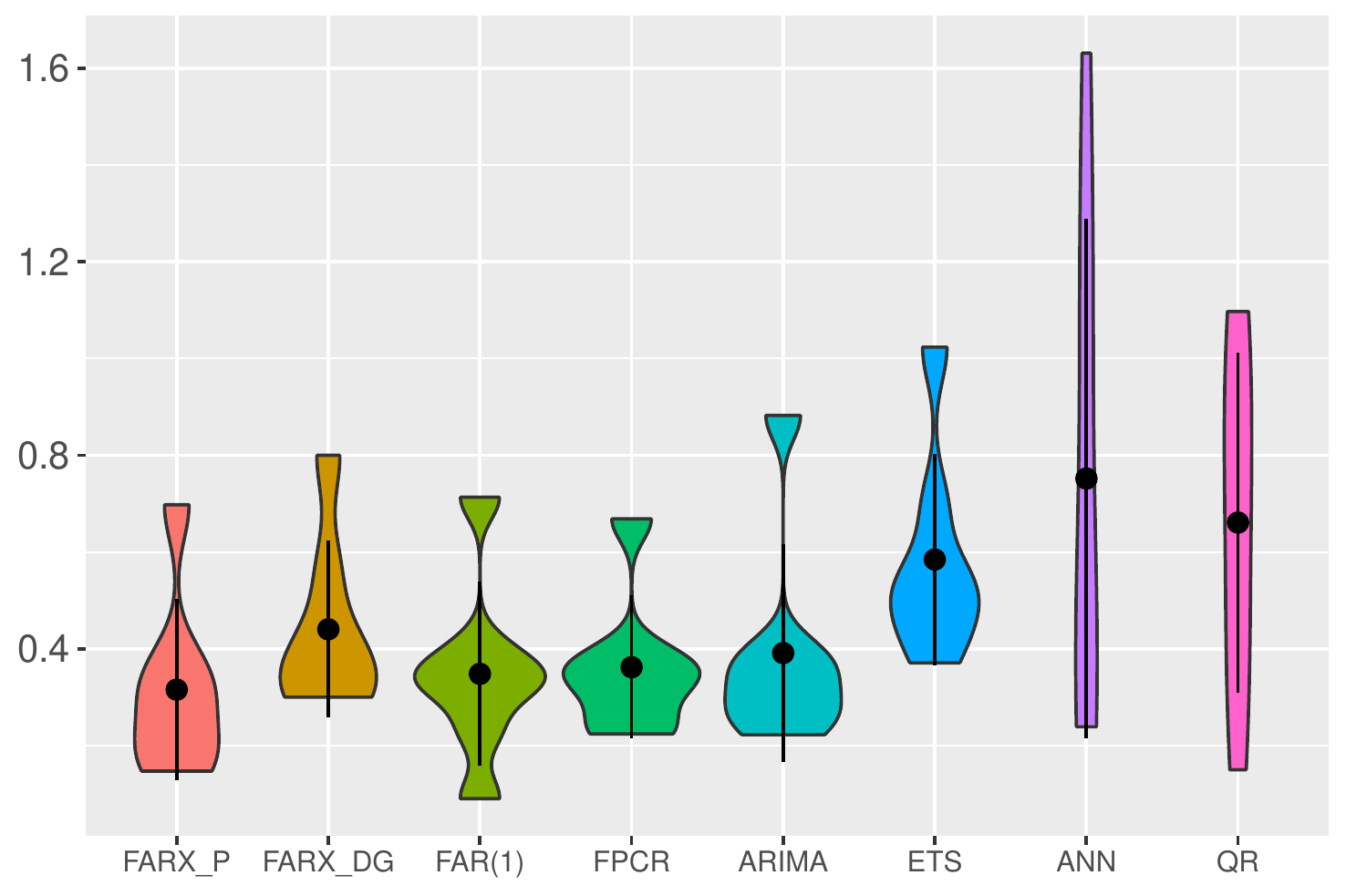}
  \\
  \includegraphics[width=4.8cm,height=4.2cm]{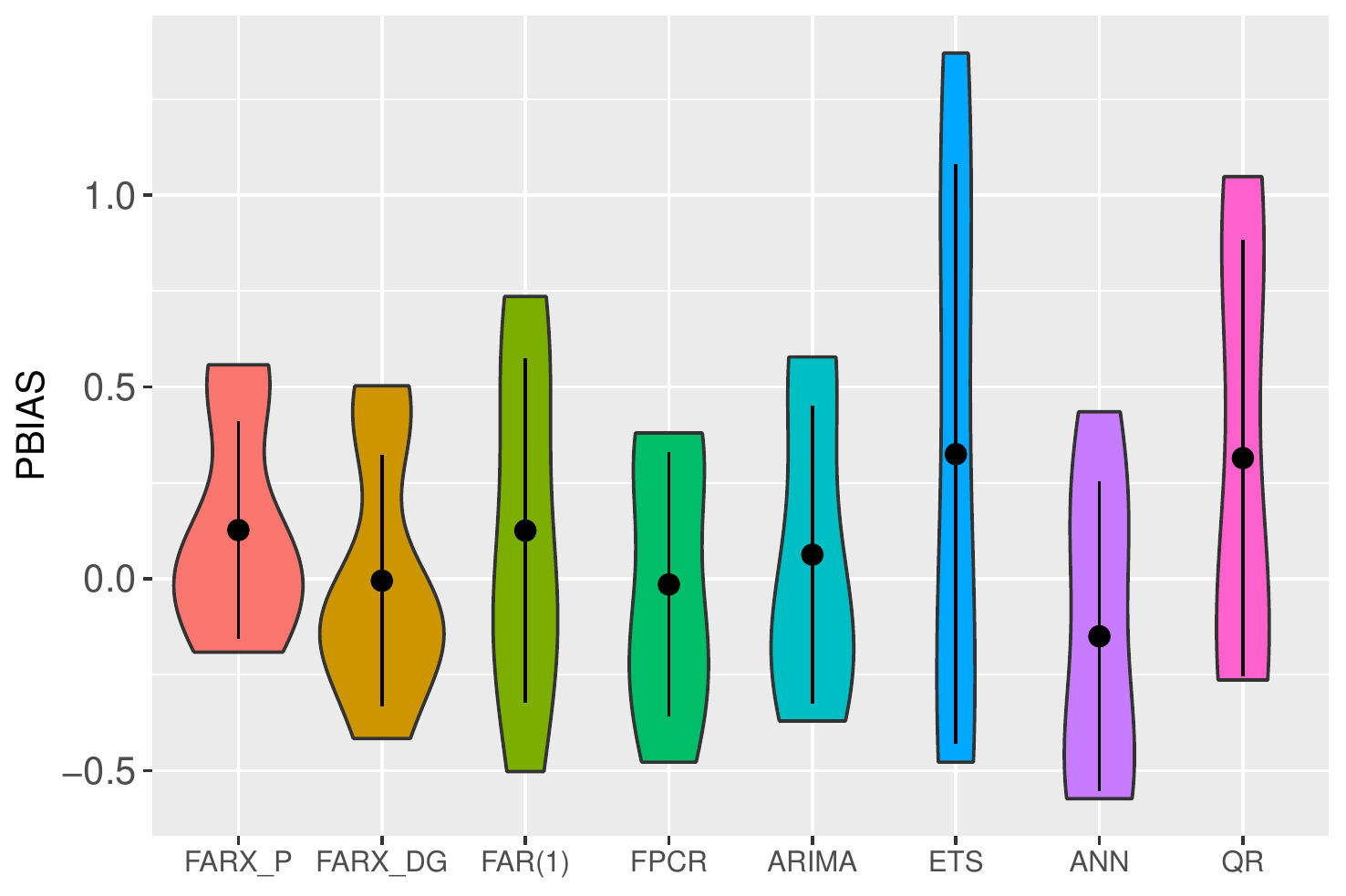}
  \includegraphics[width=4.8cm,height=4.2cm]{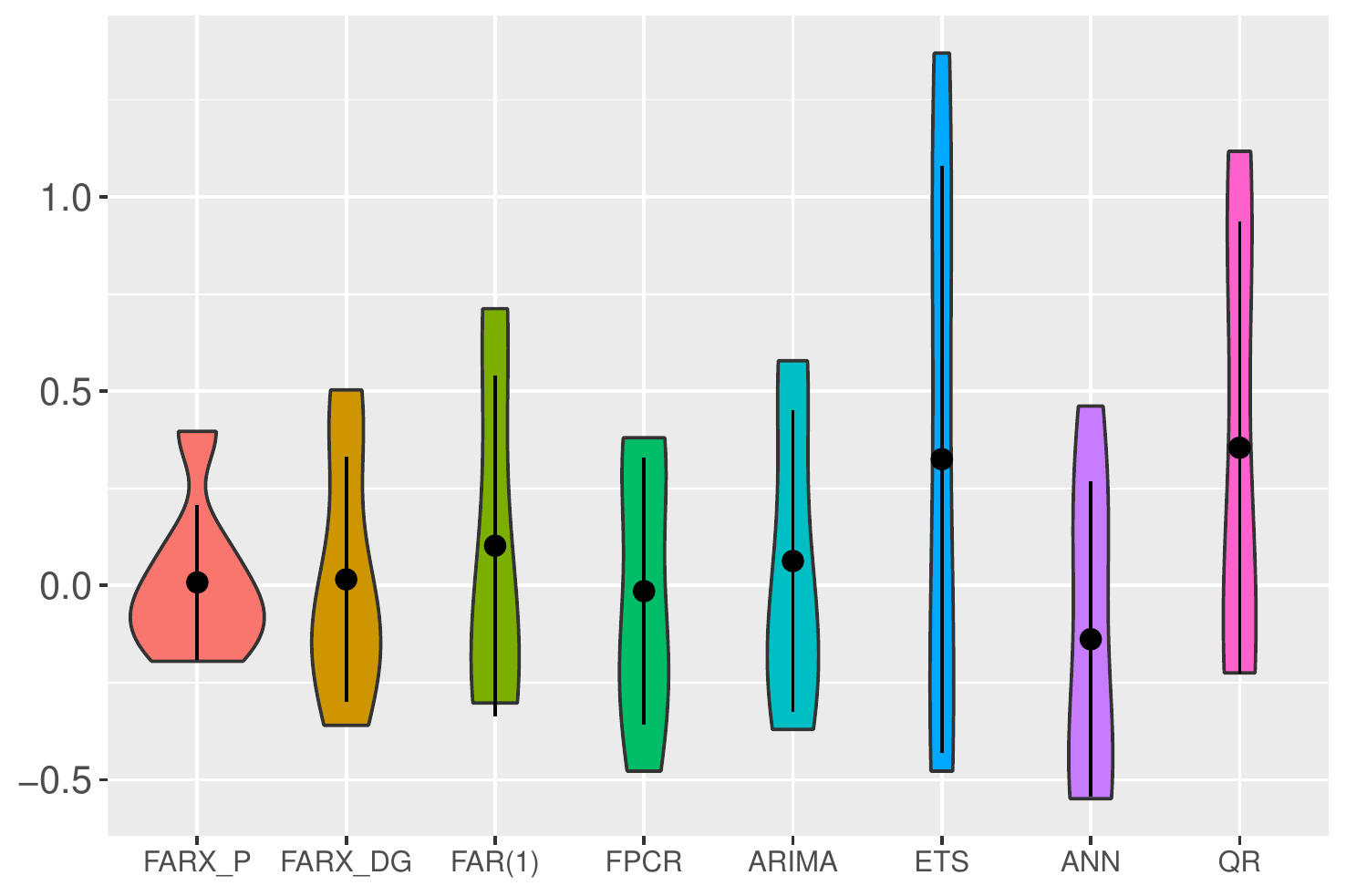}
  \includegraphics[width=4.8cm,height=4.2cm]{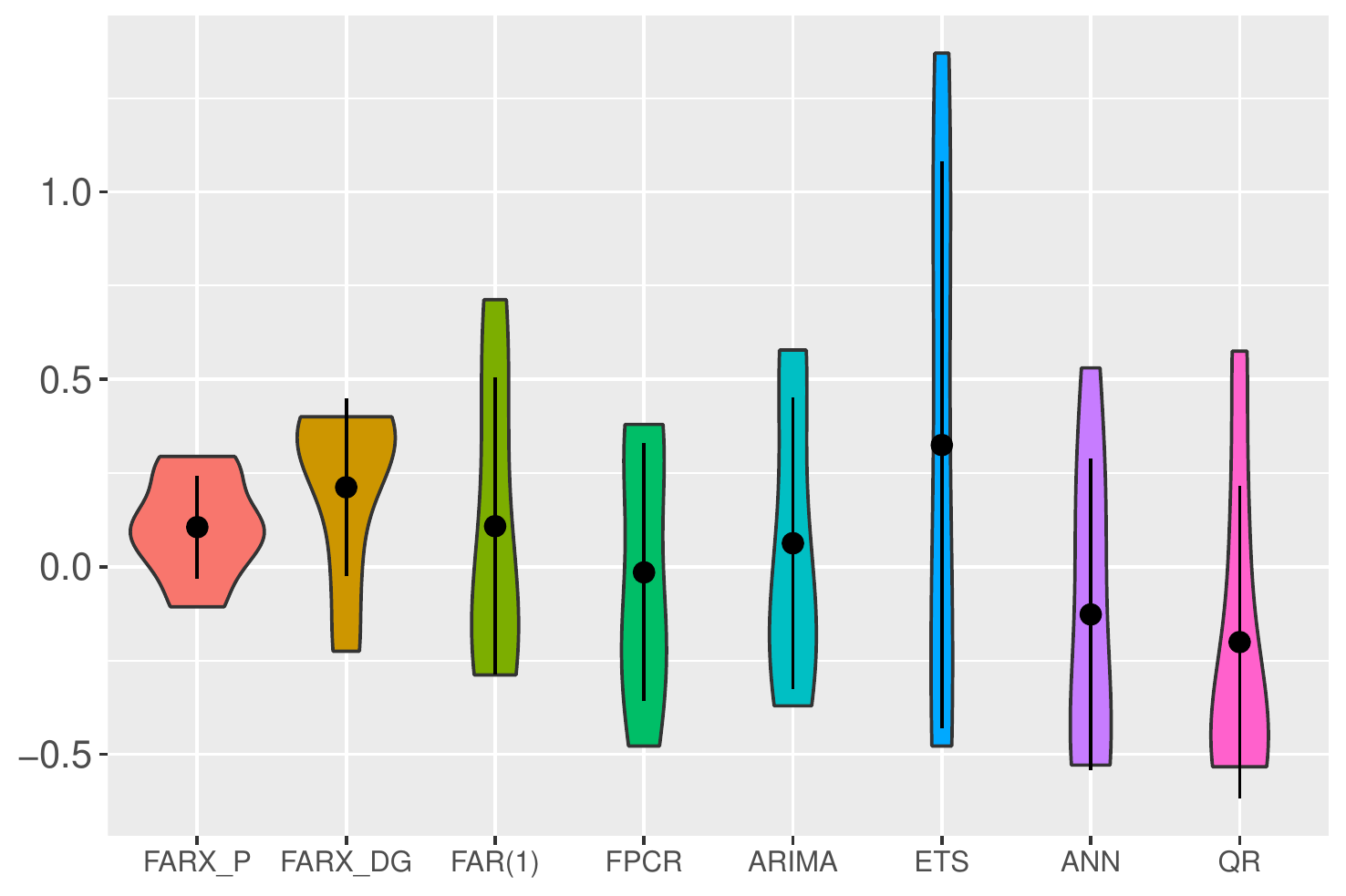}
    \\
  \includegraphics[width=4.8cm,height=4.2cm]{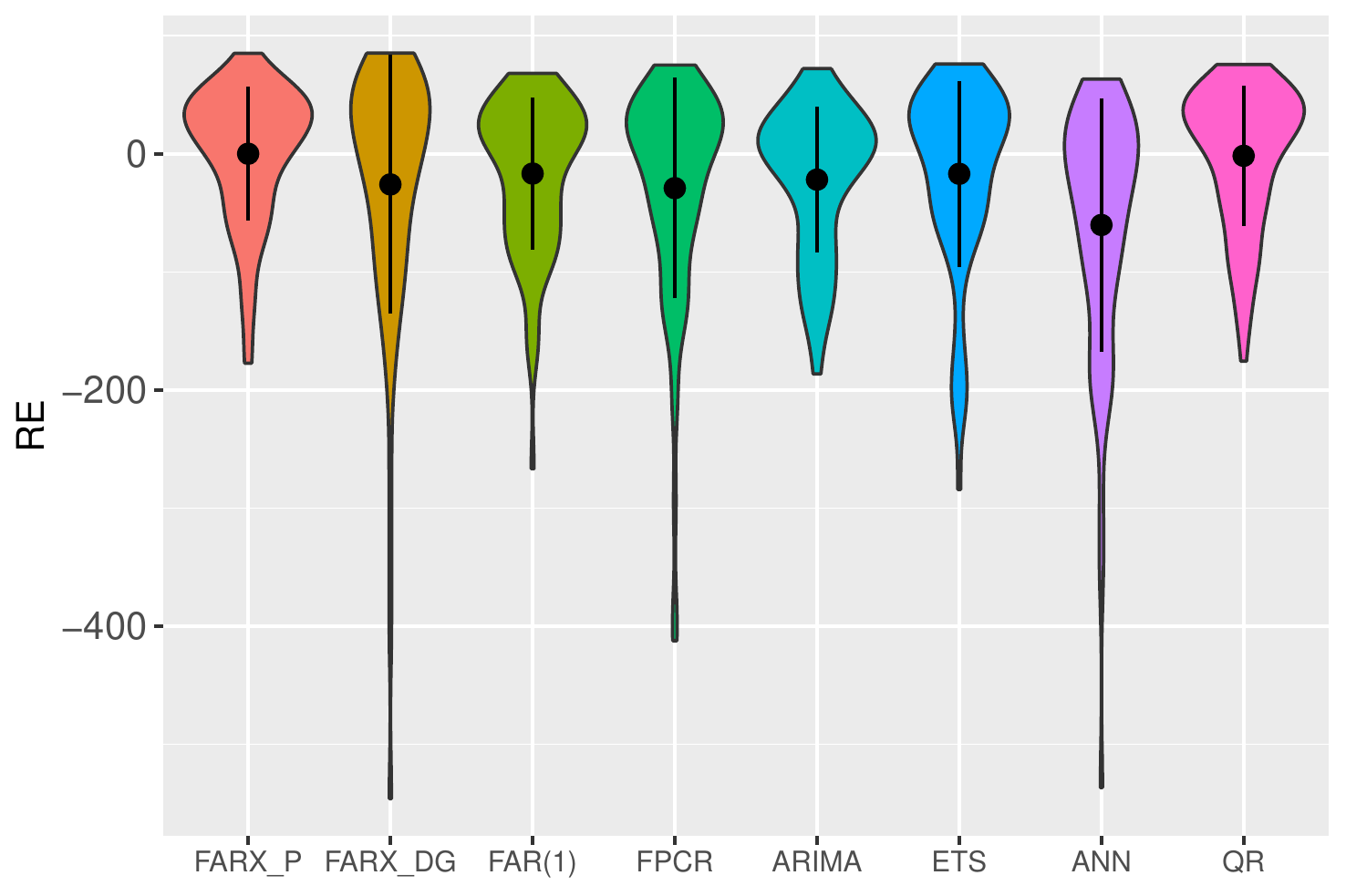}
  \includegraphics[width=4.8cm,height=4.2cm]{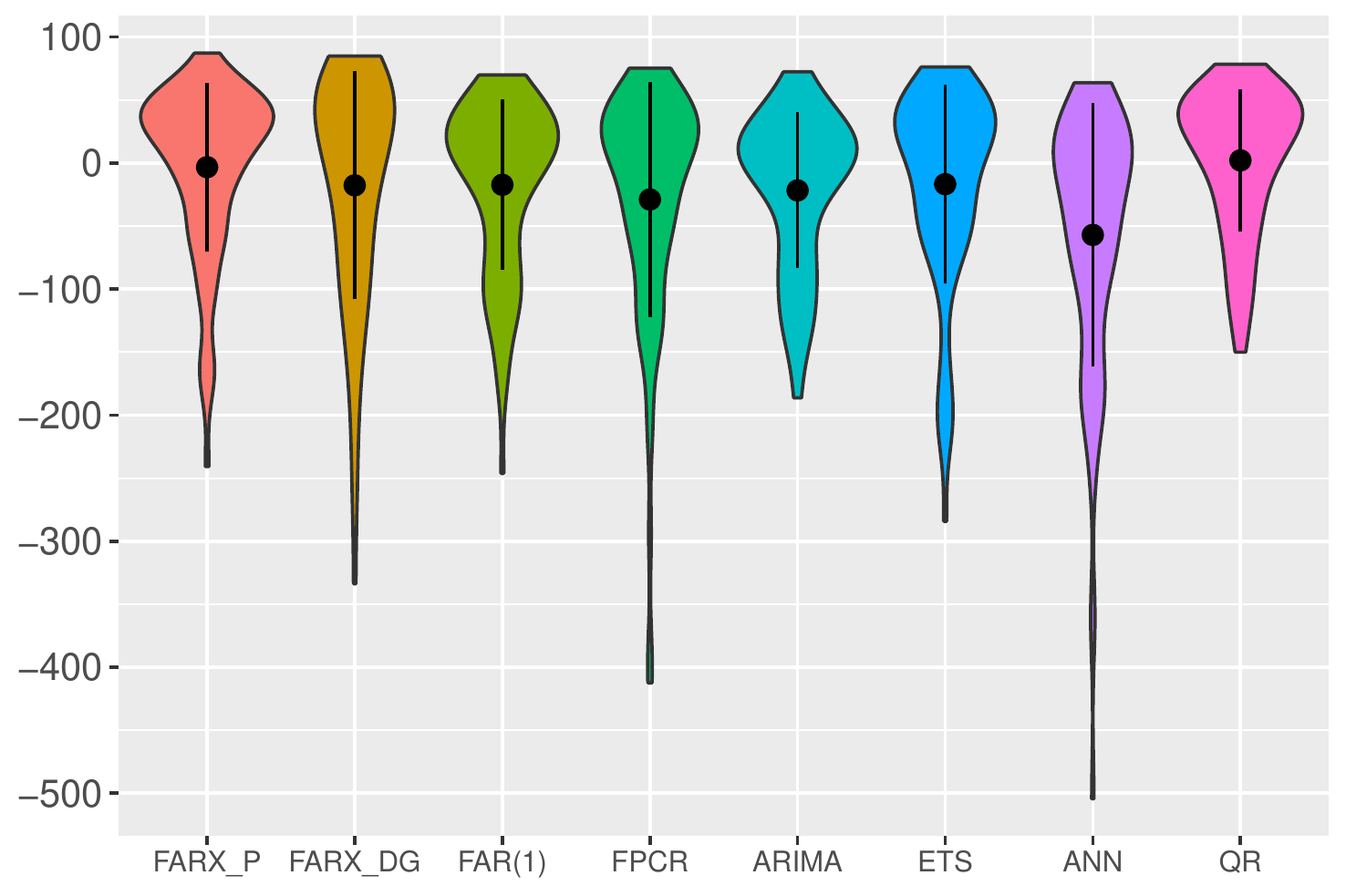}
  \includegraphics[width=4.8cm,height=4.2cm]{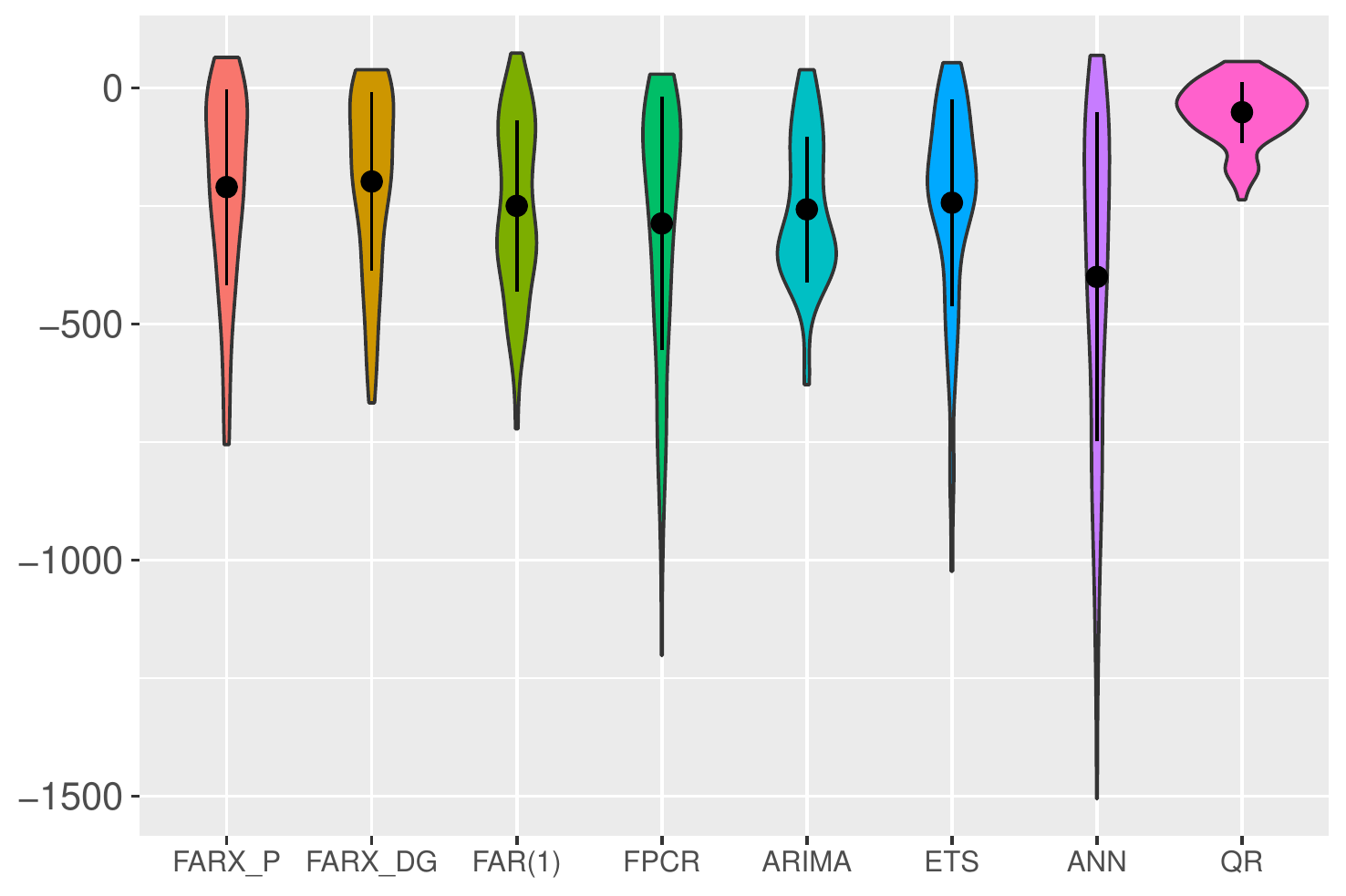}
  \caption{Violin plots of the calculated RMSPE, MAPE, RMESPE, PBIAS, and RE values for the functional and non-functional (under the first scenario) models. The columns represent the stations, while the rows represent the performance metrics. In the plots, FARX\underline{ }P and FARX\underline{ }DG denote the proposed FARX(1) model and the FARX(1) model of \cite{Damon2002}, respectively.}
  \label{fig:Fig_9}
\end{figure}

\begin{figure}[!htbp]
  \centering
  \includegraphics[width=4.8cm,height=4.2cm]{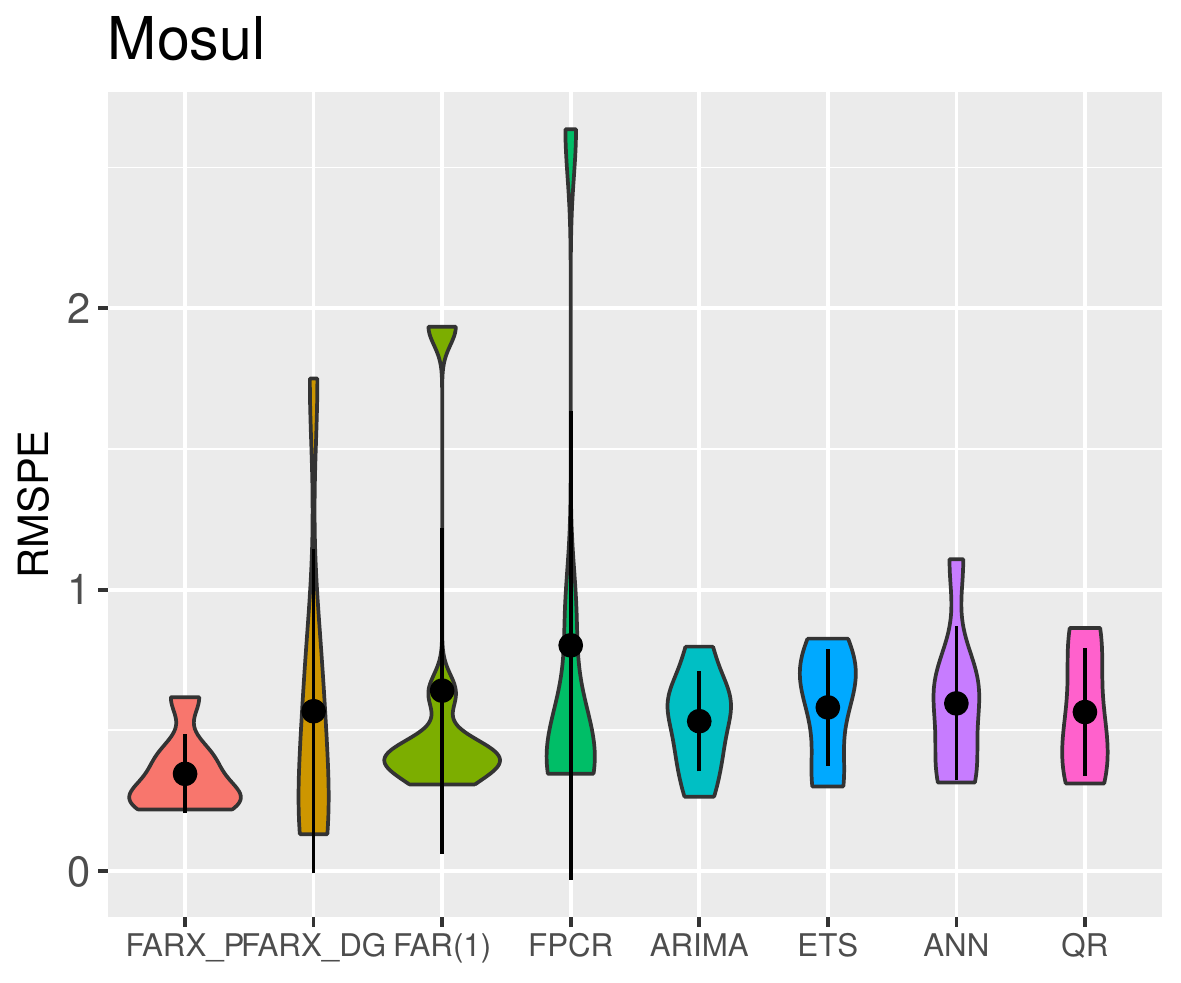}
  \includegraphics[width=4.8cm,height=4.2cm]{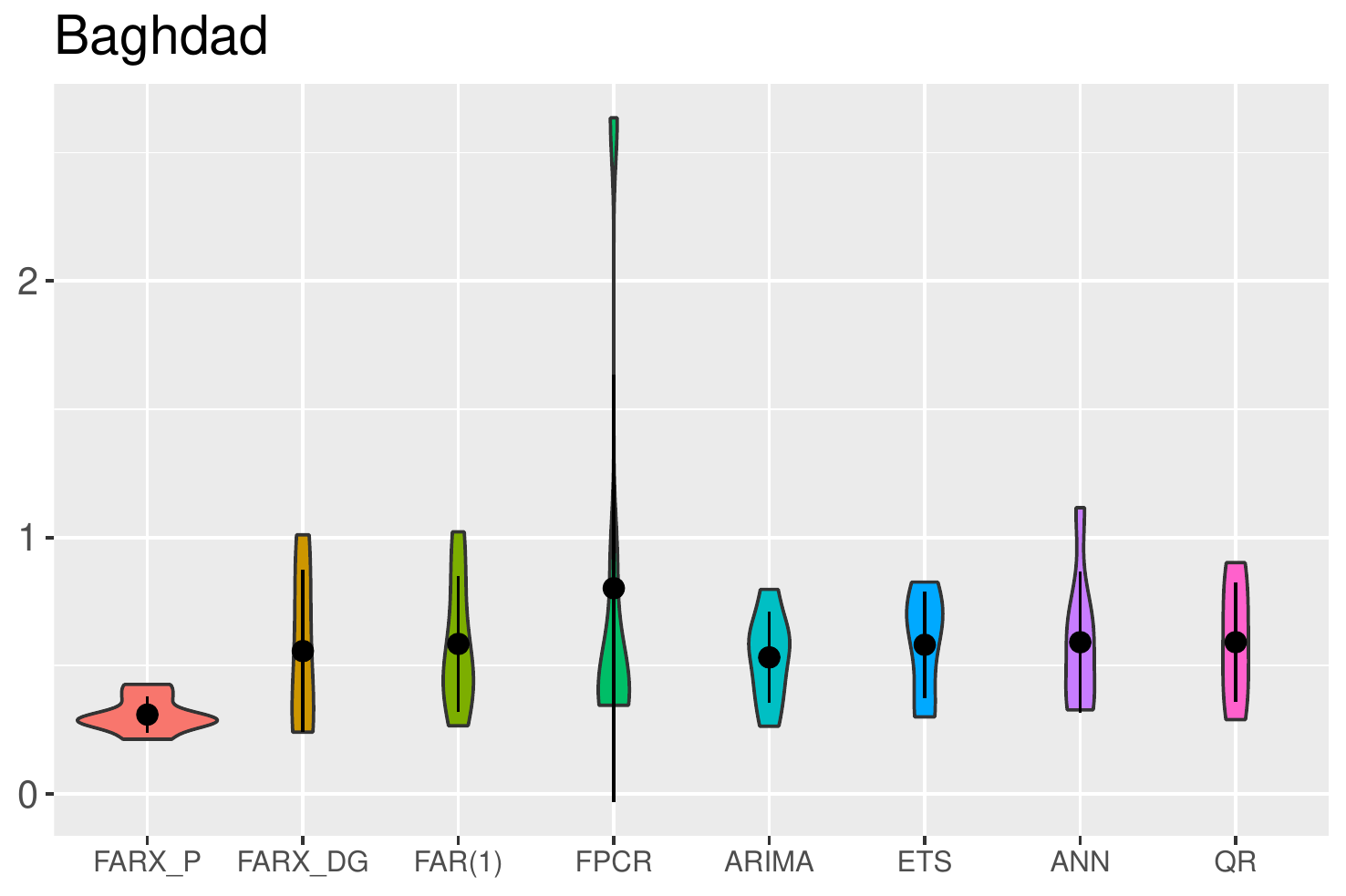}
  \includegraphics[width=4.8cm,height=4.2cm]{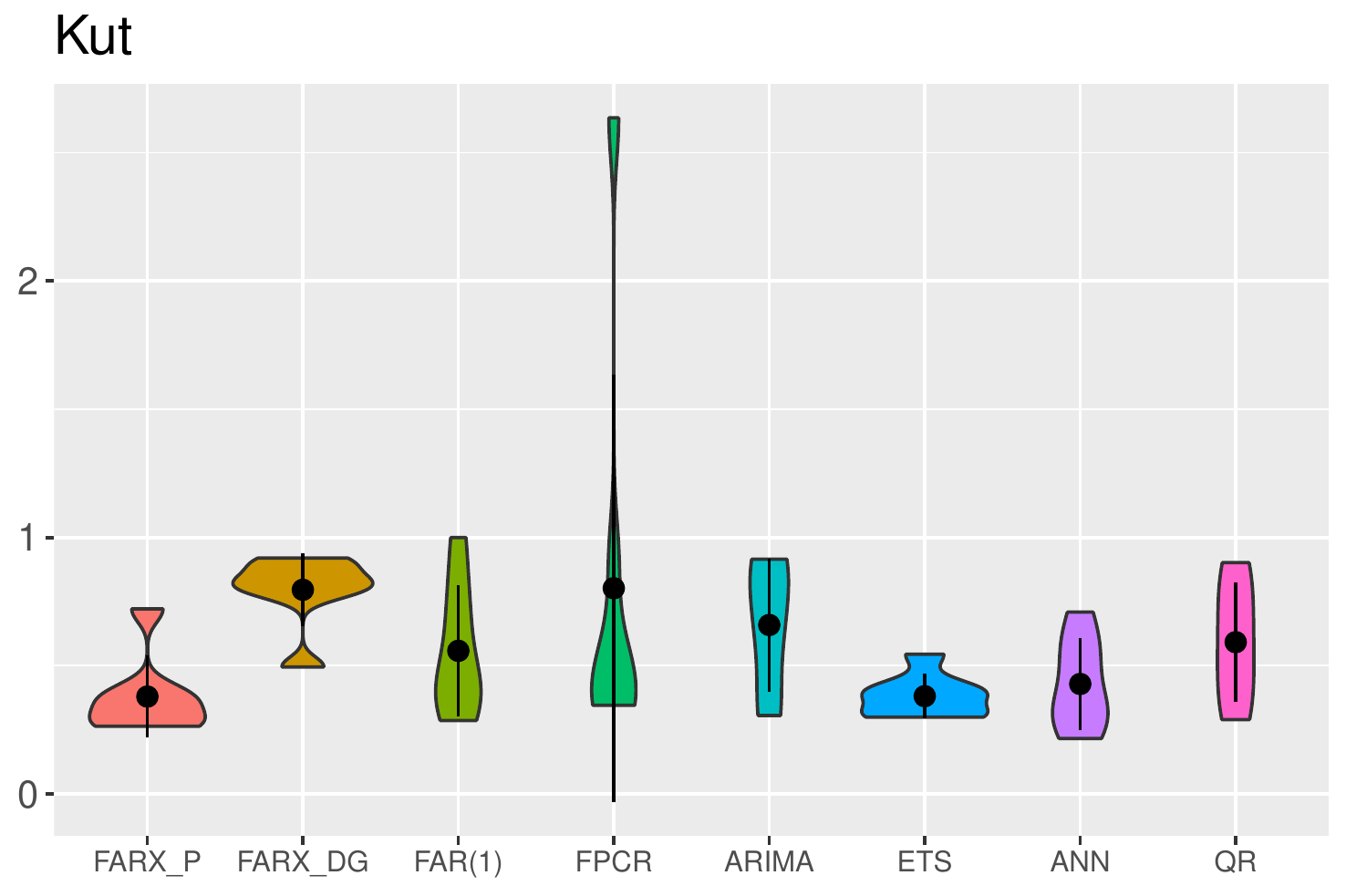}
  \\
  \includegraphics[width=4.8cm,height=4.2cm]{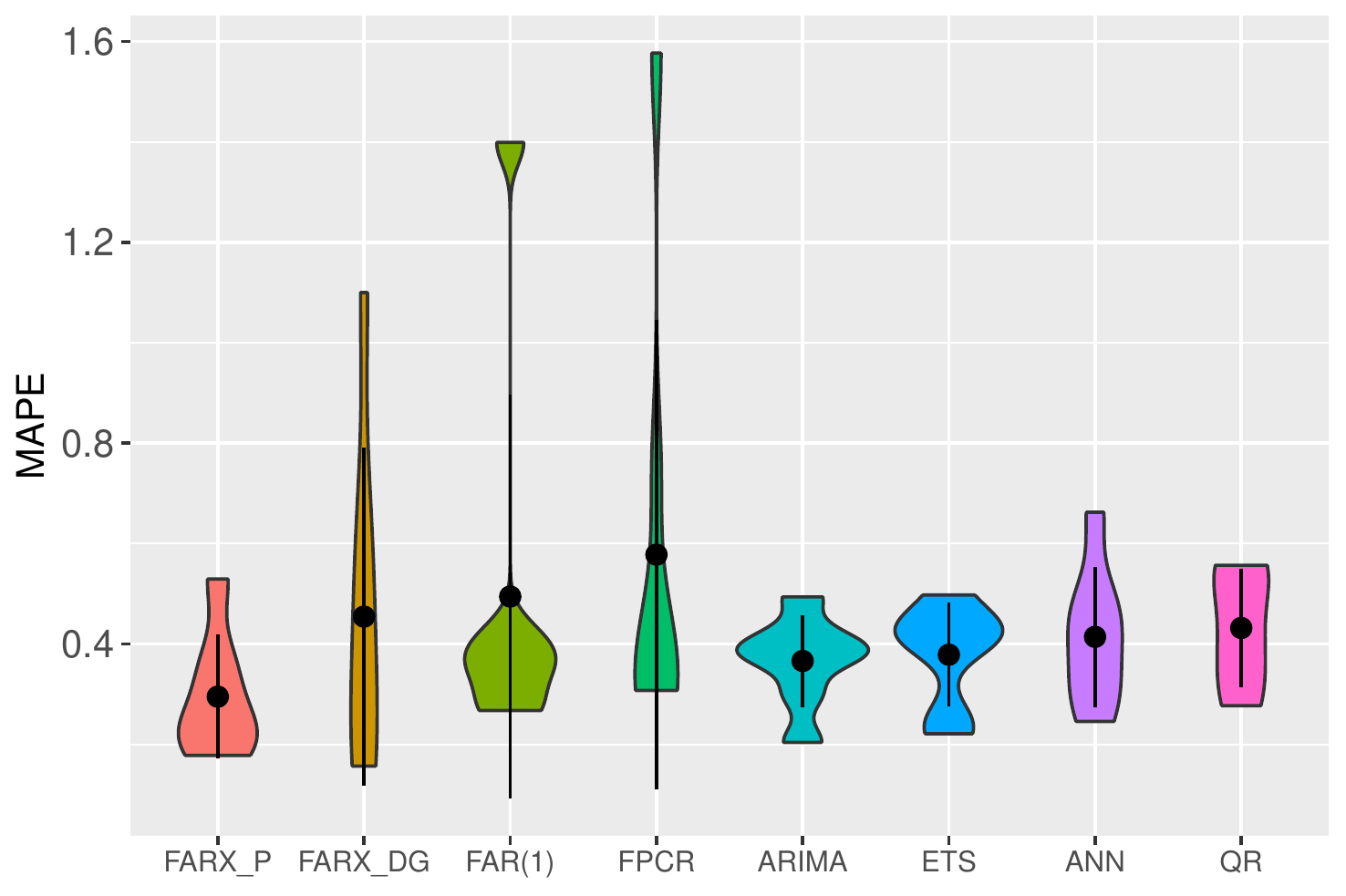}
  \includegraphics[width=4.8cm,height=4.2cm]{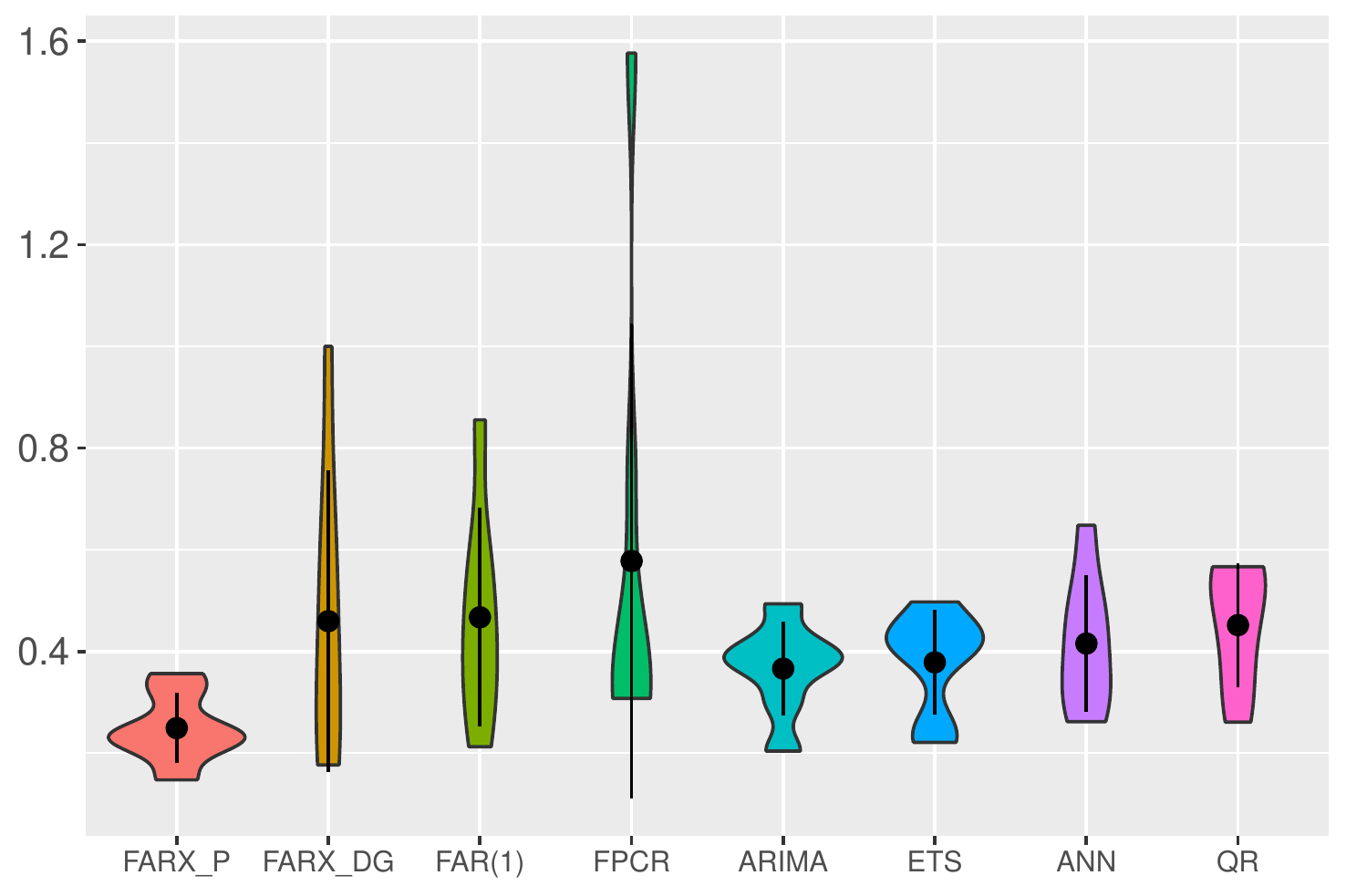}
  \includegraphics[width=4.8cm,height=4.2cm]{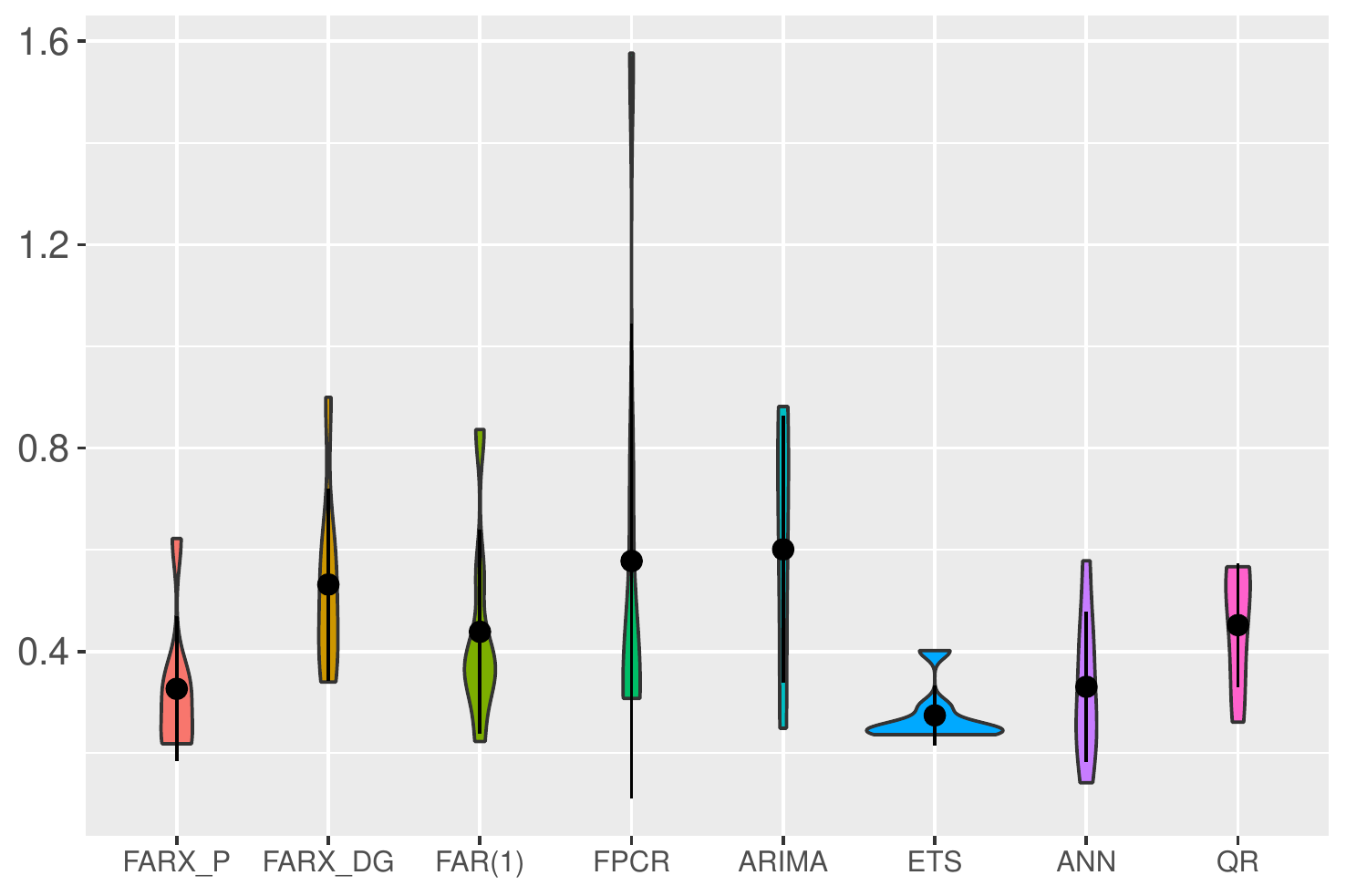}
  \\
  \includegraphics[width=4.8cm,height=4.2cm]{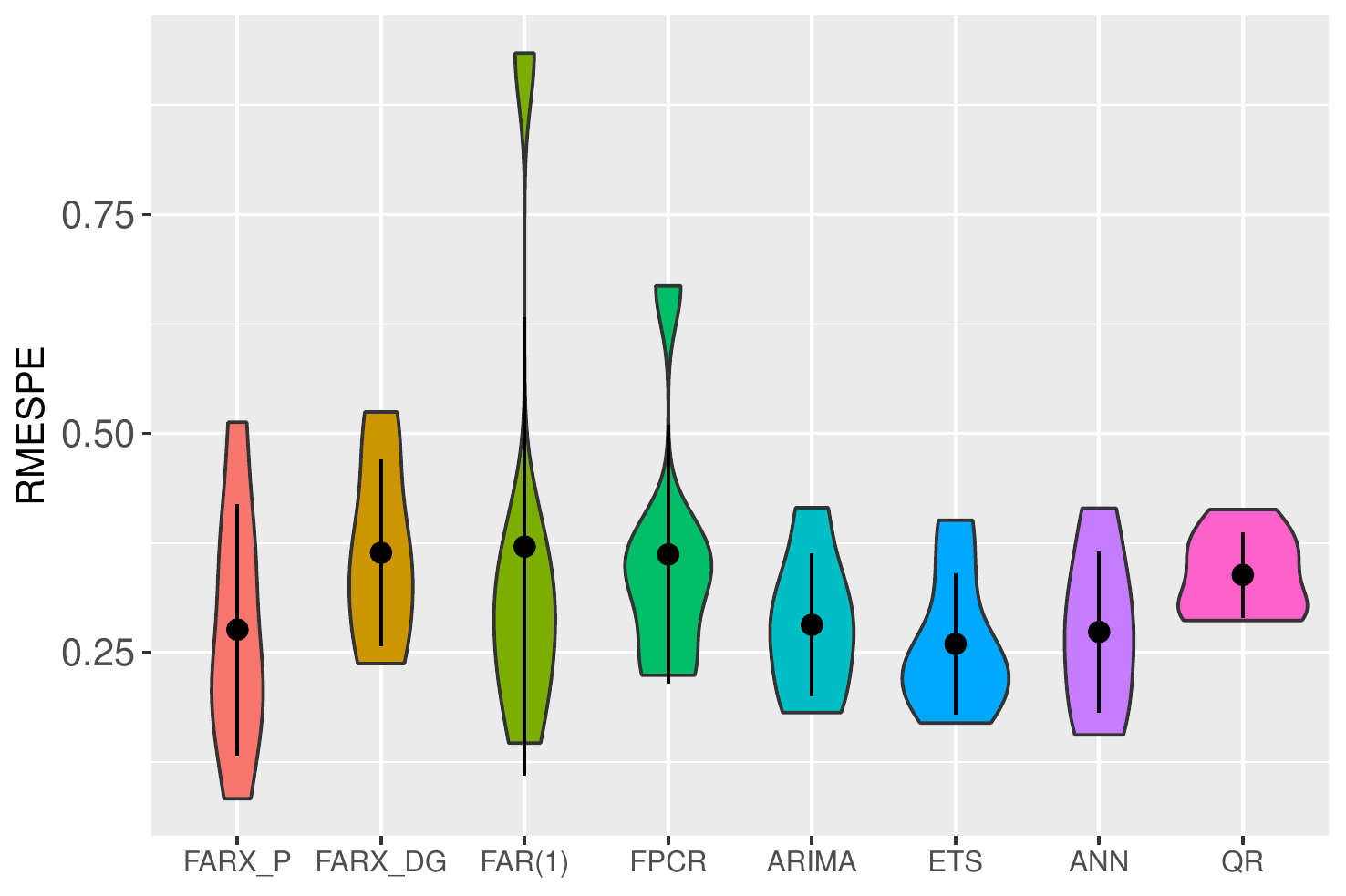}
  \includegraphics[width=4.8cm,height=4.2cm]{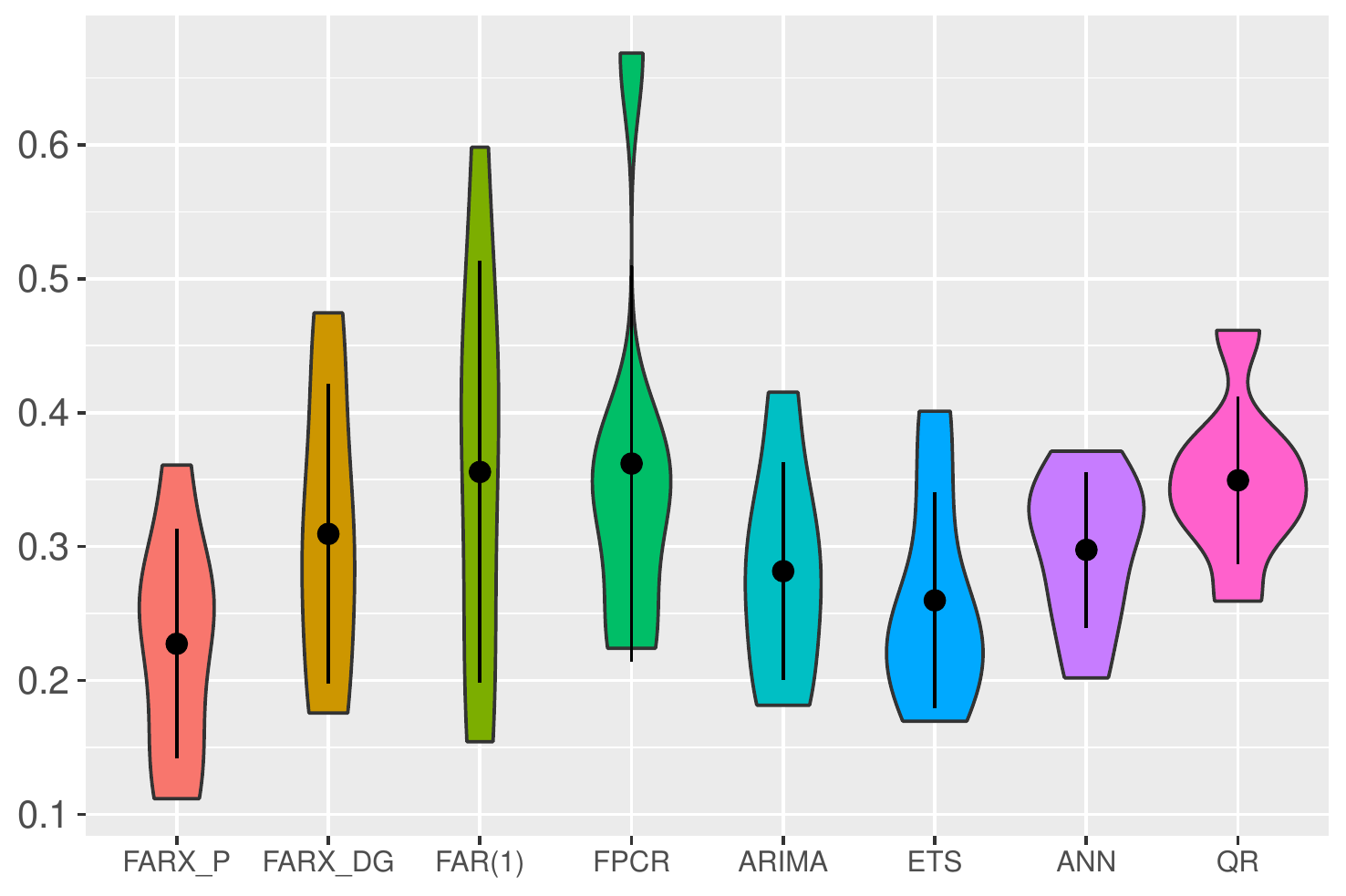}
  \includegraphics[width=4.8cm,height=4.2cm]{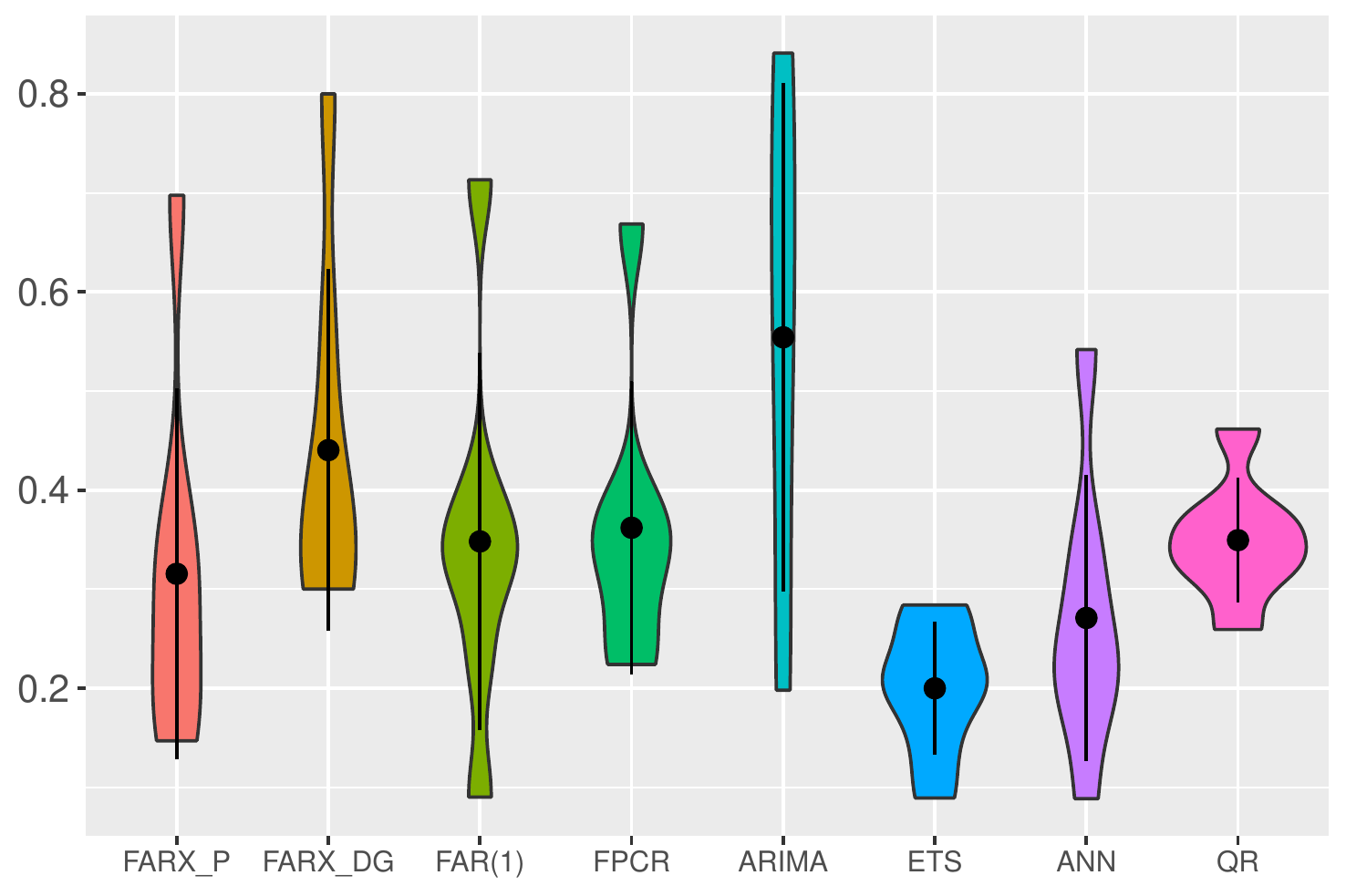}
  \\
  \includegraphics[width=4.8cm,height=4.2cm]{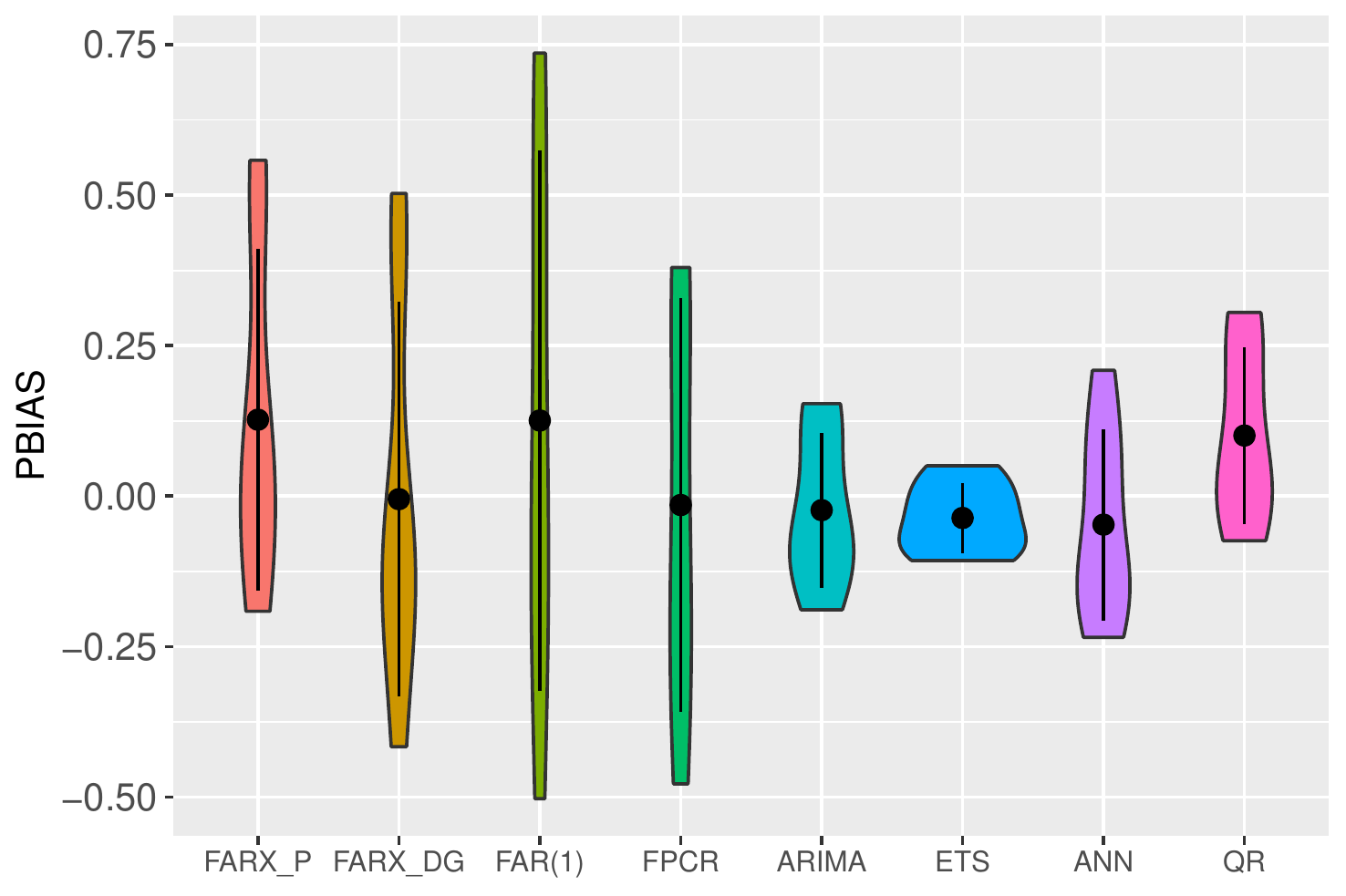}
  \includegraphics[width=4.8cm,height=4.2cm]{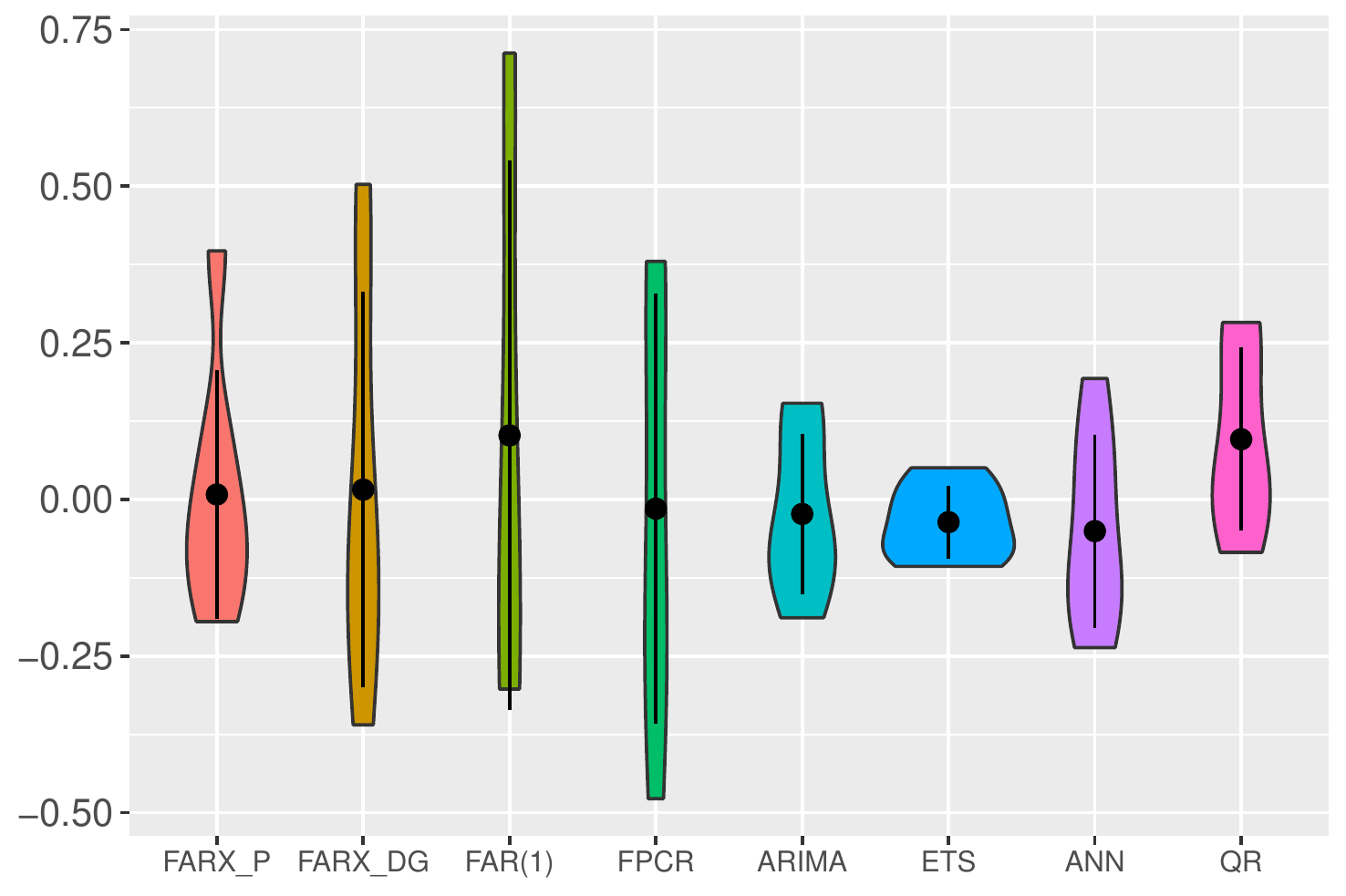}
  \includegraphics[width=4.8cm,height=4.2cm]{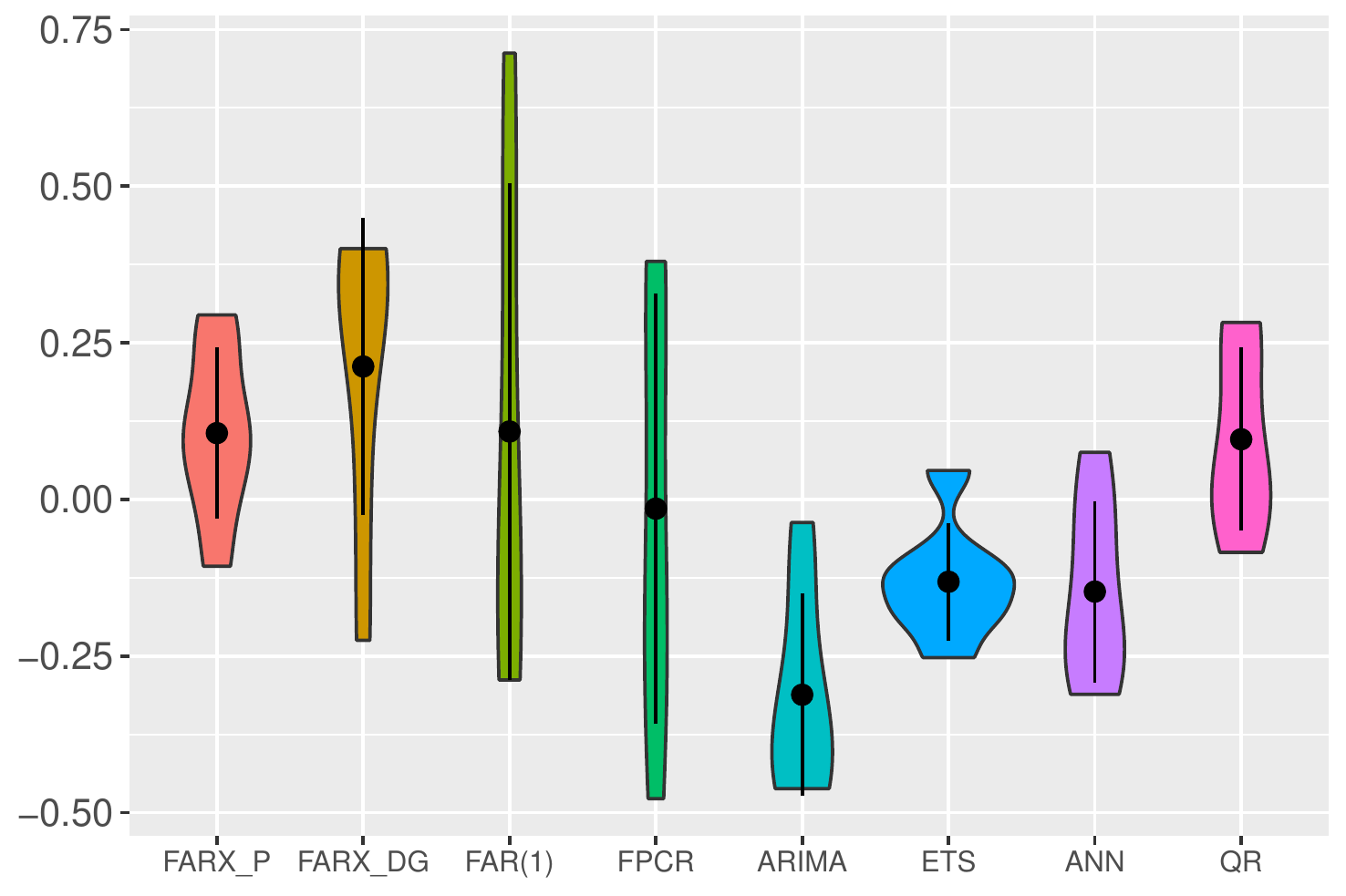}
    \\
  \includegraphics[width=4.8cm,height=4.2cm]{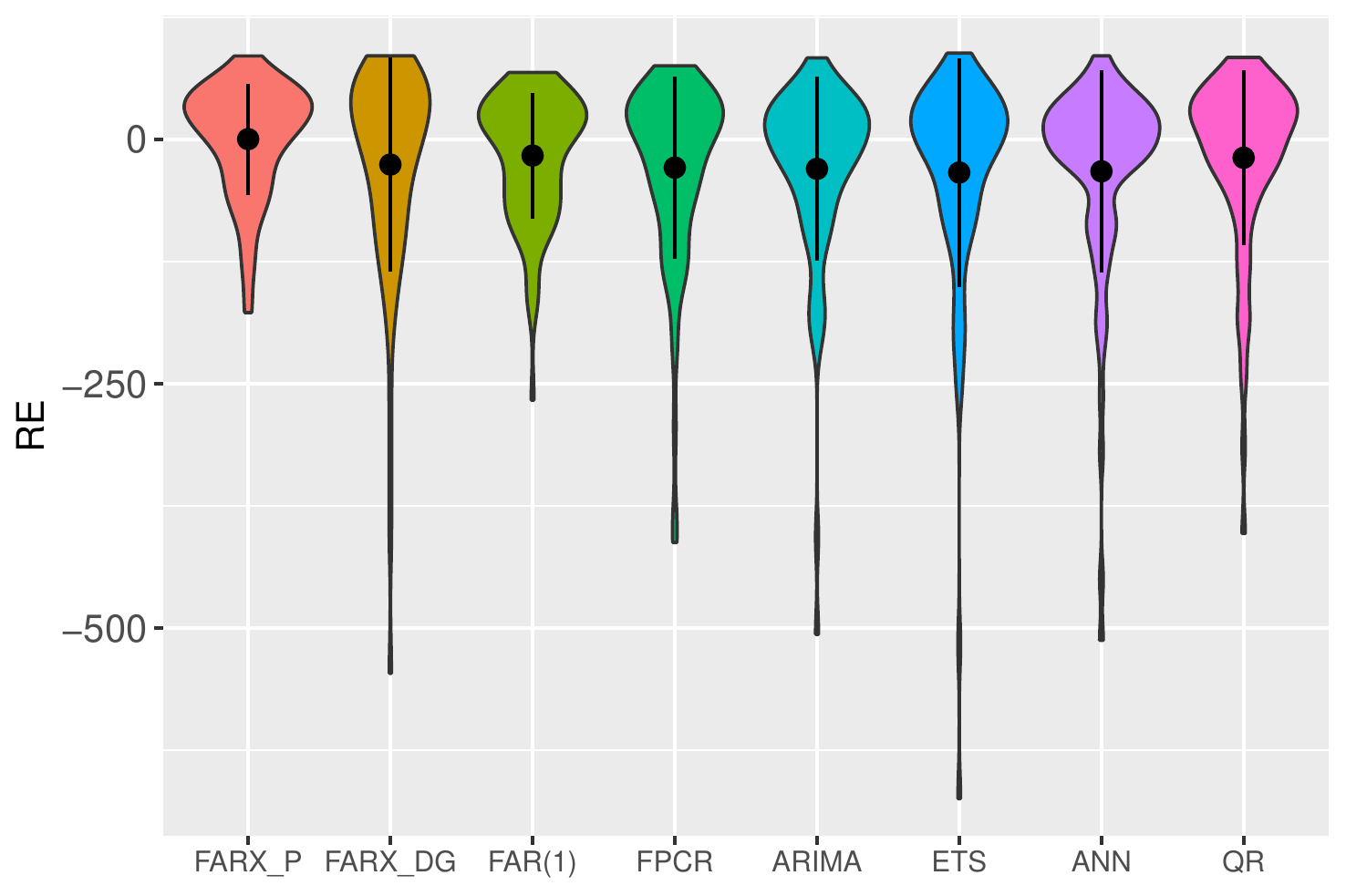}
  \includegraphics[width=4.8cm,height=4.2cm]{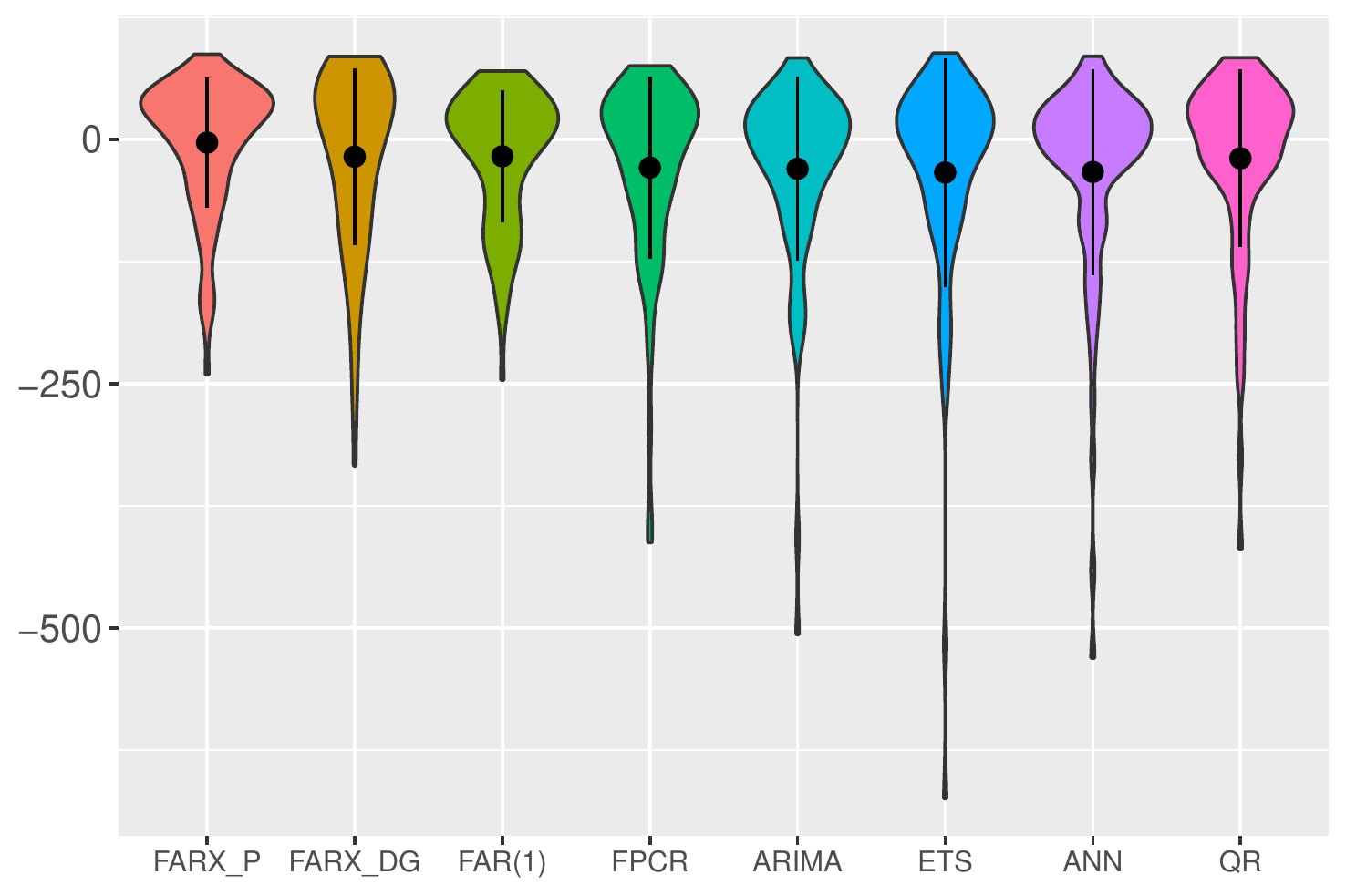}
  \includegraphics[width=4.8cm,height=4.2cm]{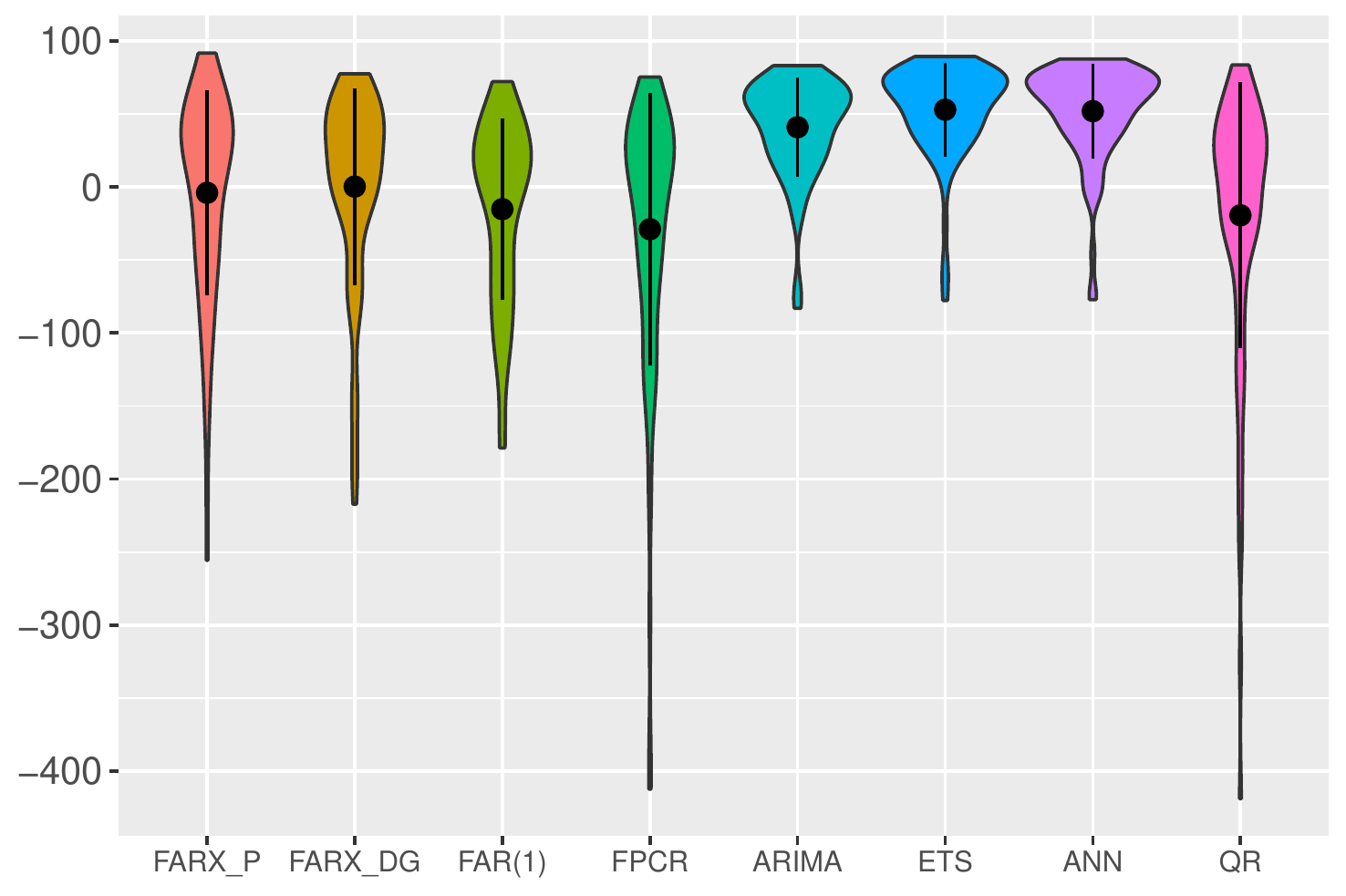}
  \caption{Violin plots of the calculated RMSPE, MAPE, RMESPE, PBIAS, and RE values for the functional and non-functional (under the second scenario) models. The columns represent the stations, while the rows represent the performance metrics. In the plots, FARX\underline{ }P and FARX\underline{ }DG denote the proposed FARX(1) model and the FARX(1) model of \cite{Damon2002}, respectively.}
  \label{fig:Fig_10}
\end{figure}

\begin{figure}[!htbp]
  \centering
  \includegraphics[width=4.8cm,height=5cm]{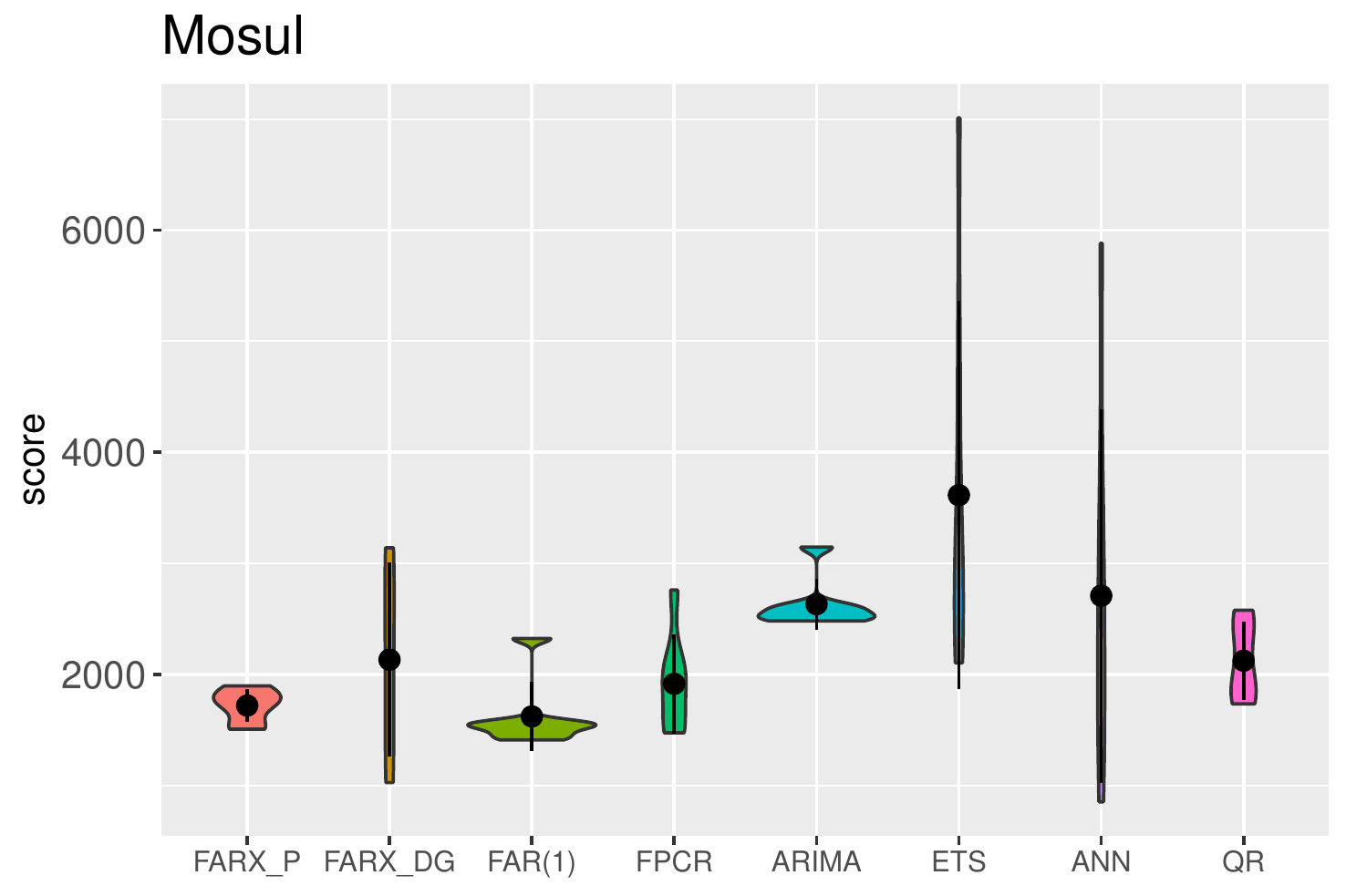}
  \includegraphics[width=4.8cm,height=5cm]{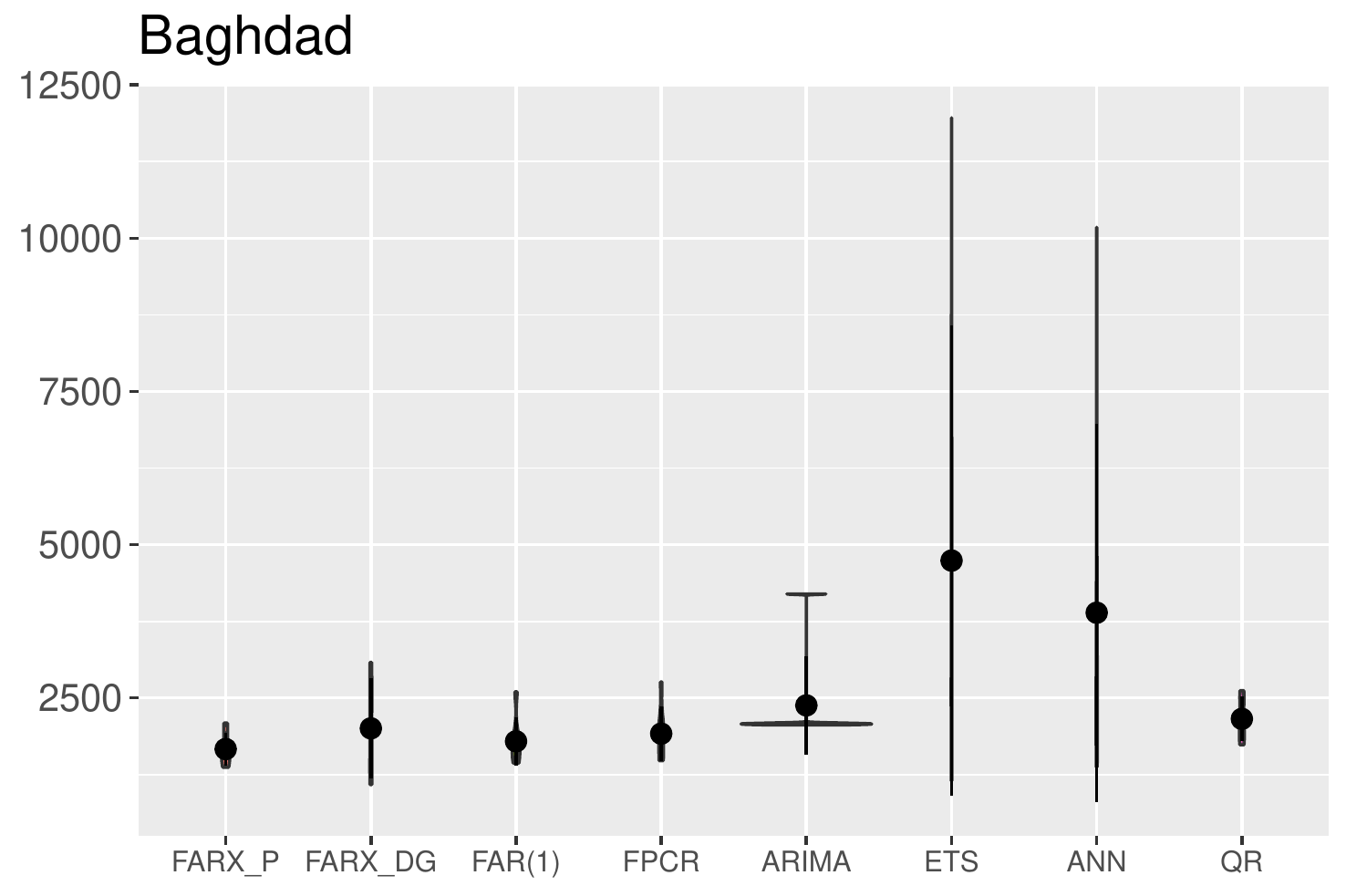}
  \includegraphics[width=4.8cm,height=5cm]{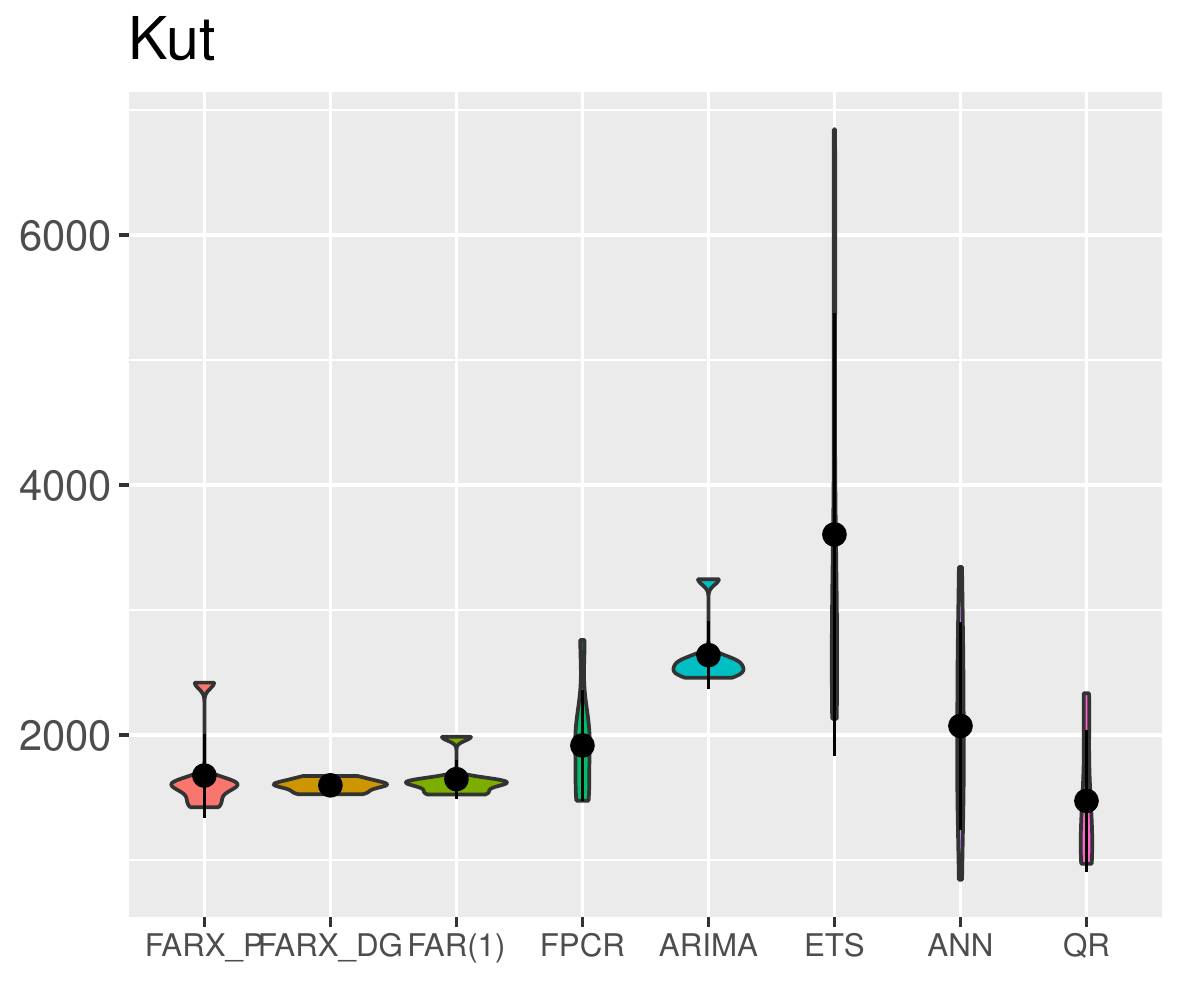}
  \\
  \includegraphics[width=4.8cm,height=5cm]{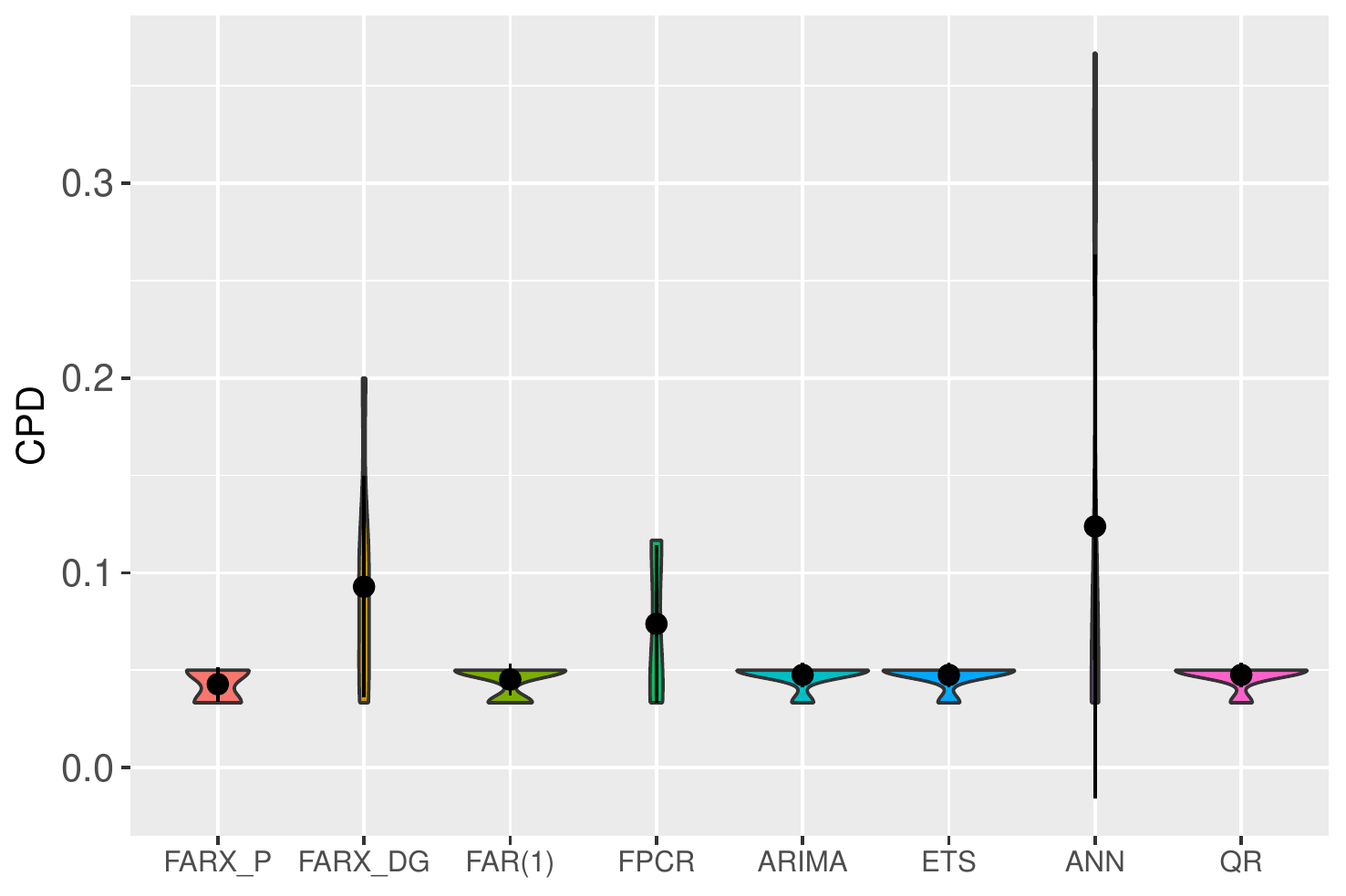}
  \includegraphics[width=4.8cm,height=5cm]{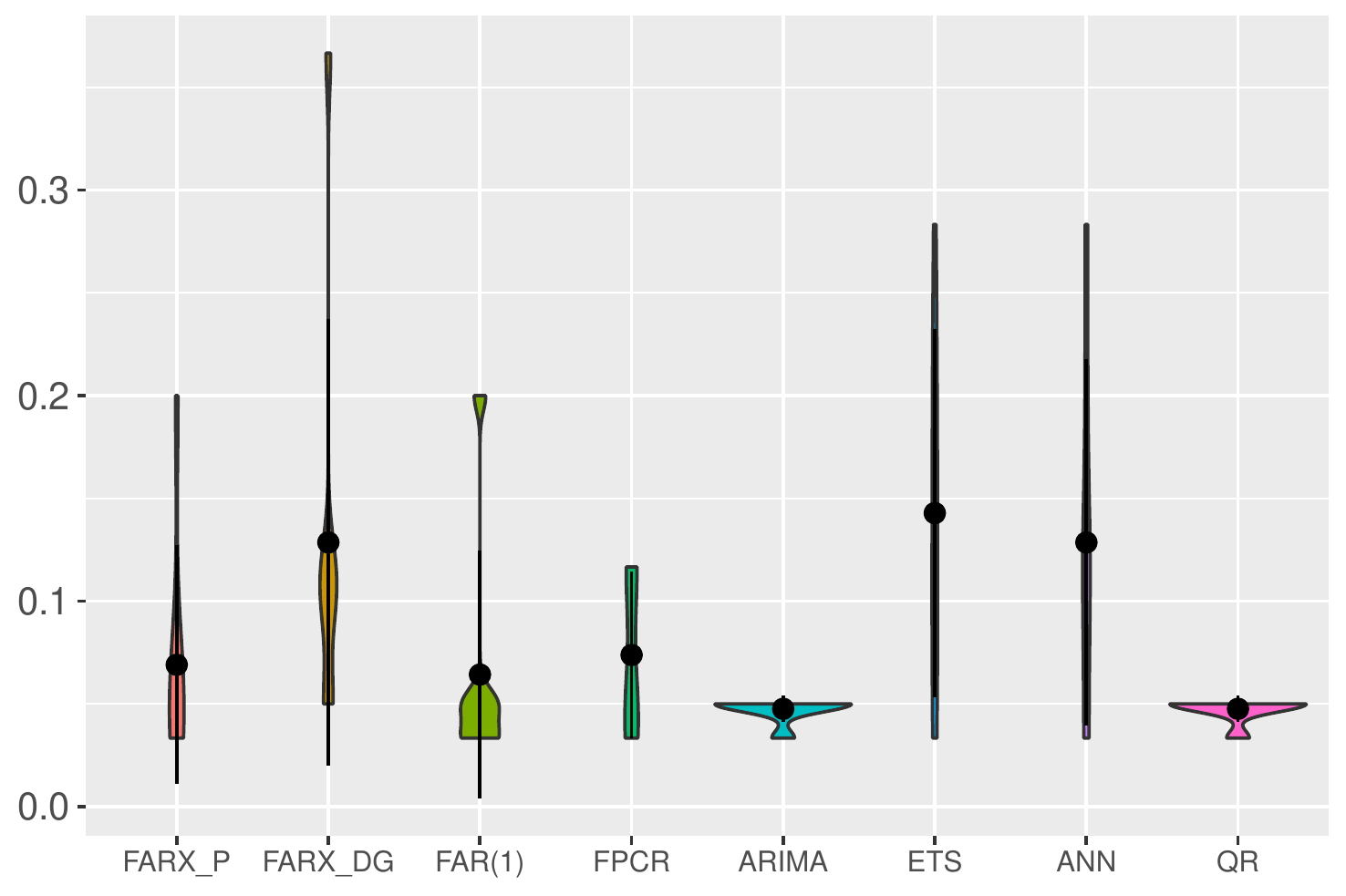}
  \includegraphics[width=4.8cm,height=5cm]{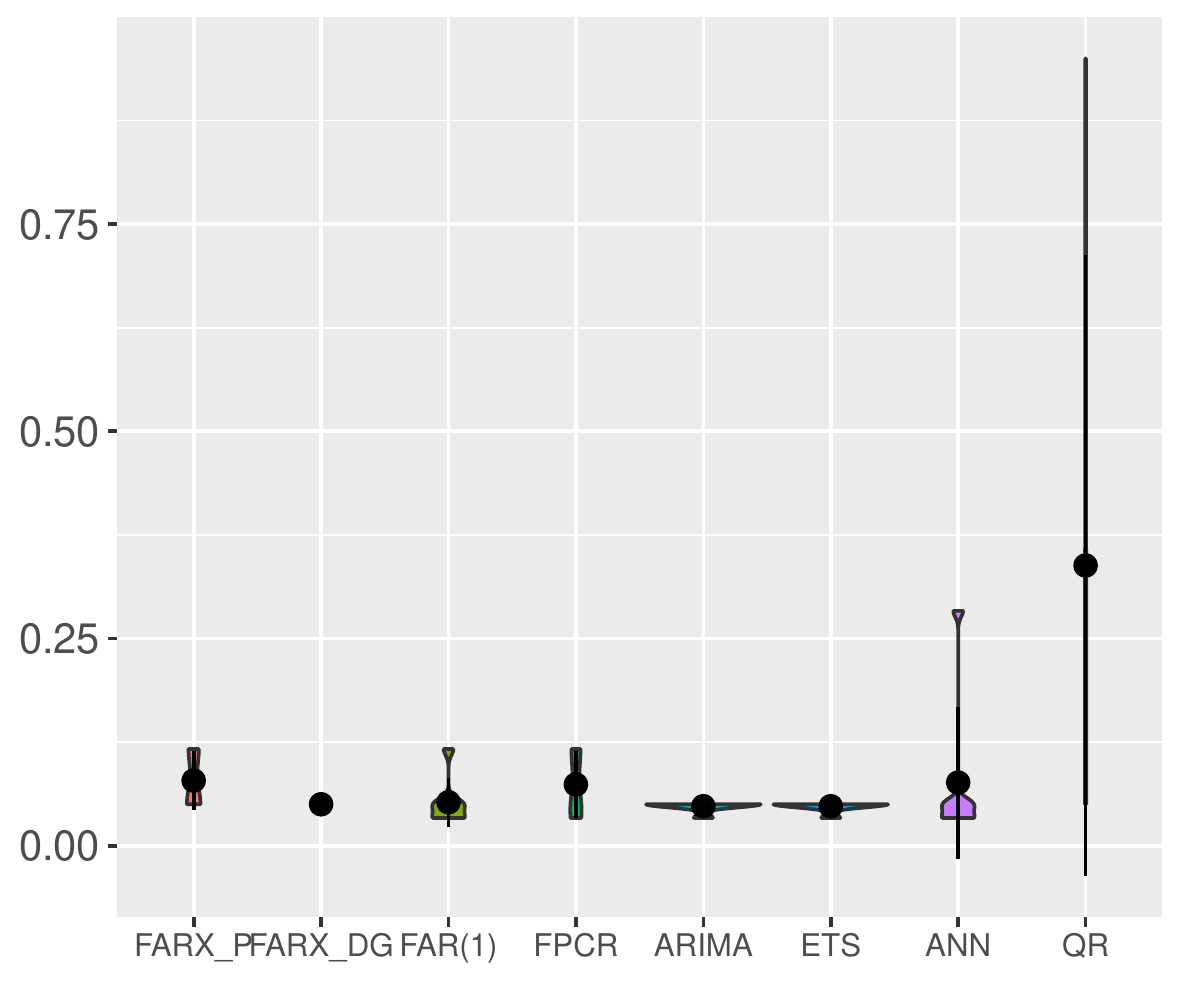}
  \caption{Violin plots of the calculated score and CPD values for the functional and non-functional (under the first scenario) models. The columns represent the stations, while the rows represent the performance metrics. In the plots, FARX\underline{ }P and FARX\underline{ }DG denote the proposed FARX(1) model and the FARX(1) model of \cite{Damon2002}, respectively.}
  \label{fig:Fig_11}
\end{figure}

\begin{figure}[!htbp]
  \centering
  \includegraphics[width=4.8cm,height=5cm]{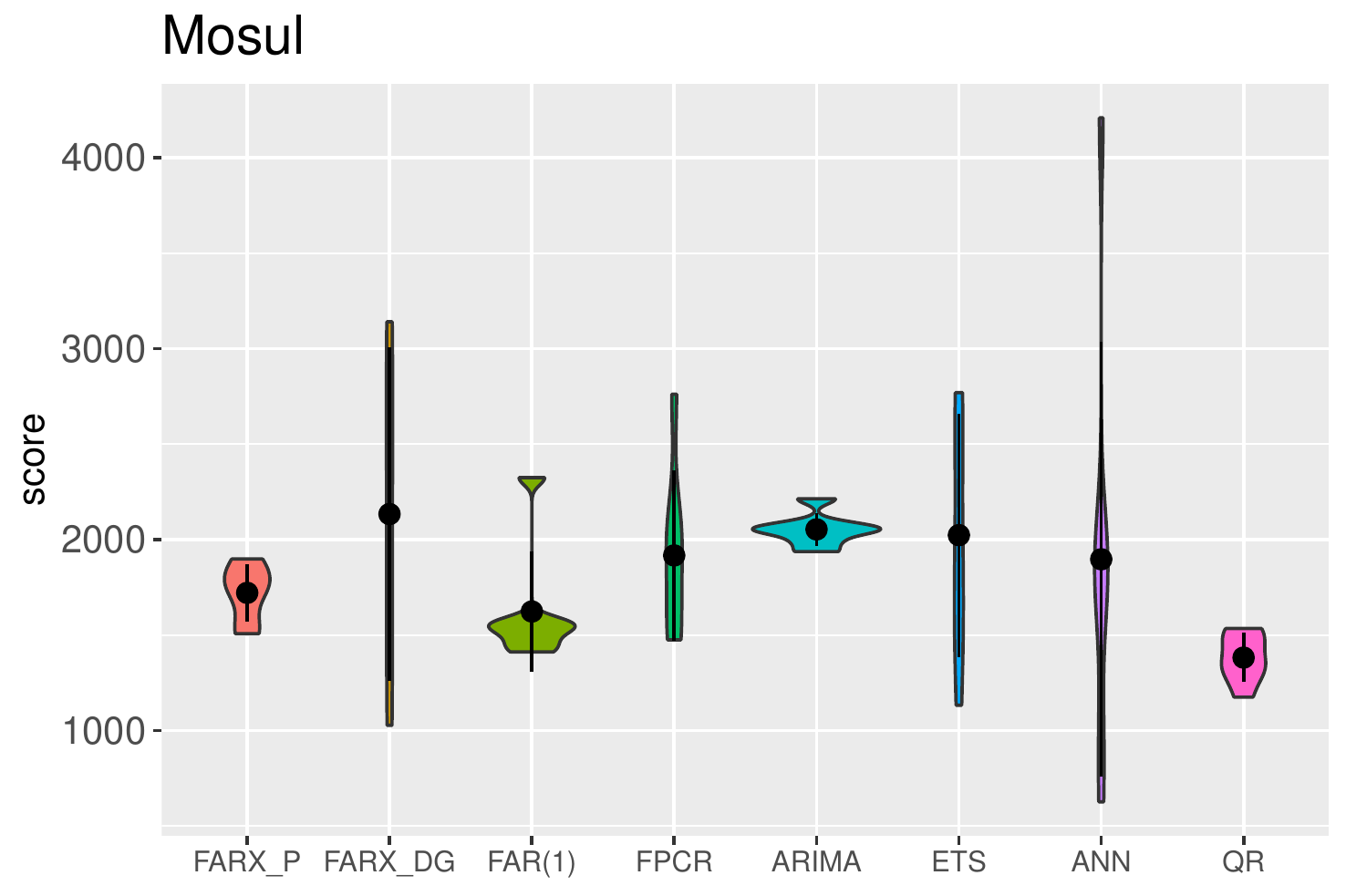}
  \includegraphics[width=4.8cm,height=5cm]{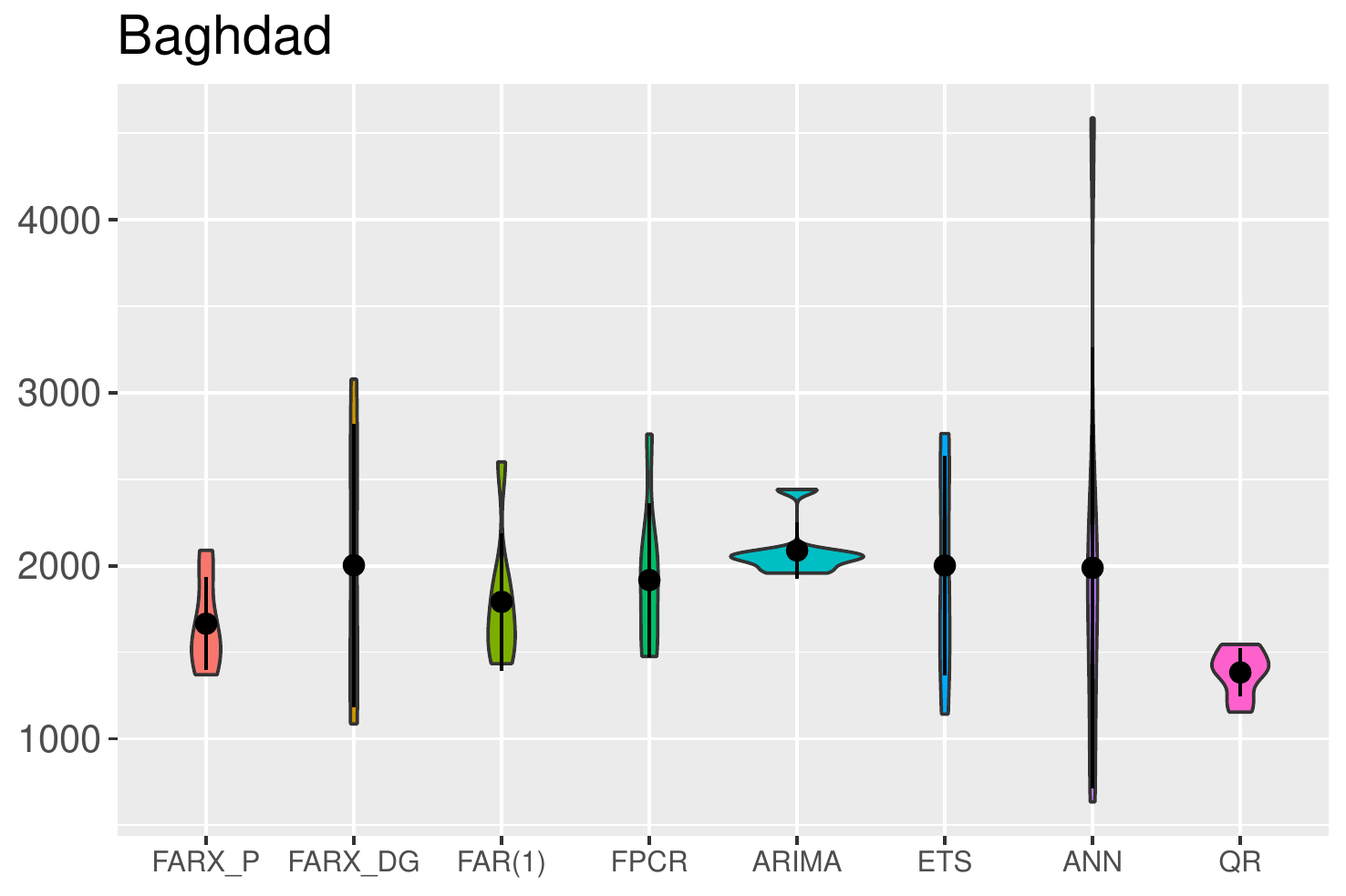}
  \includegraphics[width=4.8cm,height=5cm]{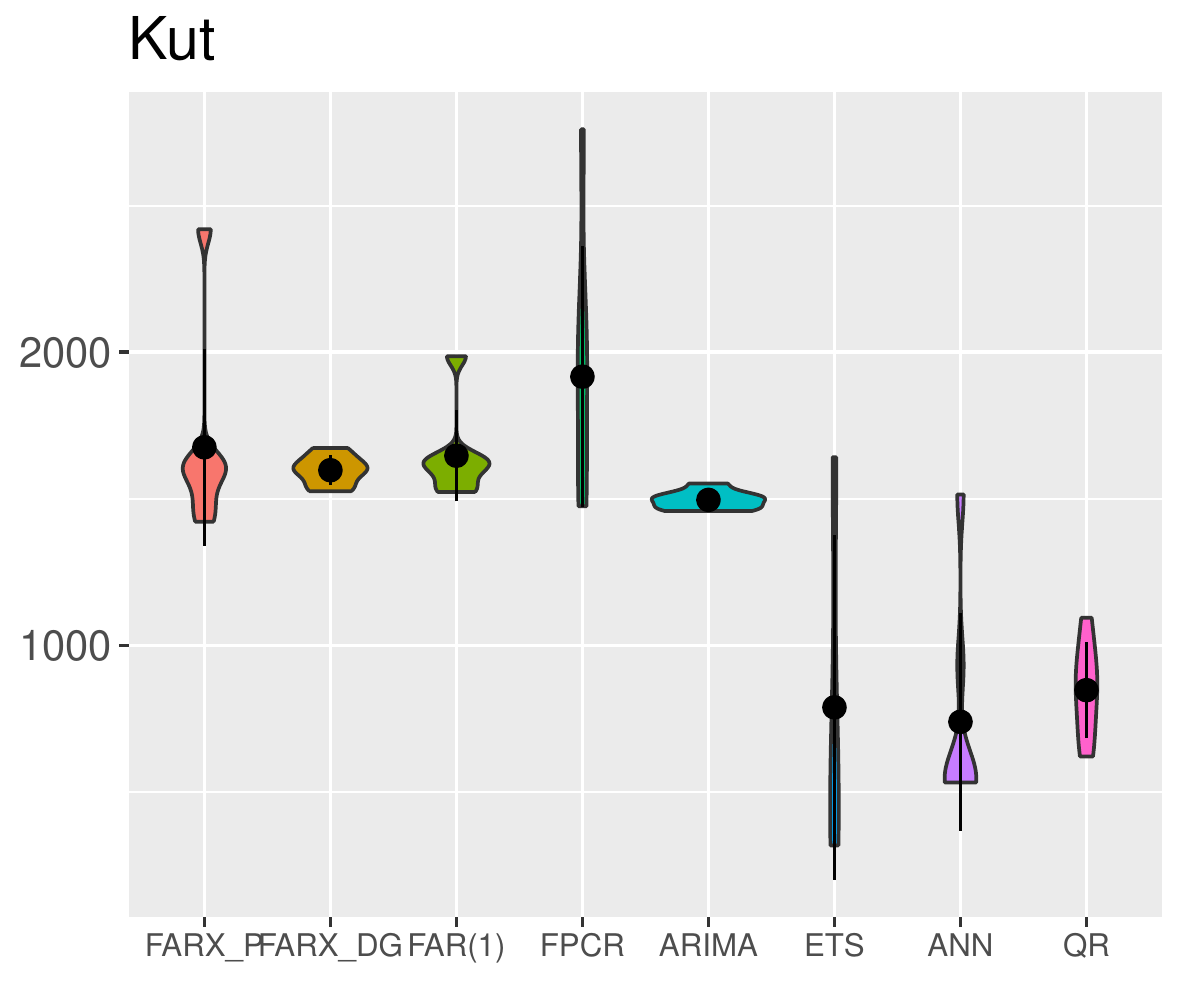}
  \\
  \includegraphics[width=4.8cm,height=5cm]{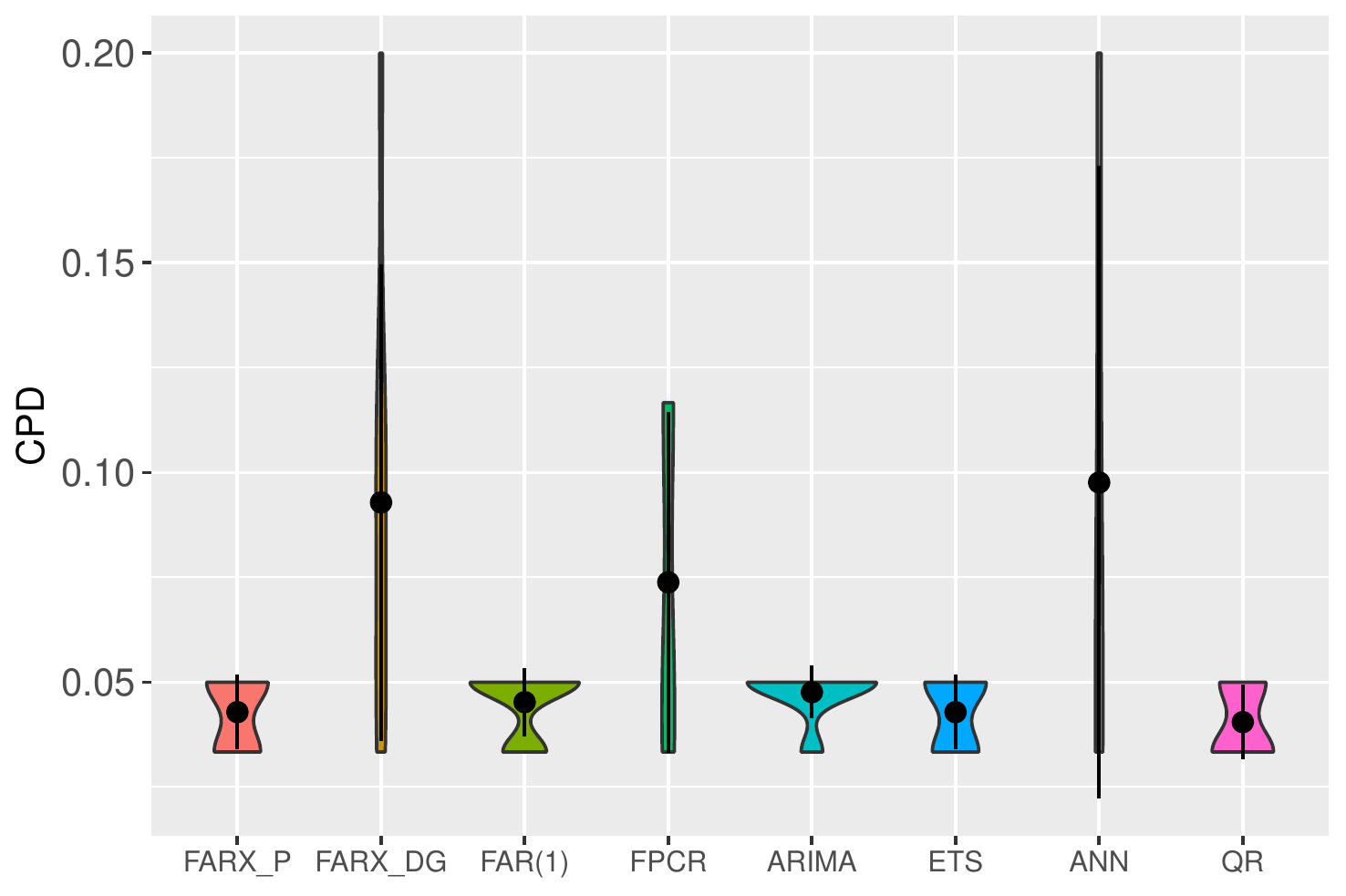}
  \includegraphics[width=4.8cm,height=5cm]{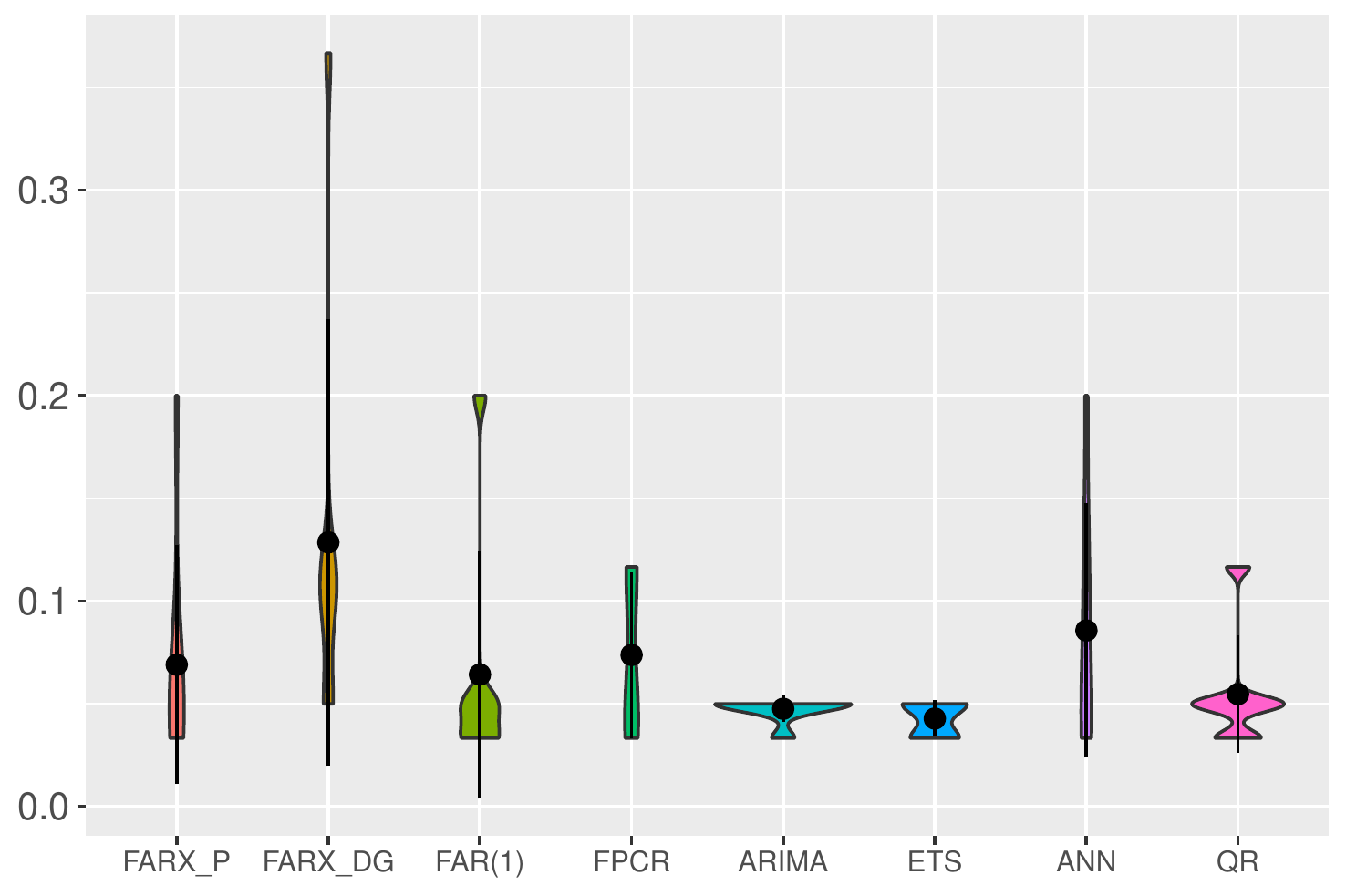}
  \includegraphics[width=4.8cm,height=5cm]{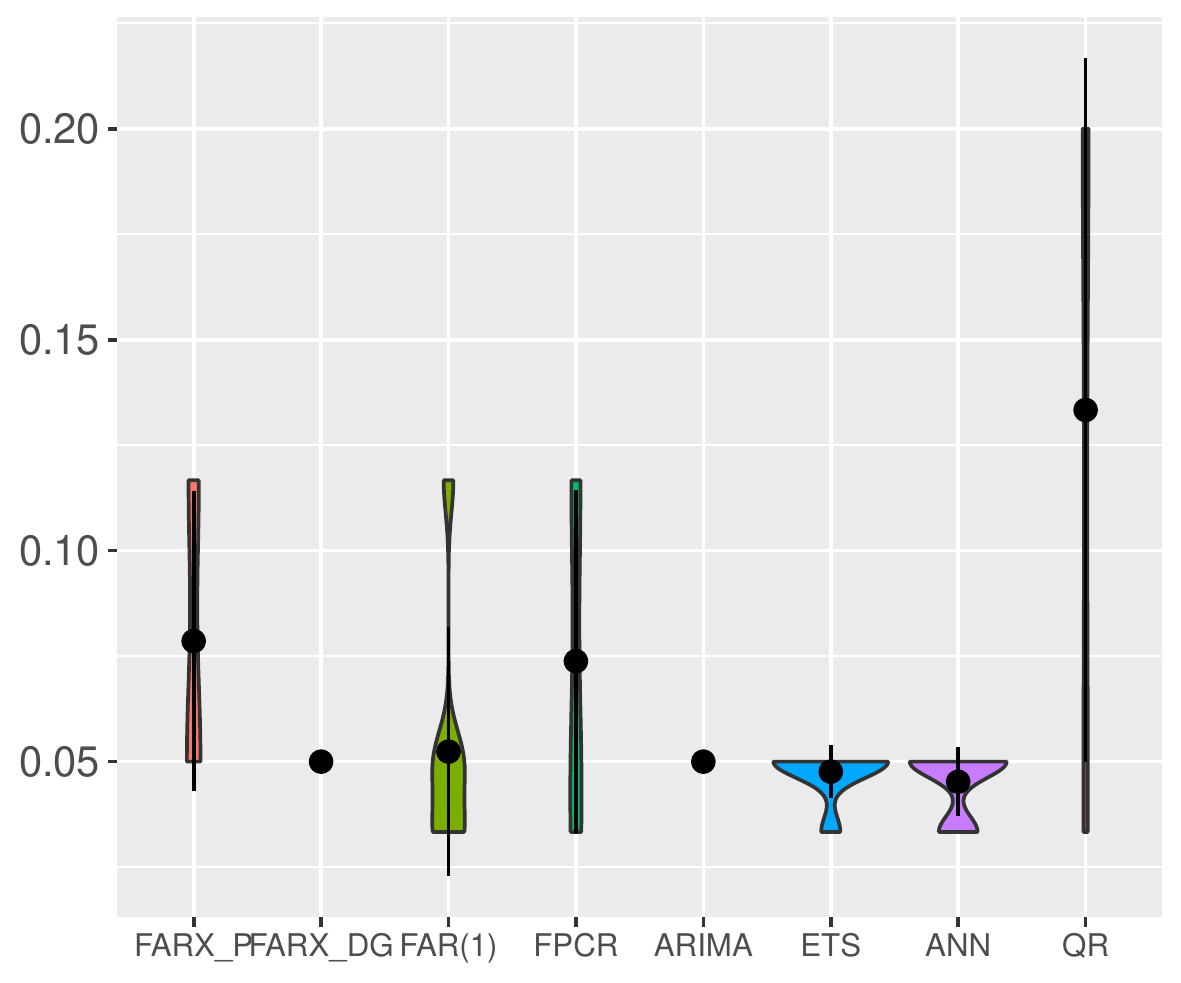}
  \caption{Violin plots of the calculated score and CPD values for the functional and non-functional (under the second scenario) models. The columns represent the stations, while the rows represent the performance metrics. In the plots, FARX\underline{ }P and FARX\underline{ }DG denote the proposed FARX(1) model and the FARX(1) model of \cite{Damon2002}, respectively.}
  \label{fig:Fig_12}
\end{figure}

The results of the computed statistical metrics are presented in Figures~\ref{fig:Fig_9}-\ref{fig:Fig_12}, in the form of violin plots. Figure~\ref{fig:Fig_9} presents the results of all prediction performance metrics (RMSPE, MAPE, RMESPE, PBIAS, and RE) obtained for both the functional and non-functional (under the first scenario) models. The results indicated that the proposed FARX(1) produced better prediction performance than the established benchmark models (for all stations). Thus, it has smaller RMSPE, MAPE, RMESPE, and PBIAS values than those of the FARX(1) model of \cite{Damon2002}, FAR(1), FPCR, ARIMA, ETS, ANN, and QR. The achieved records, presented in Figure~\ref{fig:Fig_9}, also showed that the FAR(1) model generally produced improved point forecast results over the FARX(1) model of \cite{Damon2002}. This is because the FAR(1) model parameters were estimated using the PLS method, which significantly improves the model's prediction performance. Among the traditional methods, the QR generally produced the best performance, and even it produced the best RE values among all methods. This is owing to the datasets that were used in the empirical analyses present outlying observations, and the QR with $\tau = 0.5$ (called median regression) is more robust to outliers compared with other methods (including the proposed FARX(1) model). Figure~\ref{fig:Fig_10} reports the calculated prediction performance metrics for the functional and non-functional (under the second scenario) models. This figure showed that the non-functional models produced improved prediction performance when the one-step-ahead prediction process is used compared with the first scenario. While the non-functional models produced competitive or even better performance than the functional models under the second scenario, they generally produced larger prediction error values than the proposed FARX(1) model. When considering Figures~\ref{fig:Fig_9} and~\ref{fig:Fig_10} together, the results demonstrated that the non-functional models produced improved performance over the existing functional models when they use the one-step-ahead prediction procedure. However, the proposed method produced better prediction results than the non-functional models under both scenarios.

\begin{figure}[!htbp]
  \centering
  \includegraphics[width=4.8cm,height=5cm]{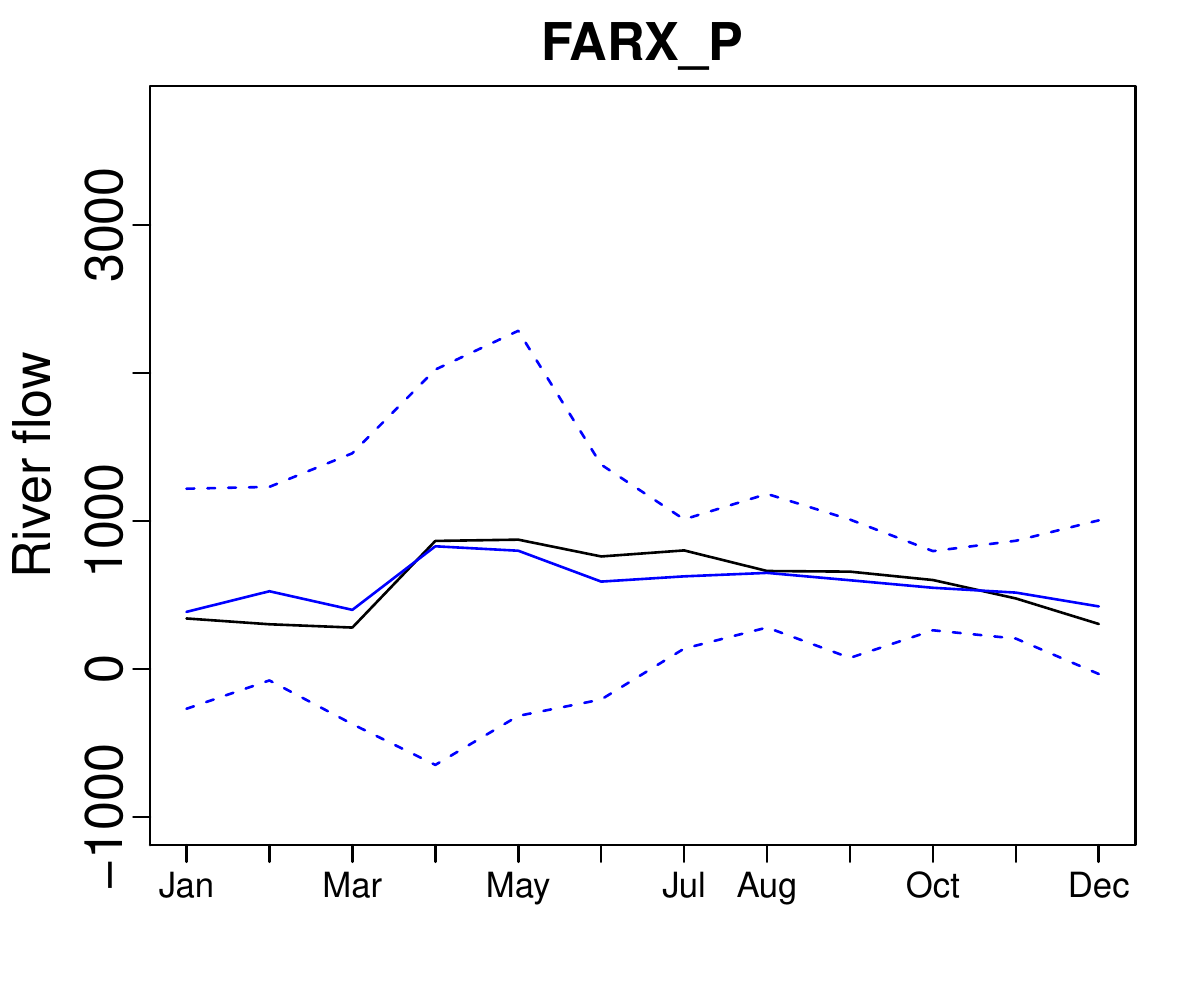}
  \includegraphics[width=4.8cm,height=5cm]{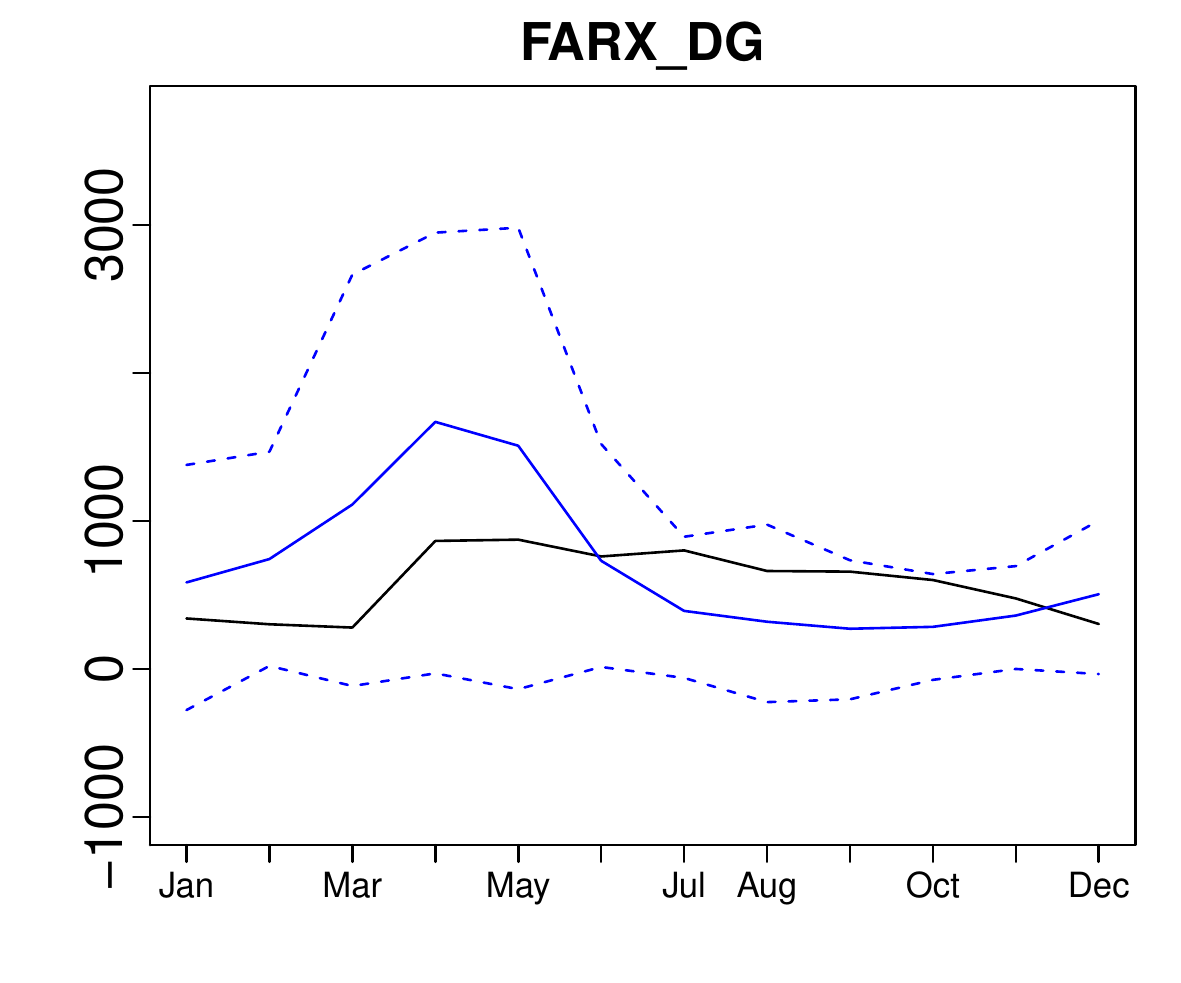}
  \includegraphics[width=4.8cm,height=5cm]{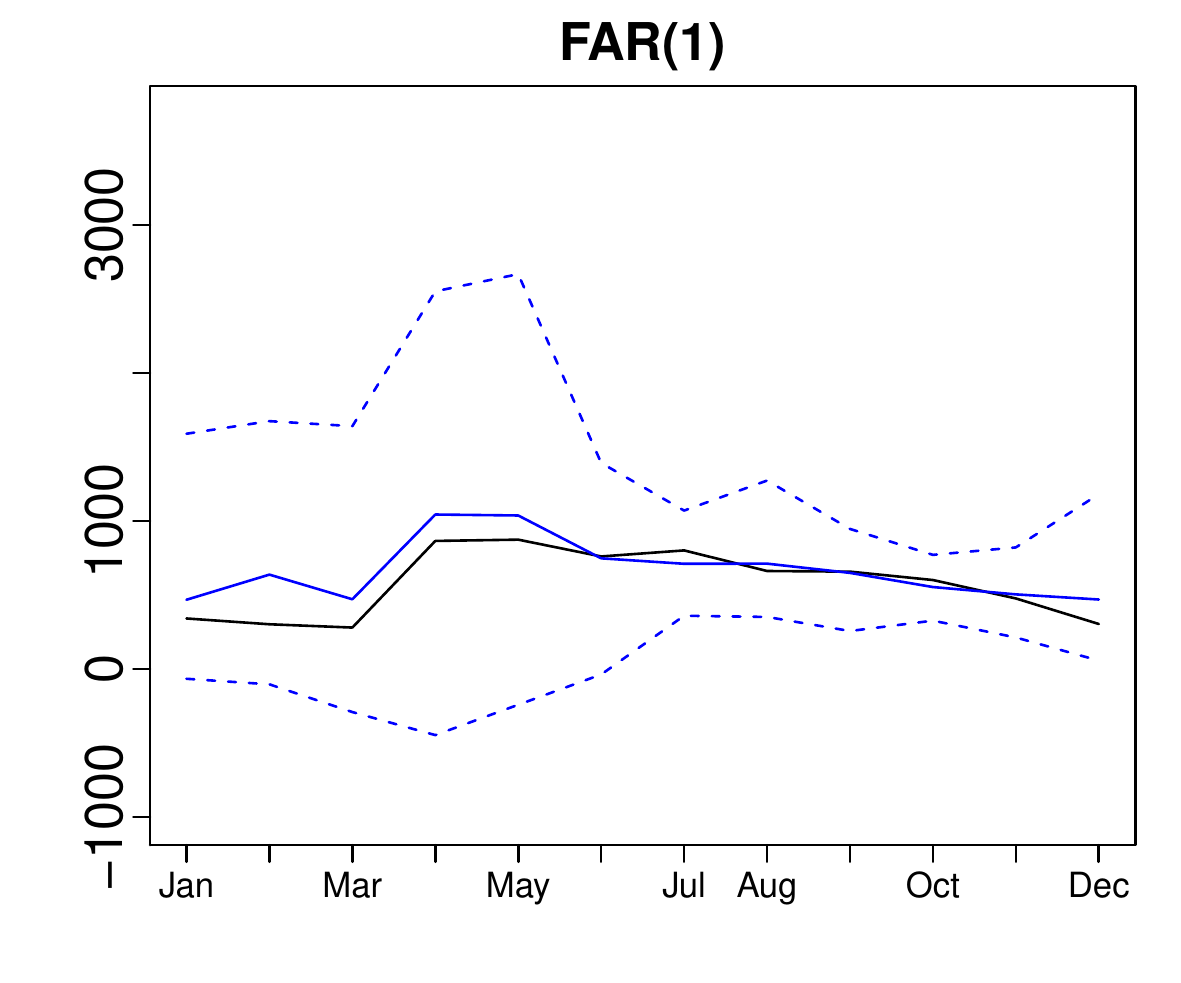}
  \\
  \includegraphics[width=4.8cm,height=5cm]{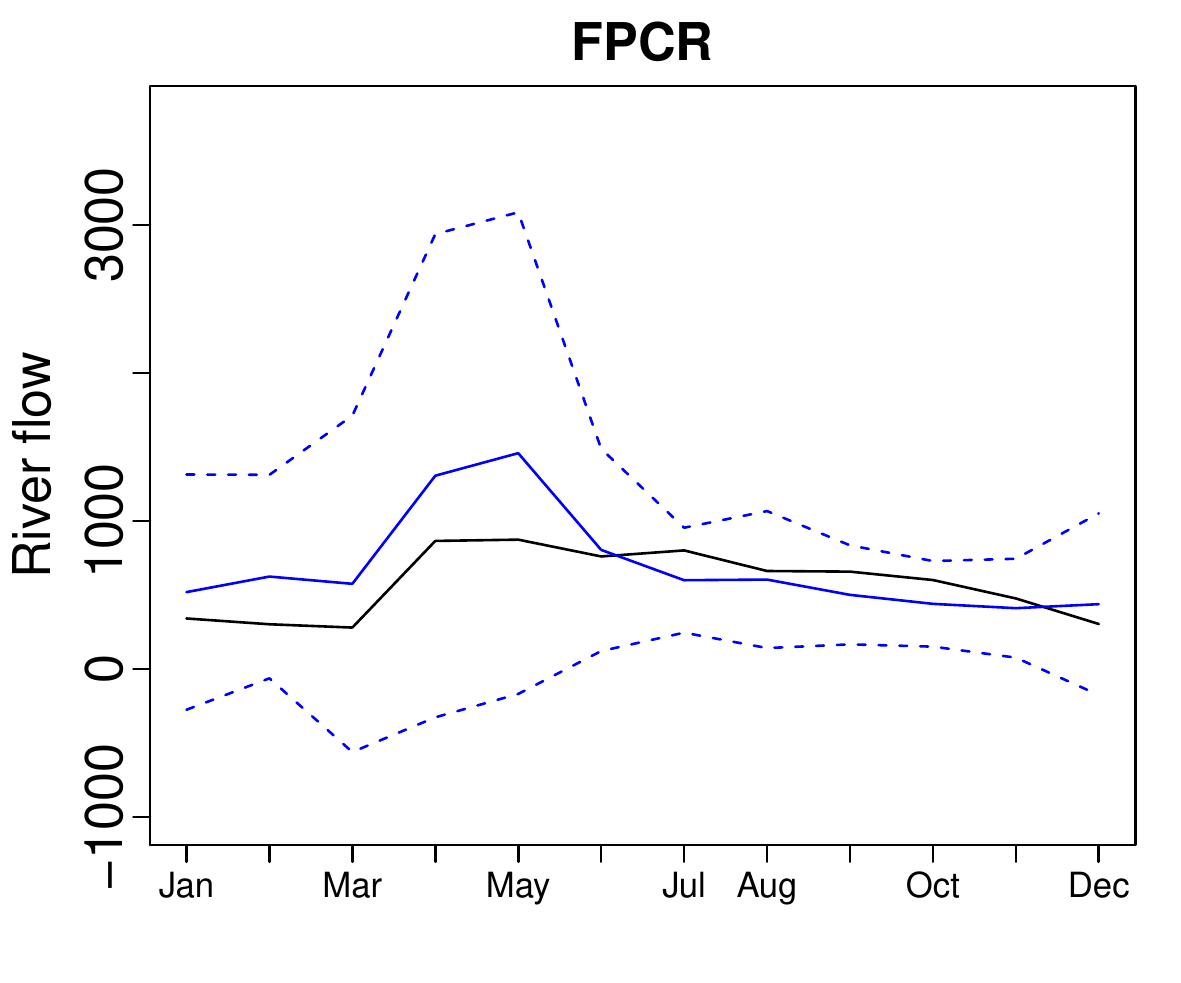}
  \includegraphics[width=4.8cm,height=5cm]{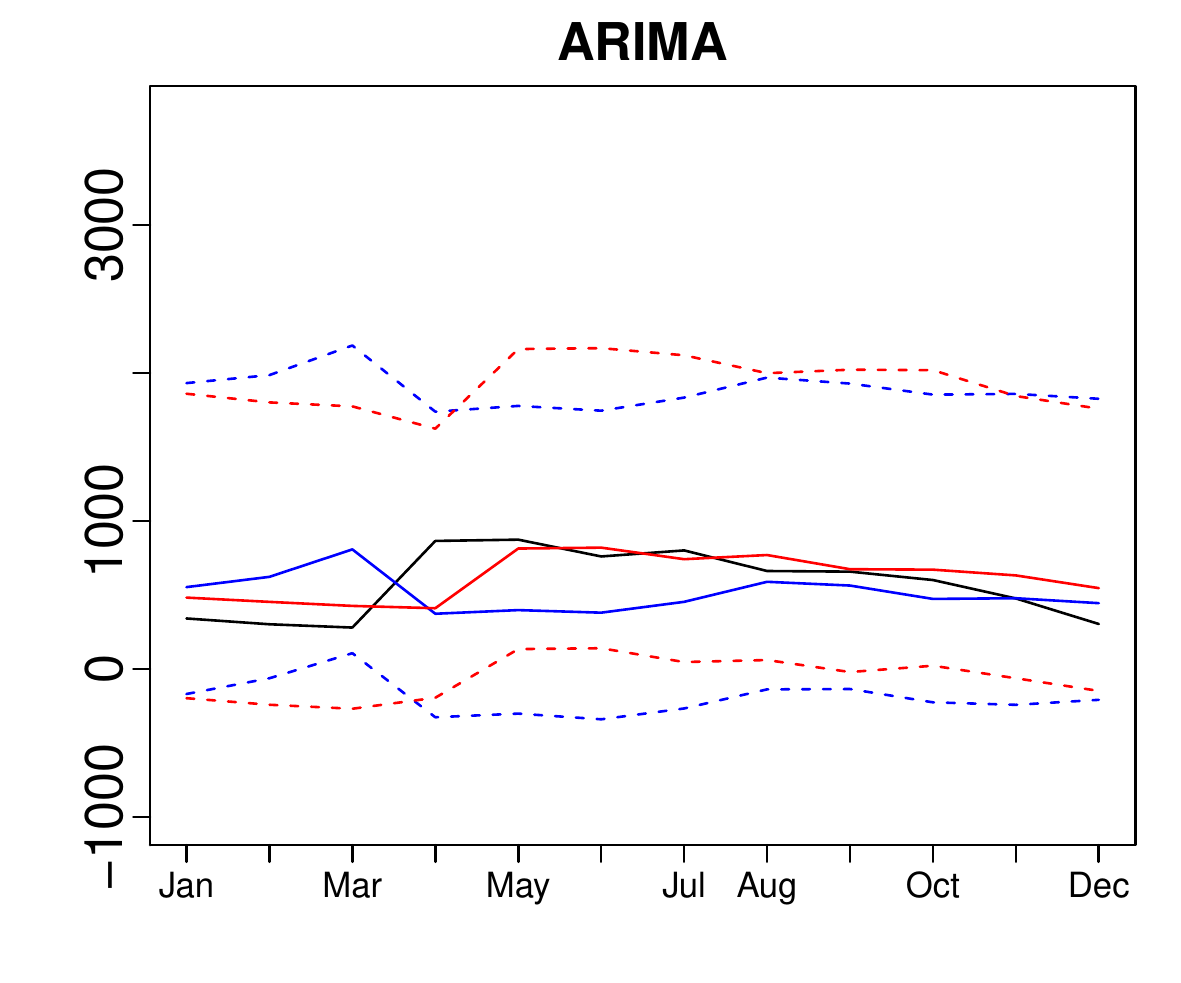}
  \includegraphics[width=4.8cm,height=5cm]{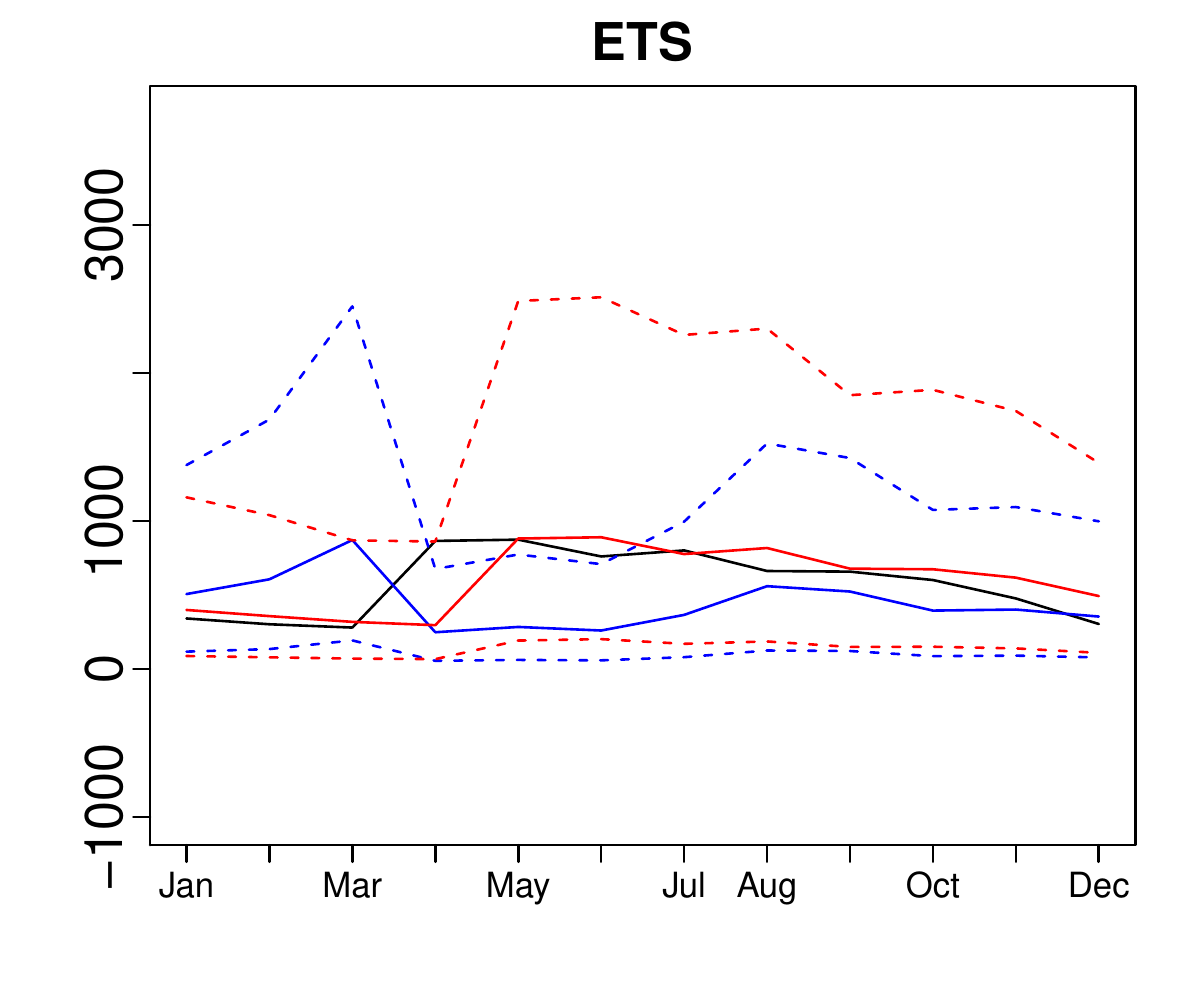}
  \\
  \includegraphics[width=4.8cm,height=5cm]{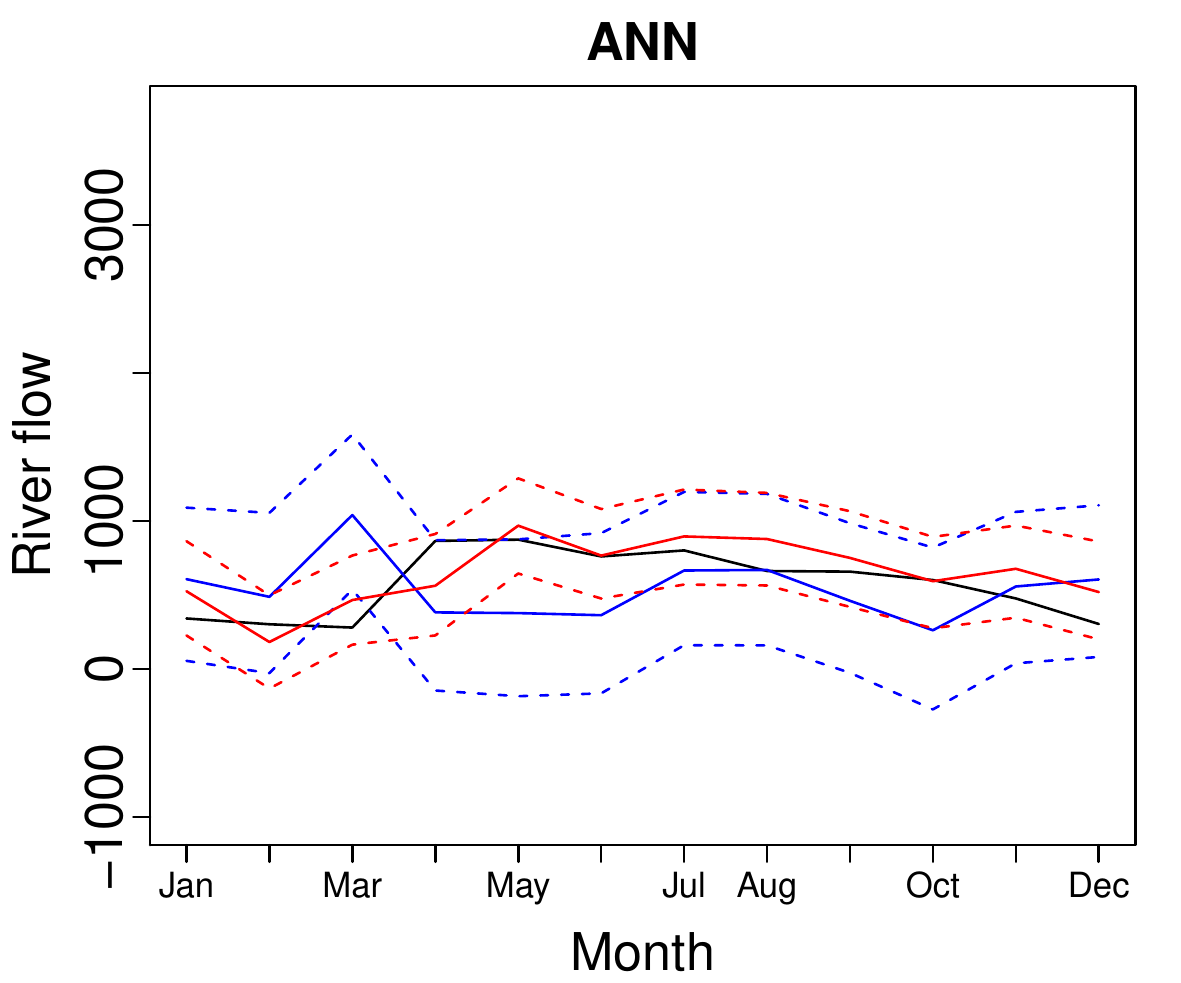}
  \includegraphics[width=4.8cm,height=5cm]{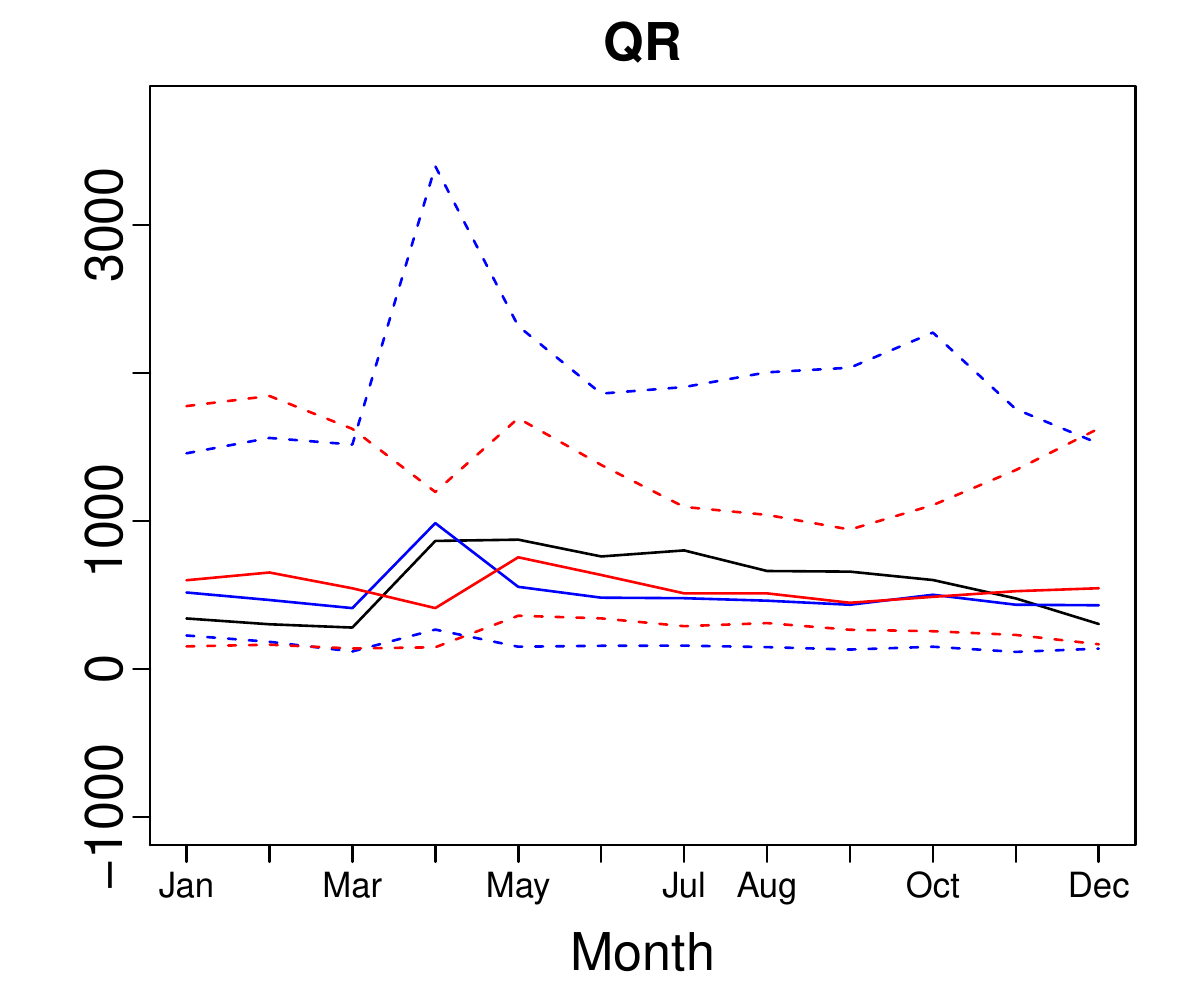}
  \caption{The one-step-ahead predictions for Baghdad station: observed function (black line), predicted function (blue (under the first scenario) and red (under the second scenario) solid lines), and 95\% bootstrap prediction intervals (blue (under the first scenario) and red (under the second scenario) dashed lines). In the plots, FARX\underline{ }P and FARX\underline{ }DG denote the proposed FARX(1) model and the FARX(1) model of \cite{Damon2002}, respectively.}
  \label{fig:Fig_13}
\end{figure}

The results of the interval score and CPD metrics for the functional and non-functional (under the first scenario) models are given in Figure~\ref{fig:Fig_11}. This figure demonstrates that the traditional ARIMA, ETS, and ANN models generally produced higher score values than functional and QR models. In other words, these models generally produced wider prediction intervals than the others. Compared with FPCR, FAR(1), and FARX(1) model of \cite{Damon2002}, the proposed model produced smaller interval score values among the functional models. Similar results were also obtained for the CPD values. However, Figure~\ref{fig:Fig_10} demonstrates that the proposed FARX(1) and FAR(1) have better prediction intervals compared with other models. Among the non-functional models, the QR produced the best score and CPD values. Figure~\ref{fig:Fig_12} presents the calculated interval score and CPD metrics for the functional and non-functional (under the second scenario). These results demonstrated that the non-functional models produced smaller interval score and CPD values than the functional models, including the proposed method. Among others, the QR tended to produce better prediction intervals (generally) since it considers the heteroscedasticity presented in the data. Under the second scenario, it is an expected result that non-functional models produce better interval scores and CPD values compared to functional models. The functional models produce long-term forecasts, which include more uncertainty compared with a one-step-ahead forecast. Figure~\ref{fig:Fig_13} presents graphical representations of the one-step-ahead prediction and 95\% bootstrap prediction intervals for the river flow curve time series in 2013 (Baghdad station).

Finally, the potential use of the proposed FARX(1) model in the field of hydrology should be emphasized. In hydrological modeling, statistical methods play an essential role in providing a practical assessment of the match between simulated and observed data. Traditional statistical models reflect the simulated data's competence using data points observed on a single time point. On the other hand, compared with traditional models, FDA models provide more reliable outputs since they use entire data observed (curve) over a continuum. In the time series context, most of the FDA models are based on only the lagged curves to model the data. However, hydrological time series are generally linked to other hydrometeorological variables. Thus, the models that use only the lagged variable/s may not produce reliable results in forecasting the data's future realizations. This paper proposes a novel FTS model that uses both lagged and other hydrometeorological variables in the model. Thus, it reflects the effects of exogenous variables on the predicted curves. Hence, the proposed modeling strategy produces improved prediction accuracy compared with other models. The proposed method can successfully predict other hydrological data that depend on other hydrometeorological variables. For example, one can predict drought occurrence using oscillation indexes as exogenous variables.

\section{Conclusion}\label{sec:5}

This study proposed a new FTS model for the prediction of river flow curve time series. The proposed model differs from the existing FTS models in terms of lagged variables used in the modeling phase. While the current models generally use only the lagged variables to construct a model, the proposed model allows for exogenous variables assumed to affect the river flow variable. In practice, the model's significant exogenous variables are unknown, and thus it was also proposed a forward procedure to select significant exogenous variables. The prediction performance of the proposed model was compared with those of existing functional and traditional models. The attained records have shown that the proposed model produces improved prediction accuracy than those of the existing models. Further, a nonparametric bootstrap procedure was proposed for further investigation of the uncertainty of predictions and to construct pointwise prediction intervals for the river flow curve time series. The results demonstrated that the proposed model produces better prediction intervals compared to its counterparts. It was observed that some of the functional datasets used in this study (see the right panels of Figures \ref{fig:Fig_6}-\ref{fig:Fig_8}) present heteroscedasticity, and thus, these datasets may not be representative for applying the proposed method because the proposed method does not model heteroscedasticity.

For future studies, the following extensions may be considered:
\begin{inparaenum}
\item[1)] The proposed model may be used to obtain forecasts of other hydro-climatic variables, such as drought, evapotranspiration, and precipitation.
\item[2)] It was considered only the $B$-spline basis expansion method to convert discretely observed data to functional form. Other basis functions, such as Fourier, wavelet, and radial basis functions, may also be considered.
\item[3)] Throughout the numerical analyses, it was considered an FPLSR approach to estimate the model parameter. Other estimation techniques, such as FPCR, may be used to estimate the model parameter.
\item[4)] The hydro-climatic time-series data, such as river flow, rainfall, and evaporation (please see Figures~\ref{fig:Fig_6}-\ref{fig:Fig_8}) may include outliers. The outliers may affect the prediction accuracy of the proposed model. A robust estimation technique may be combined with the proposed model to improve its prediction accuracy when outliers are present in the data.
\item[5)] In the numerical analyses, the proposed was applied to the seasonally differenced time series data. As a future study, a seasonal functional autoregressive time series model, such as the one proposed by \cite{Zamani}, may be adapted into the proposed method to predict seasonal FTS.
\item[6)] The empirical data analyses used only 37 years of river flow and hydrometeorological time series. However, to get more accurate results, the proposed method's finite-sample performance should be performed using large datasets. 
\item[7)] The finite-sample performance of the proposed method can also be compared with other existing methods that incorporate exogenous predictors, such as the ones studied in \cite{Tyralis}.
\end{inparaenum}





\bibliography{FARX} 

\end{document}